\documentclass[aps,prc,preprint,superscriptaddress,floatfix,nofootinbib]{revtex4}
\usepackage{graphicx}
\usepackage{dcolumn}
\usepackage{bm}

\begin{document}
\title{Single Identified Hadron Spectra from $\sqrt{s_{NN}}= 130$ GeV
Au+Au Collisions}

\newcommand{\acadsin}{Institute of Physics, Academia Sinica, Taipei 11529, Taiwan}
\newcommand{\banaras}{Department of Physics, Banaras Hindu University, Varanasi 221005, India}
\newcommand{\barc}{Bhabha Atomic Research Centre, Bombay 400 085, India}
\newcommand{\bnl}{Brookhaven National Laboratory, Upton, NY 11973-5000, USA}
\newcommand{\caucr}{University of California - Riverside, Riverside, CA 92521, USA}
\newcommand{\ciae}{China Institute of Atomic Energy (CIAE), Beijing, People\'s Republic of China}
\newcommand{\cns}{Center for Nuclear Study, Graduate School of Science, University of Tokyo, 7-3-1 Hongo, Bunkyo, Tokyo 113-0033, Japan}
\newcommand{\columbia}{Columbia University, New York, NY 10027 and Nevis Laboratories, Irvington, NY 10533, USA}
\newcommand{\fsu}{Florida State University, Tallahassee, FL 32306, USA}
\newcommand{\gsu}{Georgia State University, Atlanta, GA 30303, USA}
\newcommand{\hiroshima}{Hiroshima University, Kagamiyama, Higashi-Hiroshima 739-8526, Japan}
\newcommand{\ihepprot}{Institute for High Energy Physics (IHEP), Protvino, Russia}
\newcommand{\isu}{Iowa State University, Ames, IA 50011, USA}
\newcommand{\kek}{KEK, High Energy Accelerator Research Organization, Tsukuba-shi, Ibaraki-ken 305-0801, Japan}
\newcommand{\korea}{Korea University, Seoul, 136-701, Korea}
\newcommand{\kurchatov}{Russian Research Center ``Kurchatov Institute", Moscow, Russia}
\newcommand{\kyoto}{Kyoto University, Kyoto 606, Japan}
\newcommand{\lawllnl}{Lawrence Livermore National Laboratory, Livermore, CA 94550, USA}
\newcommand{\losalamos}{Los Alamos National Laboratory, Los Alamos, NM 87545, USA}
\newcommand{\lund}{Department of Physics, Lund University, Box 118, SE-221 00 Lund, Sweden}
\newcommand{\mcgill}{McGill University, Montreal, Quebec H3A 2T8, Canada}
\newcommand{\muenster}{Institut fuer Kernphysik, University of Muenster, D-48149 Muenster, Germany}
\newcommand{\myongji}{Myongji University, Yongin, Kyonggido 449-728, Korea}
\newcommand{\nagasaki}{Nagasaki Institute of Applied Science, Nagasaki-shi, Nagasaki 851-0193, Japan}
\newcommand{\newmex}{University of New Mexico, Albuquerque, NM, USA}
\newcommand{\nmsu}{New Mexico State University, Las Cruces, NM 88003, USA}
\newcommand{\ornl}{Oak Ridge National Laboratory, Oak Ridge, TN 37831, USA}
\newcommand{\pnpi}{PNPI, Petersburg Nuclear Physics Institute, Gatchina, Russia}
\newcommand{\riken}{RIKEN (The Institute of Physical and Chemical Research), Wako, Saitama 351-0198, JAPAN}
\newcommand{\rkrbrc}{RIKEN BNL Research Center, Brookhaven National Laboratory, Upton, NY 11973-5000, USA}
\newcommand{\saispbstu}{St. Petersburg State Technical University, St. Petersburg, Russia}
\newcommand{\saopaulo}{Universidade de S{\~a}o Paulo, Instituto de F\'isica, Caixa Postal 66318, S{\~a}o Paulo CEP05315-970, Brazil}
\newcommand{\stonybrkc}{Chemistry Department, State University of New York - Stony Brook, Stony Brook, NY 11794, USA}
\newcommand{\stonycrkp}{Department of Physics and Astronomy, State University of New York - Stony Brook, Stony Brook, NY 11794, USA}
\newcommand{\subatech}{SUBATECH (Ecole des Mines de Nantes, CNRS-IN2P3, Universit{\'e} de Nantes) BP 20722 - 44307, Nantes, France}
\newcommand{\tenn}{University of Tennessee, Knoxville, TN 37996, USA}
\newcommand{\titech}{Department of Physics, Tokyo Institute of Technology, Tokyo, 152-8551, Japan}
\newcommand{\tokyo}{University of Tokyo, Tokyo, Japan}
\newcommand{\tsukuba}{Institute of Physics, University of Tsukuba, Tsukuba, Ibaraki 305, Japan}
\newcommand{\vandy}{Vanderbilt University, Nashville, TN 37235, USA}
\newcommand{\waseda}{Waseda University, Advanced Research Institute for Science and Engineering, 17 Kikui-cho, Shinjuku-ku, Tokyo 162-0044, Japan}
\newcommand{\weizmann}{Weizmann Institute, Rehovot 76100, Israel}
\newcommand{\yonsei}{Yonsei University, IPAP, Seoul 120-749, Korea}
\newcommand{\zindkfki}{ KFKI Research Institute for Particle and Nuclear Physics (RMKI), Budapest, Hungary}
\affiliation{\acadsin}
\affiliation{\banaras}
\affiliation{\barc}
\affiliation{\bnl}
\affiliation{\caucr}
\affiliation{\ciae}
\affiliation{\cns}
\affiliation{\columbia}
\affiliation{\fsu}
\affiliation{\gsu}
\affiliation{\hiroshima}
\affiliation{\ihepprot}
\affiliation{\isu}
\affiliation{\kek}
\affiliation{\korea}
\affiliation{\kurchatov}
\affiliation{\kyoto}
\affiliation{\lawllnl}
\affiliation{\losalamos}
\affiliation{\lund}
\affiliation{\mcgill}
\affiliation{\muenster}
\affiliation{\myongji}
\affiliation{\nagasaki}
\affiliation{\newmex}
\affiliation{\nmsu}
\affiliation{\ornl}
\affiliation{\pnpi}
\affiliation{\riken}
\affiliation{\rkrbrc}
\affiliation{\saispbstu}
\affiliation{\saopaulo}
\affiliation{\stonybrkc}
\affiliation{\stonycrkp}
\affiliation{\subatech}
\affiliation{\tenn}
\affiliation{\titech}
\affiliation{\tokyo}
\affiliation{\tsukuba}
\affiliation{\vandy}
\affiliation{\waseda}
\affiliation{\weizmann}
\affiliation{\yonsei}
\affiliation{\zindkfki}
\author{K.~Adcox}	\affiliation{\vandy}
\author{S.S.~Adler}	\affiliation{\bnl}
\author{N.N.~Ajitanand}	\affiliation{\stonybrkc}
\author{Y.~Akiba}	\affiliation{\kek}
\author{J.~Alexander}	\affiliation{\stonybrkc}
\author{L.~Aphecetche}	\affiliation{\subatech}
\author{Y.~Arai}	\affiliation{\kek}
\author{S.H.~Aronson}	\affiliation{\bnl}
\author{R.~Averbeck}	\affiliation{\stonycrkp}
\author{T.C.~Awes}	\affiliation{\ornl}
% ...
\collaboration{PHENIX Collaboration} \noaffiliation

\date{\today}
\begin{abstract}
Transverse momentum spectra and yields of hadrons are measured by the
PHENIX collaboration in Au + Au collisions at $\sqrt{s_{NN}} = 130$ GeV
at the Relativistic Heavy Ion Collider (RHIC). The time-of-flight 
resolution allows identification of pions
to transverse momenta of 2 GeV/c and protons  and antiprotons to 4
GeV/c.  The yield of pions rises approximately linearly with the
number of nucleons participating in the collision, while the number of
kaons, protons, and antiprotons increases more rapidly.  The shape of
the momentum distribution changes between peripheral and central
collisions.  Simultaneous analysis of all the $p_T$ spectra indicates
radial collective expansion, consistent with
predictions of hydrodynamic models.  Hydrodynamic analysis of the
spectra shows that the expansion velocity increases with collision
centrality and collision energy.  This expansion boosts the particle
momenta, causing the yield from soft processes to exceed that for 
hard to large transverse momentum, perhaps as large as 3 GeV/c.
\end{abstract}
\maketitle

\section{\label{intro}INTRODUCTION}
Heavy ion reactions at ultrarelativistic energies provide information
on strongly interacting matter under extreme conditions. 
Lattice QCD and phenomenological predictions indicate 
that at high enough energy density a
deconfined state of quarks and gluons, the quark-gluon-plasma,
is formed.  It is expected that conditions in
ultrarelativistic heavy ion reactions may 
produce this new state of matter, the study of which 
is the major goal of the experiments at the
Relativistic Heavy Ion Collider (RHIC).

The high energy density state thus created will cool down and expand,
undergoing a phase transition to ``ordinary'' hadronic matter.
While the tools of choice to study the earliest phase of the
reactions, and thereby the new state, are probes that do not 
interact via the strong force, such as photons,
electrons, or muons, the global properties and dynamics of later 
stages in the system are best studied via hadronic observables.
Hadron momentum spectra in proton-proton reactions are often separated into
two parts, a soft part at low transverse momentum ($p_{T}$), 
where the shape is roughly
exponential in transverse mass $m_{T} = \sqrt{p_T^2 + m_0^2}$, and a high $p_T$ region where the
shape more closely resembles a power law.  
Soft production (low $p_{T}$) is 
attributed to fragmentation of a string \cite{string1,string2} 
between components of the struck nucleons, 
while hard (high $p_{T}$) hadrons are expected to originate predominantly from
fragmentation of hard-scattered partons. 
The transition between
these two regimes is not sharply defined, but is commonly believed
to be near $p_T \approx$ 2 GeV/c~\cite{owens}.

In proton-nucleus (p+A) scattering, these two regimes depend on the 
colliding system size in different ways. The soft production depends 
on the number of nucleons struck, or participating in
the collision ($N_{part}$).  The number of hard scatterings should increase 
proportionally to the
number of binary nucleon-nucleon encounters ($N_{coll}$)
since these processes
have a small elementary cross section and may be considered as 
incoherent.  Hard scattering 
also produces color strings which fragment and produce some low $p_T$ 
particles, though these are much fewer in number than those from the 
much more frequent soft scatterings.  In p+A these $N_{part}$ and $N_{coll}$ are connected by a very 
simple relation, namely $N_{part} = N_{coll} + 1$.


In nucleus-nucleus
collisions, the number of participant nucleons does not scale simply
with A, so it is more useful to study scaling with $N_{coll}$ or
$N_{part}$. Collisions are sorted according to centrality, allowing
control of the geometry and determination of   $N_{coll}$ or
$N_{part}$.

In heavy ion collisions, one expects secondary collisions of particles
(rescattering)  to take place, especially among particles with low and
intermediate transverse momentum. Rescattering may occur among 
partons early in the collision, and also among hadrons later in
the collision. Both  kinds of rescattering can lead to collective
behavior among the particles, and the presence of elliptic flow
(\cite{star_ellflow1,star_ellflow2,star_ellflow3,star_ellflow4,phenix_ellflow,phobos_ellflow}) indicates that partonic rescattering is important at RHIC.
In the extreme, rescattering can lead  to thermalization. Rescattering
has observable consequences on the final hadron momentum spectra,
causing them to be broadened as shown in this paper.  This relates to
some of the key questions regarding the evolution of the collision:
Are the size and lifetime sufficient to attain local equilibrium?  Are
the momentum distributions thermal, and if so, what are the 
chemical and kinetic freeze-out temperatures?  Can expansion be
described by hydrodynamic models?  Momentum distributions of hadrons 
as a function of centrality provide a means to
investigate these questions and permit extraction of thermodynamic
quantities  which govern the predicted phase transition.  

This paper reports semi-inclusive momentum spectra and yields
of $\pi$, K, and p from Au-Au collisions at $\sqrt{s_{NN}}=130$ GeV.  The
data are measured and analyzed by the PHENIX Collaboration
in the first year of the physics program at RHIC (Run-1).

The paper is organized as follows. In Section~\ref{expt} the PHENIX detectors used 
in the analysis are described.  The data reduction techniques
using the Time-of-Flight and Drift Chamber detectors, along with the
corrections applied to the spectra, are described in Section
~\ref{cuts}.  Functions that describe the shape of the spectra are
used to extrapolate the unmeasured portion in order to determine the
total average momentum and particle yield for each particle. The
overall systematic uncertainties in the spectra are discussed.  
The resulting minimum bias and centrality-selected particle spectra 
are presented in Section ~\ref{results}.  
In Section \ref{models} a description of the particle production within
a hydrodynamic picture is investigated.  For each centrality selection, a
hydrodynamic  parameterization of the $m_T$ distribution
is fit simultaneously to the spectra of different species.  
The data are compared to full hydrodynamic calculations.
The transition region in $p_{T}$ between hard 
(perturbative QCD) and soft (hydrodynamic behavior) physics 
is investigated by comparison of extrapolated soft
spectra to the data.  Finally, we study the 
dependence of the particle yields on the number of 
nucleons participating in the collision.

\section{\label{expt}EXPERIMENT}
\begin{figure} [ht]
\resizebox{\columnwidth}{!}{\includegraphics{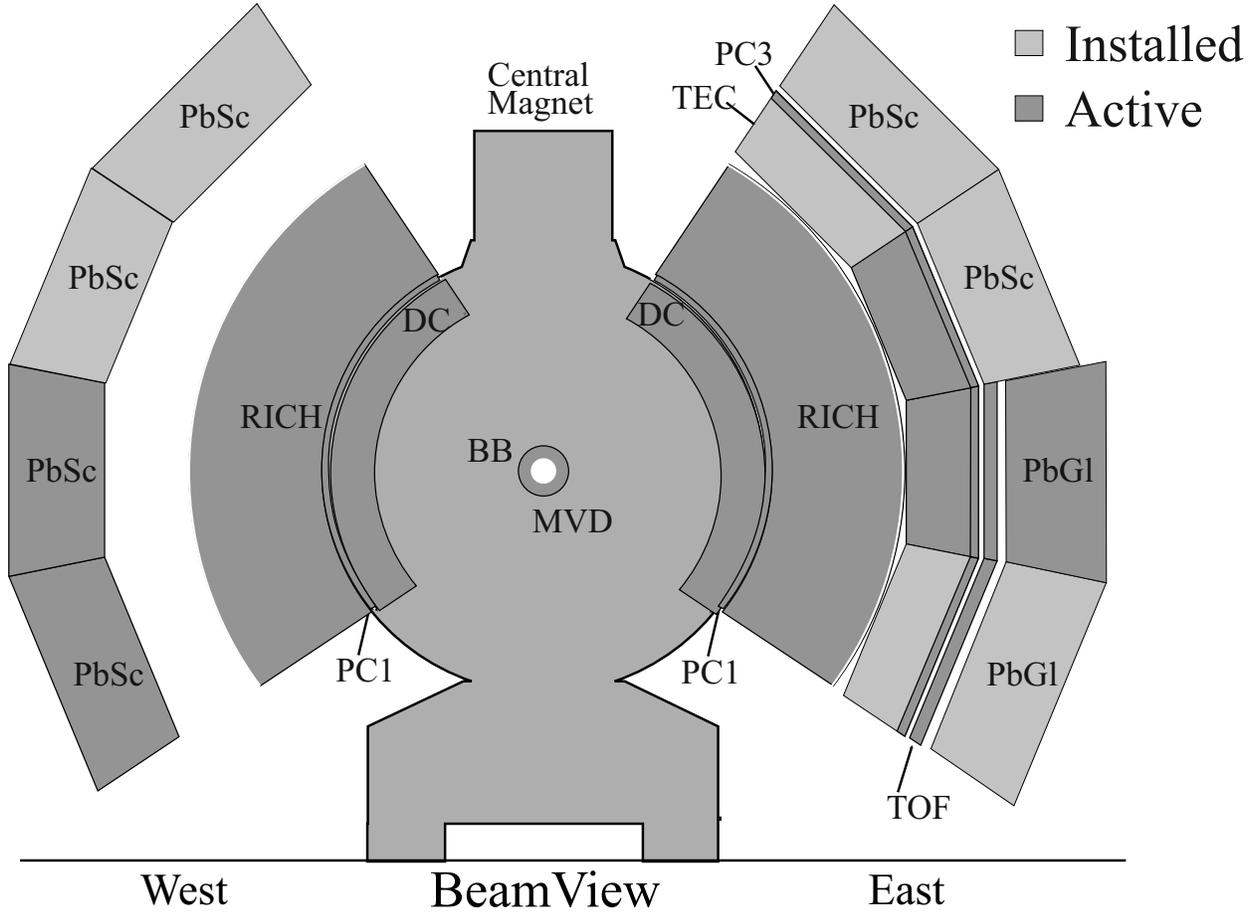}}
\caption{\label{Fig:central_arms}
A cross-sectional view of the PHENIX detector, transverse to 
the beamline.  Within the two central arm spectrometers the detectors 
that were instrumented and operational during the $\sqrt{s_{NN}} = 130$ GeV 
run are shown.}
\end{figure}

The PHENIX \cite{PHENIX,PHENIX_QM01} experiment at RHIC identifies
hadrons over a large momentum range, by the addition of excellent
time-of-flight capability to the detector suite optimized for
photons, electrons, and muons. 
PHENIX has four
spectrometer arms, two that are positioned  about midrapidity (the
central arms) and two at more forward  rapidities (the Muon Arms).
A cross-sectional view of the PHENIX detector, transverse to the beamline 
is shown in Figure~\ref{Fig:central_arms}.  Within the two central arm 
spectrometers, the detectors that were instrumented and operational during the 
$\sqrt{s_{NN}} = 130$ GeV run (Run-1) are shown.
The detector systems in  PHENIX are discussed in detail elsewhere
\cite{PHENIX_NIM}. 
The detector systems used for the measurements reported in this paper
are described in detail in the following sections.

\subsection{CENTRAL ARM DETECTORS}

\begin{figure} [ht]
\resizebox{\columnwidth}{!}{\includegraphics{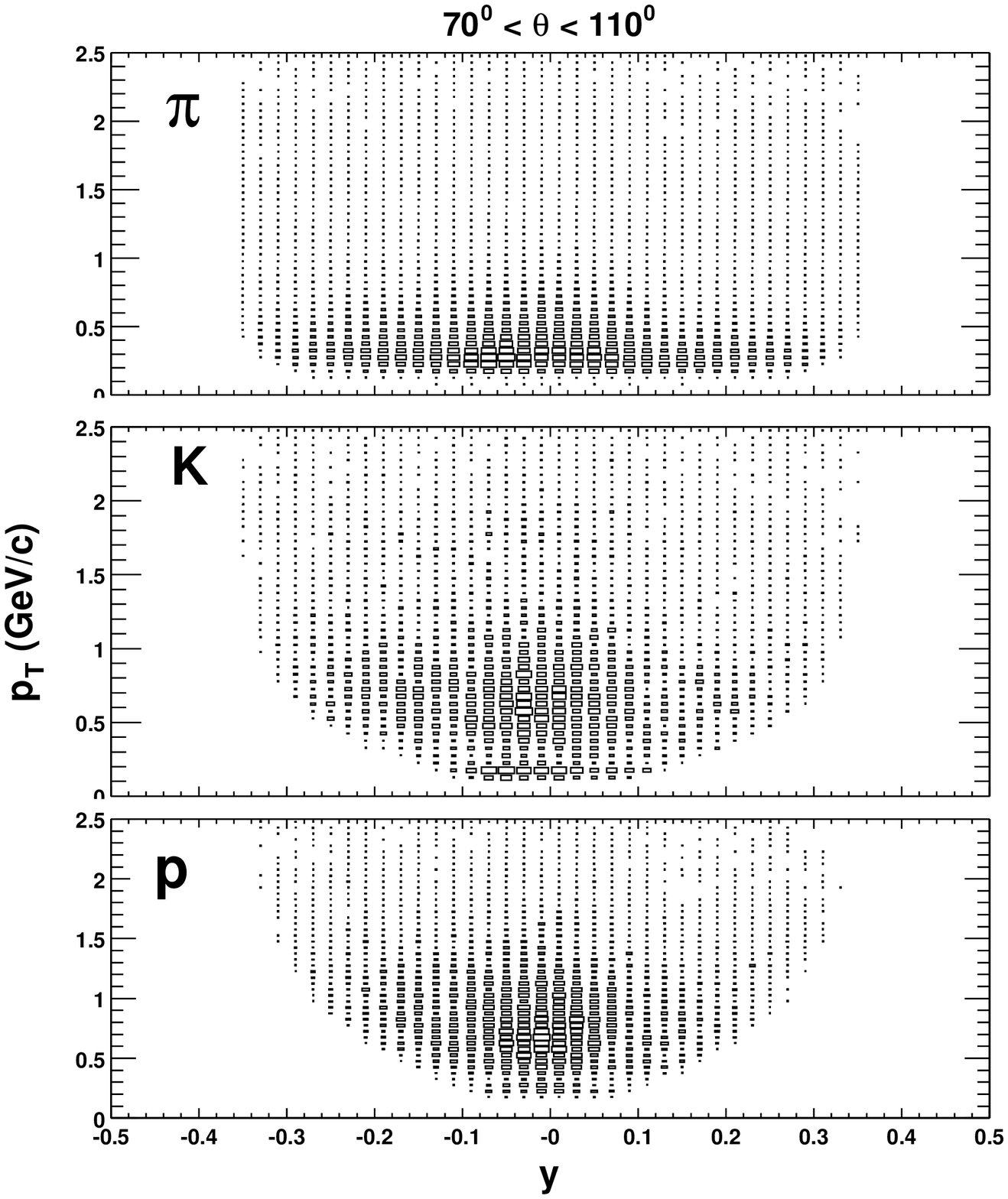}}
\caption{\label{Fig:acceptance}
The central arm spectrometer acceptance in rapidity and
transverse momentum for pions (top), kaons (middle), and protons (bottom).}
\end{figure}

The central arm spectrometers use a central magnet that produces an
approximately  axially symmetric field that focuses charged particles
into the detector acceptance.  The two central arms are labeled as
East and West Arms.  The East Arm contains the following subsystems
used in this analysis:  drift chamber (DC), pad chamber (PC), and a
Time-of-Flight (TOF) wall.  The PHENIX hadron acceptance using the TOF
system in the East Arm is  illustrated in Figure~\ref{Fig:acceptance}
where the transverse momentum  is plotted as a function of the
particle rapidity (the phase space)  within the central arm acceptance
subtending the polar angle $\theta$  from 70 to 110 degrees for pions,
kaons, and protons.  The vertical lines are the equivalent
pseudorapidity edges, corresponding to $\vert \eta \vert < 0.35$.
More details are discussed elsewhere  \cite{JEFF}.

\subsubsection{TRACKING CHAMBERS}
The charged particle tracking chambers include three layers of pad
chambers and two drift chambers.  The chambers are 
designed to operate in a high particle multiplicity environment.  

The drift chambers are the first tracking detectors that
charged particles encounter as they travel from the
collision vertex through the central arms. 
Each is 1.8 m in width in the beam direction, subtends 90-degrees 
in azimuthal angle $\phi$, centered at a radius $R_{DC} = 2.2$ m, and is filled with a 50-50 Argon-Ethane gas mixture.  It 
consists of 40 planes of sense wires arranged in 
80 drift cells placed cylindrically symmetric about the beamline.  The 
wire planes are placed in an X-U-V configuration in the following 
order (moving outward radially):  12 X planes (X1), 4 U planes (U1), 
4 V planes (V1), 12 X planes (X2), 4 U planes (U2), and 4 V planes (V2).
The U and V planes are tilted by a small $\pm5^{o}$ stereo angle to 
allow for full three-dimensional track reconstruction.  The field wire 
design is such that the electron drift to each sense wire is only from 
one side, thus removing most left-right ambiguities everywhere except 
within 2 mm of the sense wire.  The wires are divided electrically in 
the middle at the beamline center.  The occupancy for a central RHIC 
Au+Au collision is about two hits per wire.   

At the drift chamber location, the field of the central magnet is 
nearly zero, so the
DC  determines (nearly) straight-line track segments in the r-$\phi$
plane.   Each track segment is intersected with a circle at $R_{DC}$, where
it is characterized  by two angles: the angular deflection in the main
bend plane, and the azimuthal  position in
$\phi$.  A combinatorial Hough transform technique (CHT) is  used to
identify track segments by searching for location maxima in this
angular space/cite{hough}.  The DCs are calibrated with respect to the event
collision time measurement (see Section ~\ref{BBC}).  With this
calibration, the single-wire resolution in the r-$\phi$ plane is 160
$\rm{\mu}$m. The single-track wire efficiency is ~99\% and the
two-track resolution is better than 1.5 mm.  

The drift chambers are used to measure the momentum of charged 
particles and the direction vector for charged particles traversing the 
spectrometer.  The angular deflection is inversely proportional to the component of
momentum in the bend plane only.  Both the bend angle and the measured track
points are used in the momentum reconstruction  and track model, which
uses a look-up table of the measured  central magnet field grid.  For
this data set, the drift chamber momentum resolution is  $\sigma_{p}/p
= 0.6\% \oplus 3.6\%p$, where the first term is multiple scattering up 
to the drift chambers and the second is the angular resolution of the 
detector.

In Run-1, there were three pad chambers in PHENIX.
Each pad chamber measures a three-dimensional space point of
a charged track.  The pad chambers are pixel-based detectors with effective readout 
sizes of 8.45 mm along the beamline by 8.40 mm in the plane 
transverse to the beamline.  The first pad chamber layer (PC1) is 
fixed to the outer edge radially of each drift chamber at a radial 
distance of 2.49 m, while the third layer (PC3) is positioned at 
4.98 m from the beamline.   Both arms include PC1 chambers, while 
only the East Arm is instrumented with PC3.  The second layer (PC2) 
is located at an inner inscribed radius of 4.19 m in the West Arm and 
was not installed for Run-1.

The position resolution of PC1 is 1.6 mm along the
beam axis and 2.3 mm in the plane transverse to the beam axis.  The
position resolutions of PC3 are 3.2 mm and 4.8 mm,
respectively.  The PC3 is used to reject background from albedo and
non-vertex decay particles;  however, only the PC1 is used for the 
results presented here.  The PC1 is used in the global track 
reconstruction with the measured vertex position using the 
beamline detectors (see Section ~\ref{BBC}) to determine the 
polar angle of each charged track.  Both PC1 and the beamline 
detectors provide z-coordinate information with a 1.89 mm resolution.

\subsubsection{\label{tof} TIME OF FLIGHT}
The Time-of-Flight detector (TOF) serves as the primary particle
identification device for charged hadrons by the measurement of their
arrival time at the TOF wall 5.1 m from the collision vertex.  The 
TOF wall spans 30$^{\circ}$ in azimuth in the East Arm.  It consists of
10 panels of 96 scintillator slats each with an intrinsic timing resolution 
better than 100 ps.  Each slat is oriented along  the r-$\phi$ direction 
and provides timing as well as beam-axis position information for each 
particle hit recorded.  The slats are viewed by two photomultiplier tubes, 
attached to either end of the scintillator.
A $\pm\rm{2} \sigma$ $\pi$/K separation at momenta up 
to 2.0 GeV/c, and a $\pm\rm{2} \sigma$ ($\pi$+K)/proton separation up to 4.0 GeV/c can be achieved.

For each particle, the time, energy loss in the scintillator, and
geometrical position are determined. The total time
offset is calibrated slat by slat.  A particle hit in the
scintillator is defined by a measured pulse height which is  also used
to correct the time recorded at each end of the slat (slewing
correction).  After calibration, the average of the times at either
end  of the slat is the measured time for a particle.  The azimuthal
position  is proportional to the time difference across the slat and
the known velocity  of light propagation in the scintillator (for
Bicron BC404, this is 14 cm/ns).  The slat position along the beamline
determines the longitudinal coordinate  position of the particle.  The
total time of flight is measured relative to the Beam-Beam counter
initial time (see  Section ~\ref{BBC}), the measured time in the
Time-of-Flight detector, and a global time offset from the RHIC
clock. Positive pions in the momentum range  $1.4<p_{T}<1.8$ GeV/c are
used to determine the TOF resolution.  The timing calibration in this
analysis results in a resolution of $\sigma = 115$ ps.\footnote{Ultimately, 
96 ps results  after further calibration, as reported in \cite{PHENIX_NIM}.}

\begin{figure}[ht]
\resizebox{\columnwidth}{!}{\includegraphics{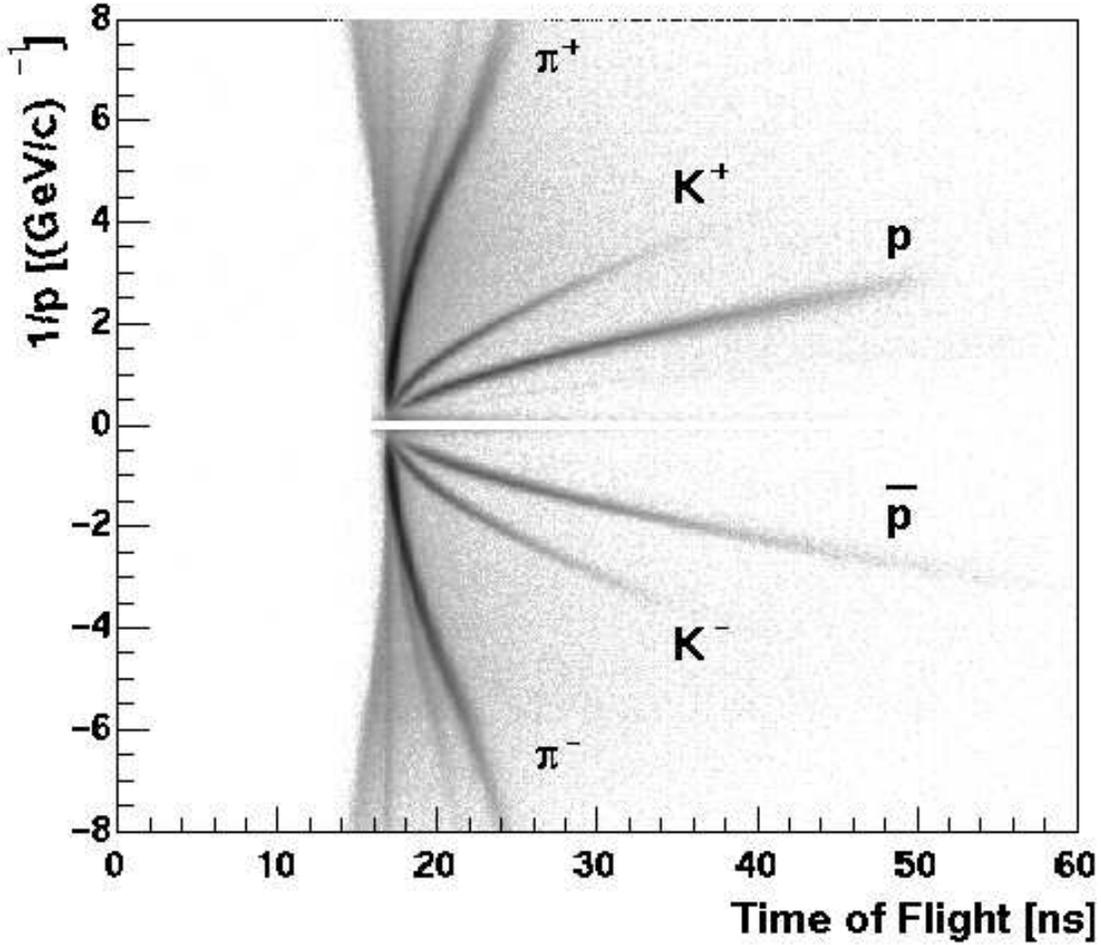}}
\caption{\label{fig:tofPID}
Scaled Time-of-Flight versus reciprocal momentum  in
minimum-bias Au+Au collisions at $\sqrt{s_{NN}}$ =  130 GeV.  The
distribution demonstrates the particle identification  capability
using the TOF for the Run-1 data taking period.}
\end{figure}

Particle identification for
charged hadrons is performed by combining the information from the
tracking system with the timing information from the  BBC and the
TOF. Tracks at 1 GeV/c in momentum point to the TOF with 
a projected resolution $\sigma_{proj}$ of 5 mrad in azimuthal angle and 
2 cm along the beam axis.  Tracks that point to the TOF with less than 
2.0 $\sigma_{proj}$ were selected.  Figure~\ref{fig:tofPID} shows 
the resulting time-of-flight as a function of the reciprocal momentum in  
minimum-bias Au+Au collisions.  

\subsection{\label{BBC}BEAMLINE DETECTORS}
\begin{figure} [ht]
\resizebox{\columnwidth}{!}{\includegraphics{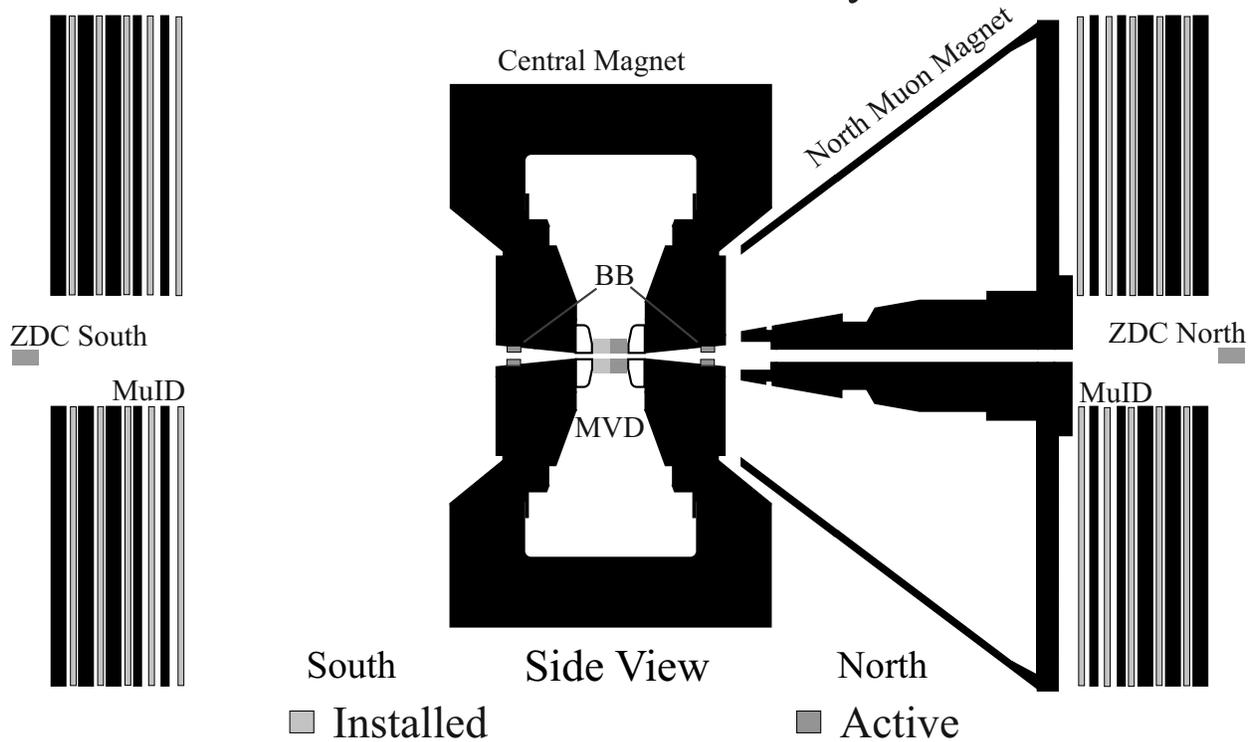}}
\caption{\label{Fig:beamline}
A side view of the PHENIX detector, parallel to 
the beamline.  The beamline detectors determine the collision 
vertex position along the beam direction, and the trigger 
and timing information for each event.}
\end{figure}

The beamline detectors determine the collision vertex position along
the beam direction, and the trigger and timing information for each 
event.  These detectors include the Zero Degree Calorimeters (ZDCs), 
the Beam-Beam Counters (BBC), and the Multiplicity Vertex Detector 
(MVD) and are positioned in PHENIX as shown in Figure~\ref{Fig:beamline}.  

The Zero Degree Calorimeters are small transverse area  hadron
calorimeters that are installed at each of the four RHIC
experiments. They measure the fraction of the energy deposited by
spectator neutrons from the collisions and serve as an event
trigger for each RHIC experiment.  The ZDCs  measure the unbound
neutrons in small forward cones ($\theta<$2 mrad)  around each beam
axis.  Each ZDC is positioned 18 m up and downstream from the
interaction point  along the beam axis. A single ZDC
consists of 3 modules each  with a depth of 2 hadronic interaction
lengths and read out by a  single PMT.  Both time and amplitude are
digitized for each of the  3 PMTs as well as an analog sum of the PMTs
for each ZDC. \cite{nim_ZDC}

There are two Beam-Beam counters each positioned 1.4 m from
the interaction point, just behind the central magnet poles along the
beam axis (see Figure~\ref{Fig:beamline}). The BBC consists of two
identical sets of  counters installed on both sides of the interaction
point along  the beam.  Each counter consists of 64 Cherenkov
telescopes, arranged radially about the collision  axis and situated
north and south of the MVD.  The BBCs 
measure the fast secondary particles produced in each collision  at forward
angles, with $3.0 \le \eta \le 3.9$, and full azimuthal coverage.


For both the ZDC and the BBC, the time and vertex position are
determined using the measured time  difference between the north and 
the south
detectors and the known distance between the two detectors.  The start
time ($T_0$) and the vertex position along  the beam axis
($Z_{vertex}$)  are calculated as $T_0 = (T_1 + T_2)/2$ and  $Z_{vertex} =
(T_1 - T_2)/2c$, where $T_1$ and $T_2$ are the average  timing of
particles in each counter and $c$ is the speed of light.  With an 
intrinsic timing resolution of 150 ps, the ZDC vertex is 
measured to within 3 cm.  In Run-1, the BBC timing resolution of 70 ps 
results in a vertex position resolution of 1.5 cm.

\begin{figure} [ht]
\resizebox{\columnwidth}{!}{\includegraphics{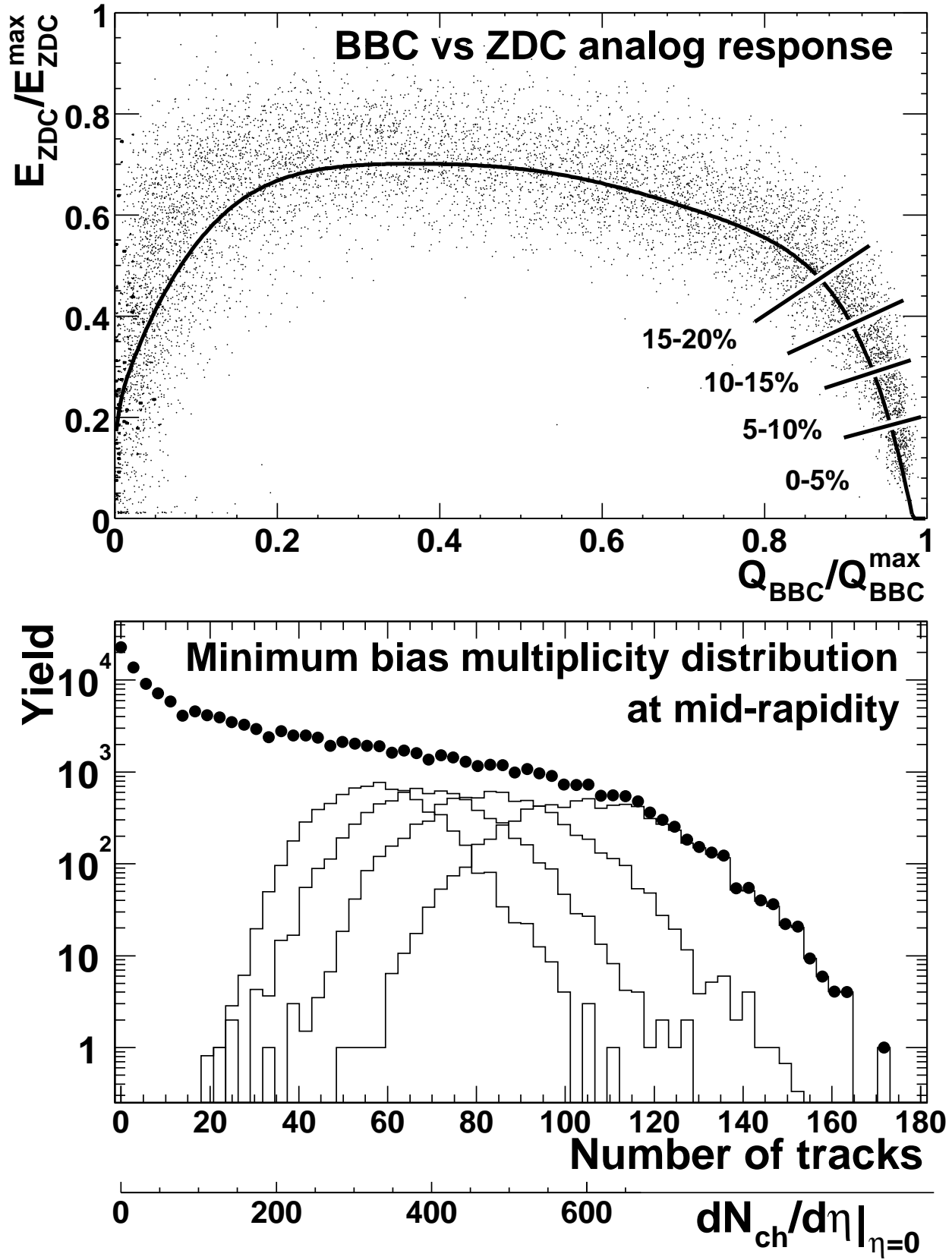}}
\caption{\label{cent}
The event centrality (upper plot) is determined using a
correlation measurement of the fraction of neutron energy recorded in the
ZDCs (vertical scale) and the fractional charge measured in the BBCs
(horizontal scale).  The equivalent track multiplicity in each centrality
selection is shown in the lower plot.}
\end{figure}

Event centrality is determined using a correlation measurement between
neutral energy deposited in the ZDCs and fast particles recorded in
the  BBCs as shown in Figure~\ref{cent}.  The spectator nucleons are 
unaffected by  the interaction and travel at their initial 
momentum from each respective  ion.  The number of neutrons 
measured by the ZDC is proportional to the  number of spectators, 
while the BBC signal increases with the number of participants.    

\section{\label{cuts}DATA REDUCTION AND ANALYSIS}

\subsection{DATA REDUCTION}
The PHENIX Level-1 trigger selected events with hits coincident
in both the ZDC and BBC detectors, and in time with the RHIC clock.
A total of 5M events were recorded at $\sqrt{s_{NN}}=$ 130 GeV in
the ZDCs~\cite{PHENIX_QM01}.
The collision position along the beam direction was required to be
within $\pm$ 30 cm of the center of PHENIX, using the collision vertex
reconstructed by the BBC.

The trigger on both BBC and both ZDC counters includes $92\pm4\%$ of 
the total inelastic cross section ($6.8\pm0.4$ barns).  
A Monte Carlo Glauber model \cite{Glauber} is
used with a simulation of the BBC and ZDC responses to determine the 
number of nucleons participating in the collisions for the minimum 
bias events.  The Woods-Saxon parameters determined from
electron  scattering experiments are: radius $ = 6.38 
\pm 0.06$
fm, diffusivity $ = 0.54\pm0.01$~fm~\cite{glauber_parameters},
and the nucleon-nucleon inelastic cross-section, 
$\sigma^{inel}_{N+N} = 40\pm3$ mb.  An additional
systematic uncertainty enters the radius parameter since the radial
distribution of neutrons  in large nuclei should be larger than for
protons and is not well determined~\cite{glauber_syst}.

The centrality selections used in this paper are 0-5\%, 5-15\%,
15-30\%, 30-60\%, and 60-92\% of the total geometrical cross section,
where 0-5\% corresponds to the  most central collisions.

Only tracks which are reconstructed in all three dimensions
are included in the spectra.  These tracks are then matched within 2$\sigma_{proj}$
to the measured positions in the TOF detector.  For each
TOF hit, the time, position, and energy loss are measured in the TOF
detector.  The widths of residual distance distributions between 
projected tracks and TOF hit positions, $\sigma_{proj}$, 
increase at lower momentum due
to multiple scattering.  Therefore, a momentum-dependent hit
association criterion was defined.  

Finally, a requirement on 
energy loss in the TOF is applied to each track to exclude false 
hits by requiring the energy deposit of  at least minimum ionizing 
particle energy.  A $\beta$-dependent energy loss cut whose 
form is a parameterization
of the Bethe-Bloch formula\cite{PDB} is used, where
\begin{equation}
dE/dx \approx \beta^{-5/3}
\end{equation}
and $\beta = L/ct$, where $L$ is the pathlength of the particle's
trajectory from the BBC vertex to the TOF detector, t is the
particle's time-of-flight, and c is the speed of light.
The approximate Bethe-Bloch formula is scaled by a factor to fall
below the data and thereby serve as a cut.  The resulting equation  is
$\Delta E = A\beta^{-5/3}$ where A is a scaling factor 
equal to $1.6$ MeV.  The energy loss cut reduces low momentum
background under the kaon and proton mass peaks.  The fraction of
tracks excluded after the energy loss cut is less than $5.5\%$.

The measured momentum ($p$), pathlength ($L$), and time of flight ($t$) in
the  spectrometer are used to calculate the particle mass, which  is
used for particle identification:
\begin{equation}
m^2 = \frac{p^2}{c^2} \left[ {{\left(\frac{1}{\beta}\right)}^2 - 1} \right].
\label{m2}
\end{equation}

The width of the peaks in the mass-squared distribution depend on both
the momentum and time-of-flight resolutions.  An analytic form for the
width of $m^2$ as a function of momentum resolution $\sigma_p$ and
time of flight resolution is determined using Equation \ref{m2}.  The
error in the particle's pathlength L results in an effective time
width that is included with the TOF resolution, $\sigma_T$,
\begin{equation}
\sigma^2_{m^2}= 4 m^4 {\frac{\sigma_p}{p}}^2 + 4p^4
\frac{1}{\beta^2}{\frac{\sigma_T}{t}}^2.
\label{m3}
\end{equation}

The momentum resolution of the drift chambers is expressed in the
following form
\begin{eqnarray}
\sigma_p^2 = { \left( C_1 p \frac{1}{\beta} \right) }^{2} + { \left(
C_2 p^2 \right) }^{2}, \label{m4}\\
C_1 = \frac{\delta\phi_{ms}}{K_{1}}, \label{m4a} \\
C_2 = \frac{\delta\phi_{\alpha}}{K_{1}}, \label{m4b} 
\end{eqnarray}
where $C_1$ and $C_2$ are the multiple scattering and angular
resolution terms, respectively.  The units of $\delta\phi_{ms}$ are
mrad GeV/c.  The constant $K_1$ is the momentum kick on the particle from
the magnetic field and is equal to $87.3$ mrad GeV/c.  
The constant $C_1$ is the width in $\phi$ due to the multiple
scattering (ms) of a charged particle with materials of the spectrometer up
to  the drift chambers.  The $C_2$ term is the angular resolution of the bend 
angle ($\alpha$), which is the angular deflection in $\phi$ of the track segment 
relative to the radius to the collision vertex.  

Equation \ref{m4} is used in Equation \ref{m3} with $\beta =p/
\sqrt{p^2+m^2_0}$, where $m_0$ is the mass centroid of the
particle's mass-squared distribution.  
The mass centroid is close to the 
rest mass of the particle; however due to residual misalignments and 
timing calibration, the centroid of the distribution is a fit
parameter in 
order to avoid cutting into the distribution.  The $m^2$ width for each 
particle is written as follows:
\begin{eqnarray}
\lefteqn{\sigma^2_{m^2} =} \\ 
\nonumber & C_1^2\cdot4m^4(1+\frac{m^2_0}{p^2})
+ C_2^2\cdot 4 m^4_0 p^2  
+ C_3^2 \cdot (4p^2(m_0^2+p^2)) 
\label{m5}
\end{eqnarray}
where the coefficient $C_3$ is related to the combined TOF,
\begin{equation}
C_3 = \frac{\sigma_{T} c }{L},
\label{m5a}
\end{equation}
and pathlength contributions to the time width, $\sigma_{T}$ in
Equation ~\ref{m5a}.  From the measured drift chamber momentum
resolution, $C_{1} = 0.006$ and  $C_{2} = 0.036$ c/GeV.  While the 
TOF resolution is 115$\pm$5 ps, the pathlength uncertainty 
introduces a width of 
$\approx$ 20-40 ps, so 145 ps is used for $\sigma_{T}$ in $C_{3}$.

\begin{figure} [ht]
\resizebox{\columnwidth}{!}{\includegraphics{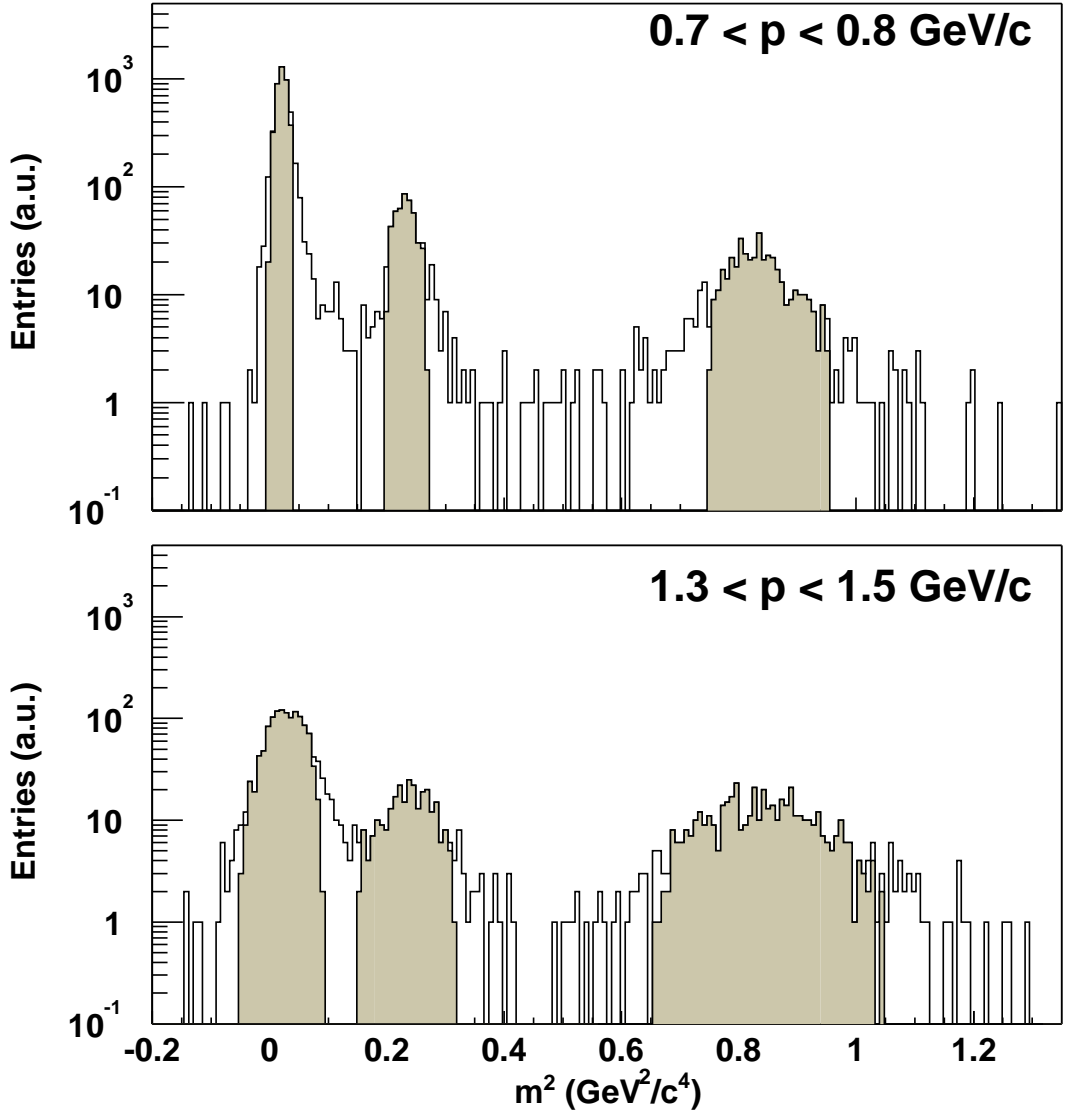}}
\caption{\label{Fig:PID}
The mass-squared distributions of positive pions, kaons, and protons
for two different momentum slices.  The momentum slice $\rm{0.7}<p<\rm{0.8}$
GeV/c is
the upper panel and $\rm{1.3}<p<\rm{1.5}$ GeV/c is the lower panel.  The shaded
regions correspond to the 2$\sigma$ particle identification bands
based on the calculated mass-squared width, the measured mass-squared
centroids, and the known detector resolutions.}
\end{figure}

The pions, kaons, and protons are identified using the measured  peak
centroids of the $m^2$ distribution and selecting $2\sigma$ bands; 
shown as shaded regions in Figure~\ref{Fig:PID} for two different 
momentum slices.
The 2$\sigma$ bands for pions and kaons do not overlap up to $p_{T} = $2 GeV/c.
The protons are identified up to $p_{T} = $ 4 GeV/c.  By studying 
variations in the $m^{2}$ centroid and width
before the  particle identification cut is applied, the uncertainty
in the particle identification is estimated to be 5$\%$ for all
particles.

Kaons are depleted by decays in flight and geometrical acceptance.  For the low 
momentum protons, energy loss and geometrical acceptance cause a drop in the 
raw yield for $p_T < \rm{0.5}$ GeV/c,
as seen in Figure~\ref{Fig:acceptance}.

The remaining background contribution was determined by reflecting
the track about the midpoint of PHENIX along the beamline and repeating the
association and PID cuts used in the TOF detector. This random background
was evaluated separately for each particle type.
The background contribution is $\approx$ 30\% for the kaon spectra at
$\rm{0.2} < p_T < \rm{0.4}$ GeV/c and defines the low $p_T$ limit in the spectra. The
background is $< \rm{5}\%$ in all other cases, and negligible above 0.8 GeV/c in the 
measured momentum range in this analysis.  The background was not subtracted 
but is instead treated as a systematic uncertainty.  This uncertainty is 2, 5, and 3$\%$ for
pions, kaons, and protons, respectively, at $p_{T} < \rm{0.6}$ GeV/c
and is negligible at higher momenta.

\subsection{\label{corr}ANALYSIS}
The raw spectra include inefficiencies from detector acceptance,
resolution, particle decays in flight and track reconstruction.
The baseline efficiencies are determined
by simulating and reconstructing single hadrons. Multiplicity
dependent effects are then evaluated by embedding simulated
single hadrons into real events and calculating the degradation of
the reconstruction efficiency.

\subsubsection{CORRECTIONS:  ACCEPTANCE, DECAYS IN FLIGHT, AND DETECTOR RESPONSE}
The corrections for the finite detector aperture, pion and kaon decays in flight, 
and the detector response are determined using single particles in the
the GEANT \cite{geant} simulation of the detector. 
All details of each detector are modeled, including dead channels in the 
drift chambers, pad chambers, and Time-of-Flight detector.
All physics processes are automatically taken into account, resulting
in corrections for multiple scattering, anti-proton annihilation, pion and 
kaon decays in flight, finite geometrical acceptance of the detector,
and momentum resolution, which affects the spectral shape above 2.5 GeV/c.

The drift chamber simulated response is tuned to describe the
response of the real drift chambers on the single-wire level.  This
is done using a simple geometrical model of the drift chamber  and
the straight-line trajectories of particles from the  zero-field data.
This simple model of the drift cell in the drift chamber  is
sufficient to describe the observed drift distance distribution,
the pulse width, the single wire efficiency, and  the detector
resolution.  The TOF response is simulated  by smearing
the true time of flight using a Gaussian distribution with a width as measured
in the data.

\begin{figure} [ht]
\begin{center}
\resizebox{\columnwidth}{!}{\includegraphics{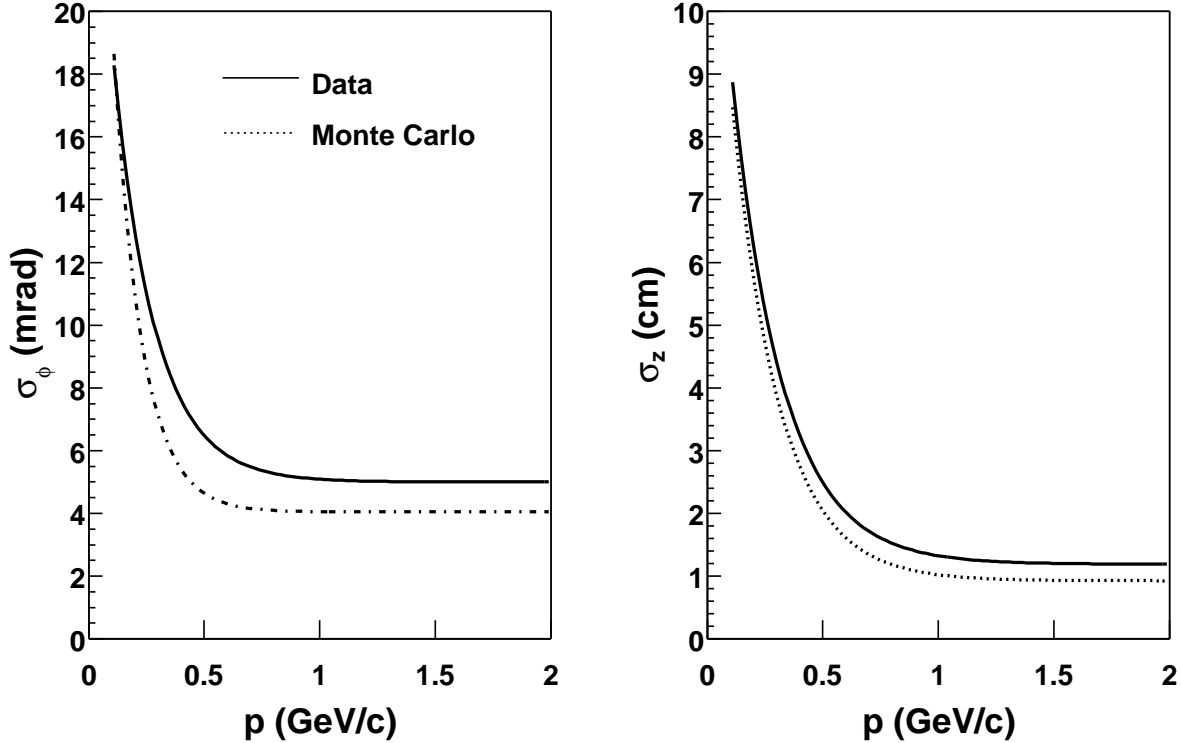}}
\end{center}
\caption{\label{sigMatch}
Comparison of the momentum-dependent residuals of DC tracks
matched to TOF hits in azimuthal angle $\phi$ (left) and z (right)
between data (solid) and  simulation (dashed).}
\end{figure}

Figure~\ref{sigMatch} shows the momentum dependence of the residual distance 
between projected tracks and TOF hits for the real (solid line) and simulated 
(dashed) events. These residuals are parameterized in the azimuthal angle $\phi$ 
and the beamline direction $z$, separately for data and simulation.
For each case, tracks that fall outside 2$\sigma$ of the parameterized width 
are rejected, thus allowing use of the Monte Carlo to evaluate the correction
for the 2$\sigma$ match requirement for real tracks.

A fiducial cut is made in both the simulation and in the 
data to ensure the same fiducial volume.  The systematic uncertainty
in the acceptance correction is approximately 5$\%$.

The simulated distributions are generated uniformly in $p_T$, $\phi$,
and  $y$.  For each hadron, sufficient Monte Carlo events are
generated to obtain the correction factor for every measured
$p_T$ bin.  The statistical errors from  the correction factors were
smaller than those in the data and both  are added in quadrature.

The distribution of the number of particles generated in each $p_T$
slice,  dN/d$p_T$, is the ``ideal" input
distribution  without detector and reconstruction effects.  This
distribution is normalized  to 2$\pi$ and 1 unit of rapidity.  After
detector response and track reconstruction, the output 
distribution is the number of particles found in each  $p_T$ slice.
The final corrections  are determined
after an iterative weighting procedure.  First, the flat input  and
output distributions are weighted by exponential functions for all
particles  using an inverse slope of 300 MeV.  The ratio of input to
output distributions is determined as a function of momentum.  In each
$p_T$ slice, the corresponding  ratio is applied to the data.  The
corrected data are next fitted with exponentials  for kaons and
protons (see Equation~\ref{ptexp}), and a power-law for the pions (see
Equation~\ref{pow}).  The original flat  input and output distributions
are weighted by these resulting functions.  The  procedure is repeated
until the functions remain constant in their parameters.  The weighted
input and output distributions are divided to produce acceptance
correction factors.  The corrections are larger for kaons due to the
decays in flight.  The statistical error in determination of
the correction factor is added in  quadrature to the statistical error
in the data.
  
\subsubsection{HIGH TRACK-DENSITY EFFICIENCY CORRECTION}
A final multiplicity dependent correction is determined using simulated 
single-particles embedded into real events.  This  correction depends
on both the quality of the track reconstruction  in a high
multiplicity environment and the type of particle measured.

\begin{figure} [ht]
\resizebox{\columnwidth}{!}{\includegraphics{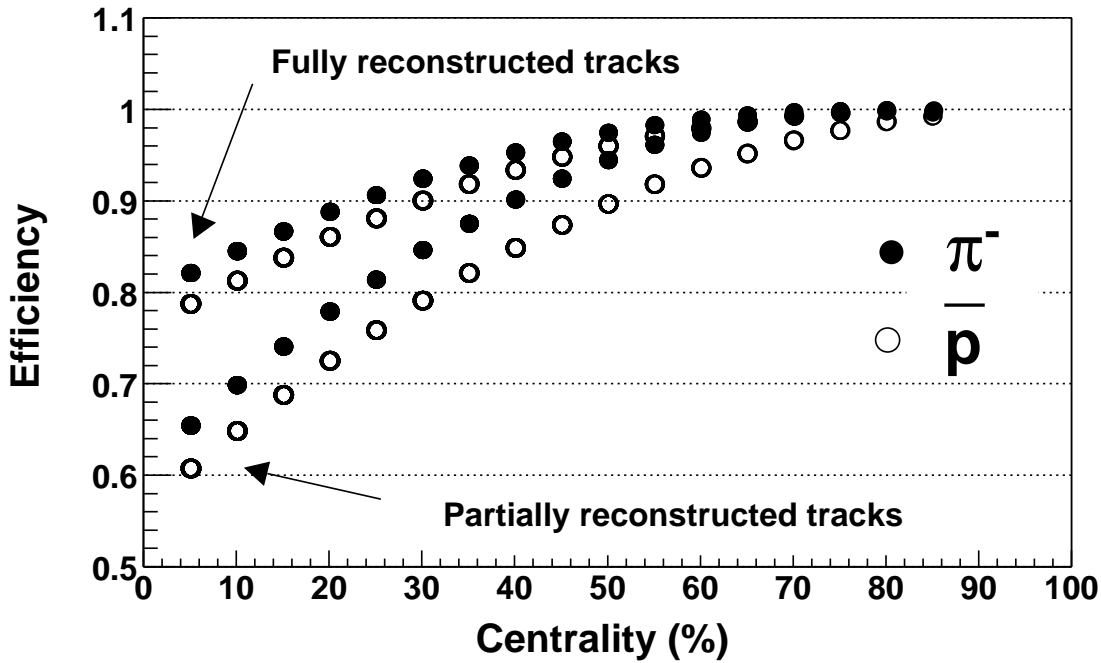}}
\caption{\label{efficiencyCorrection}
The multiplicity dependent efficiency correction for pions
(solid) and (anti)protons (open) for two types of tracks.  The upper
set of points correspond to fully reconstructed tracks in the drift
chambers; while the lower set of points correspond to partially
reconstructed tracks in the drift chambers.}
\end{figure}

Depending on the centrality of the  event, the correction factor is
determined for each particle  in the raw transverse momentum
distribution and is applied as a weight.  The final efficiency
corrections are shown in Figure~\ref{efficiencyCorrection}, where the
correction for  pions is shown as solid circles and for (anti)protons 
as open circles. The horizontal  axis ranges from the most central to the
most peripheral events  in increments of 5\%. The systematic
uncertainty in the multiplicity efficiency correction is 9\%.

The difference between pions (solid) and (anti)protons (open)  is due
to the different TOF efficiencies for each  particle (protons are
slower than pions).  In a small fraction of cases two particles  may
hit the same TOF slat at different times, and the slower particle  is
assigned an incorrect time.  The particle will then fall outside  the
particle identification cuts.  This effect depends on the type of
particle.

For each particle, two curves are shown, representing the DC tracking
inefficiency for two types of tracks:  fully reconstructed and
partially reconstructed tracks.  Fully reconstructed tracks include
X1 and X2 sections. In a high track-density environment,
tracks may be partially reconstructed  or hits may be incorrectly
associated.  There are two cases when this  incorrect hit association
occurs.  In the first case, the direction  vector in the azimuth
prevents the track from pointing properly  to the PC1 detector, and the
correct hit cannot be associated.  In the  second case, the track is
reconstructed properly, but there are two possible  PC1 points.  If no
UV hits are found, then the wrong PC1 point can be associated  to the
track and the track's beamline coordinate is mis-reconstructed.  In
both of  these cases, the track fails the matching criteria in the TOF
detector  and is lost.

\subsubsection{DETERMINING THE YIELD AND MEAN $p_T$}
The dN/dy and $\langle p_{T} \rangle$ are determined 
using the data in the measured region and 
an extrapolation to the unmeasured 
region after integrating a functional form fit to the data.
A function describing the spectral
shape is fit to the data, with varying $p_T$ ranges to control
systematic uncertainties in the fit parameters.  The fitted shape
is extrapolated, 
integrated over the unmeasured range, and then combined with the
measured data to get the full yield.
Two different functions are used to estimate
upper and lower bounds for each spectrum.  The average between the
upper
and lower bounds is used for dN/dy and $\langle p_T \rangle$.
The statistical error is determined from the data, and the 
systematic uncertainty is taken as 1/2 the difference
between the upper and lower bounds.

For pions, a power-law in $p_{T}$ (Equation ~\ref{pow}) and an
exponential in $m_{T} ( = \sqrt{p^2_T + m^2_0})$ (Equation
~\ref{mtexp})  are fit to the data.  For kaons and (anti)protons, two
exponentials, one in $p_{T}$ (Equation ~\ref{ptexp}) and the other  in
$m_{T}$ are used.  The $p_{T}$ exponential provides an upper limit for  
the extrapolated yield, which is  
most important for the (anti)protons.  The power-law function has
three parameters labeled $A$, $p_0$, and $n$ in Equation ~\ref{pow}.  The
exponentials have two parameters, $A$ and $T$.
\begin{equation}
\frac{d^{2}N}{2\pi p_{T} dp_{T} dy} = A \left( \frac{p_{0}}{p_{0} +
p_{T}} \right) ^{n}\label{pow}
\end{equation}
\begin{equation}
\frac{d^2 N}{2\pi m_{T}dm_{T} dy} = A e^{-m_{T}/T} \label{mtexp}
\end{equation}
\begin{equation}
\frac{d^{2}N}{2\pi p_{T}dp_{T}dy} = A e^{-p_{T}/T} \label{ptexp}
\end{equation}

\subsection{\label{syst}SYSTEMATIC UNCERTAINTIES}
In Table ~\ref{Tab:syst}, the sources of systematic uncertainties in
both $\langle p_T \rangle$ and dN/dy are tabulated.  The sources
of uncertainty include the extrapolation in $p_T$, the background, and the
Monte Carlo corrections and cuts.  The uncertainty in the 
Monte Carlo corrections
is 11\% and includes:  the multiplicity efficiency correction
of 9\%, the particle identification cut of 5\%, and the fiducial cuts
of 5\%.  The uncertainties in the correction functions are added
in  quadrature to the statistical error in the data.  Background is
only relevant for $p_T < 0.6$ GeV/c in the spectra.

The total systematic uncertainty in the $\langle
p_{T} \rangle$ depends on the extrapolation and background
uncertainties;  the uncertainties are 7$\%$, 10$\%$, and 8$\%$ for
pions, kaons, and protons,  respectively.  The overall uncertainty 
on dN/dy includes the uncertainties on $\langle p_{T} \rangle$ in 
addition to the uncertainties from the corrections and 
cuts; the uncertainties are  13$\%$, 15$\%$, and 14$\%$  for 
pions, kaons, and protons, respectively \cite{phd}.  

The hadron yields and $\langle p_T \rangle $ values include an
additional uncertainty arising from the fitting function used for
extrapolation to the unmeasured region at low and high $p_T$.
The magnitude of the extrapolation is
$30\pm6\%$ of the spectrum for pions, $40\pm8\%$ for kaons, and
$25\pm7.5\%$ for protons \cite{phd}.  The systematic uncertainty 
quoted here is taken as 1/2 the difference between the results from the 
two different functional forms.

\begin{table} [ht]
\caption{\label{Tab:syst}
The sources of systematic uncertainties in $\langle p_T
\rangle$ and dN/dy.}
\begin{ruledtabular}
\begin{tabular}{lccc}
  & $\pi$ (\%)  & K (\%) & (anti)p (\%)\\ \hline
Extrapolation & 6 & 8 & 7.5\\
Background ($p_T<0.6$ GeV/c) & 2 & 5 & 3 \\ 
$\langle p_T \rangle$ total& 7 & 10 & 8 \\ 
Corrections and cuts & 11 & 11 & 11 \\ 
dN/dy total & 13 & 15  & 14 \\
\end{tabular}
\end{ruledtabular}
\end{table}

The momentum scale is known to better than 2\%, and the momentum 
resolution affects the spectra shape, primarily for protons, 
above 2.5 GeV/c.  
The momentum resolution is corrected by the Monte Carlo.
As other sources of uncertainty on the number of
particles at any given momentum are much larger, momentum resolution 
effects are neglected in determining the overall systematic uncertainty 
from the data reduction.

\section{\label{results} RESULTS}
\subsection{TRANSVERSE MOMENTUM DISTRIBUTIONS}
\begin{figure*}[hbt]
\includegraphics[width=1.0\linewidth]{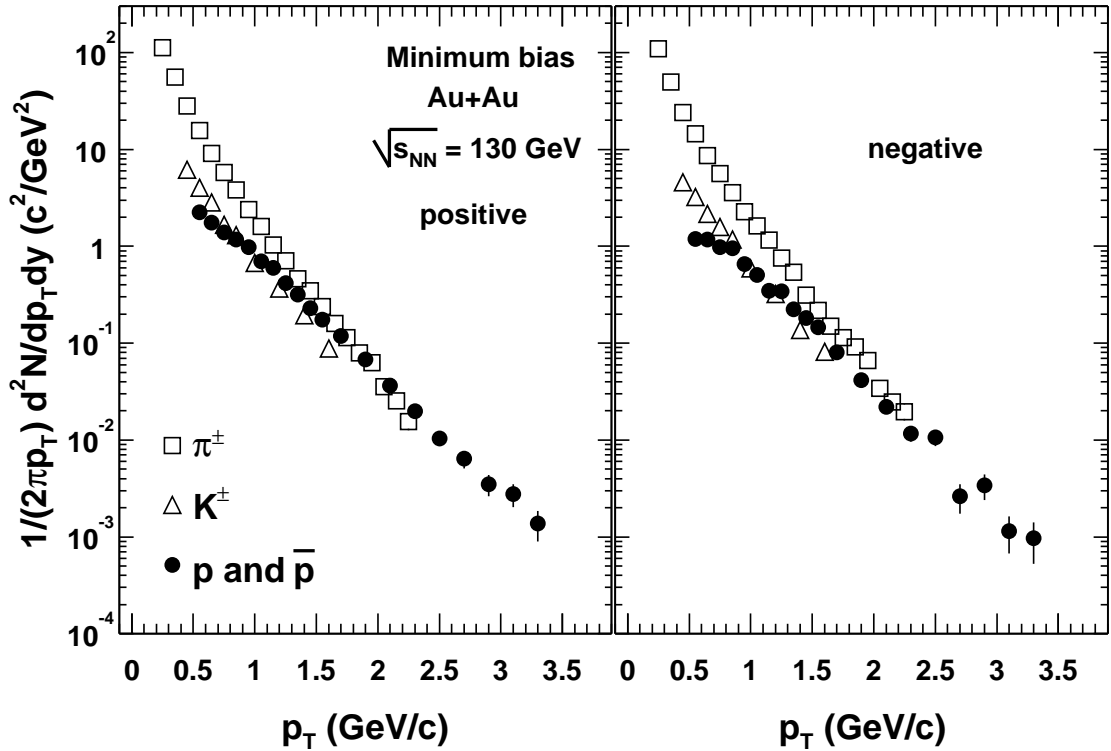}
\caption{\label{fig1}
The spectra of positive particles (left) and negative
(right) in minimum-bias collisions from Au+Au collisions at
$\sqrt{s_{NN}}=$130 GeV.  The errors include both statistical 
and systematic errors from the corrections.}
\end{figure*}

The invariant yields as a function of $p_T$ for identified hadrons are shown in
Figure~\ref{fig1}, while 
Figure~\ref{piKpCent} provides the centrality dependence of the spectra.  
The spectra are tabulated in Appendix~\ref{xsec}.
The $\pi^{\pm}$, $K^{\pm}$, p, and $\overline{p}$ invariant yields for the most central, mid-central, and the most peripheral
collisions,  were reported previously \cite{PPG006}.  Pion and 
(anti-)proton invariant yields are comparable for 
$p_T>$1 GeV in the most central collisions.  

\begin{figure*} [hbt]
\includegraphics[scale=0.6]{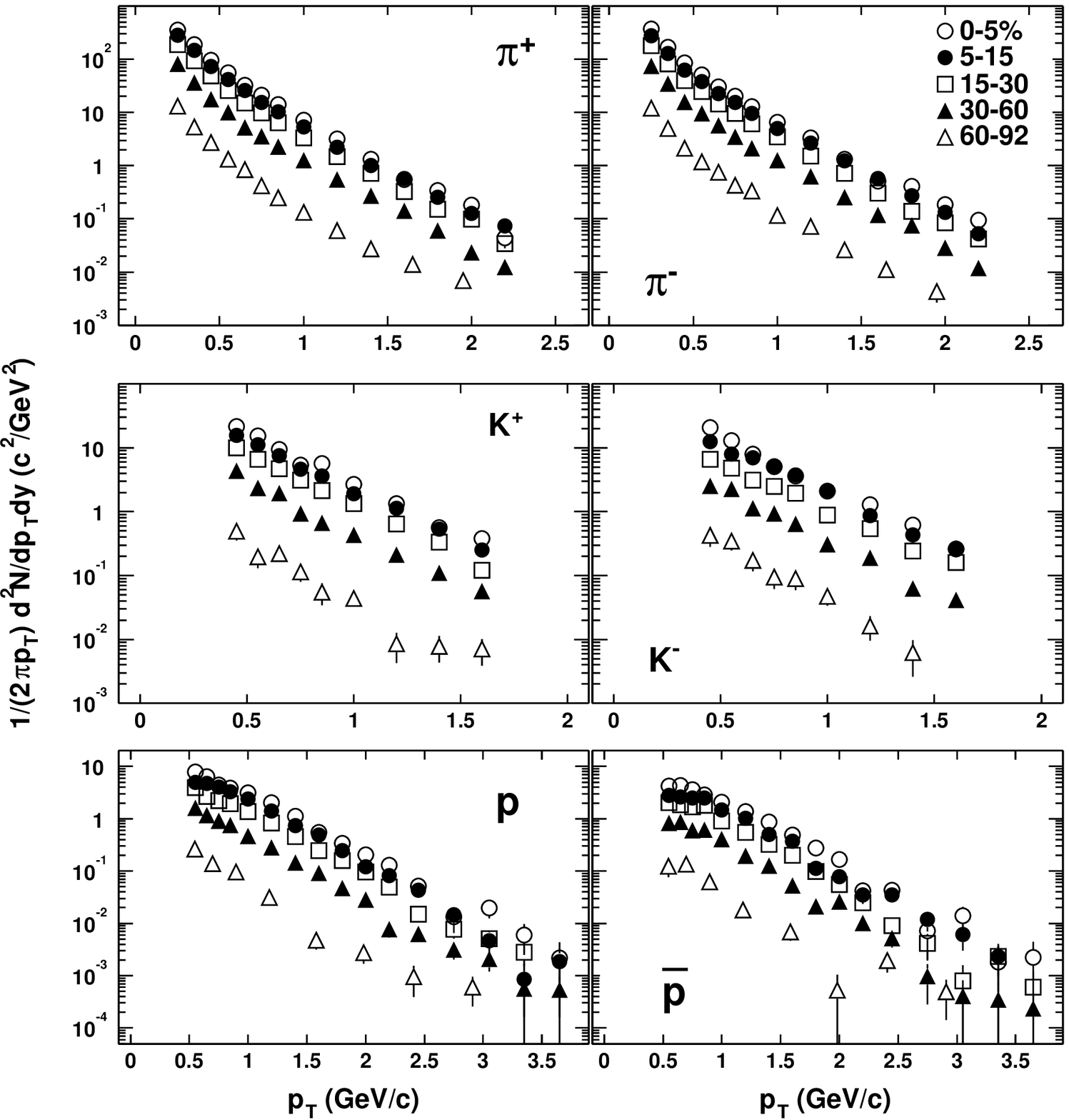}
\caption{\label{piKpCent}
The hadron spectra for five centralities from the most
central 0-5\% to the most peripheral 60-92\% at $\sqrt{s_{NN}}=$ 130 GeV. 
Errors include both the statistical and point-by-point error in the
corrections added in quadrature.}
\end{figure*}

\begin{figure*} [htb]
\includegraphics[scale=0.6]{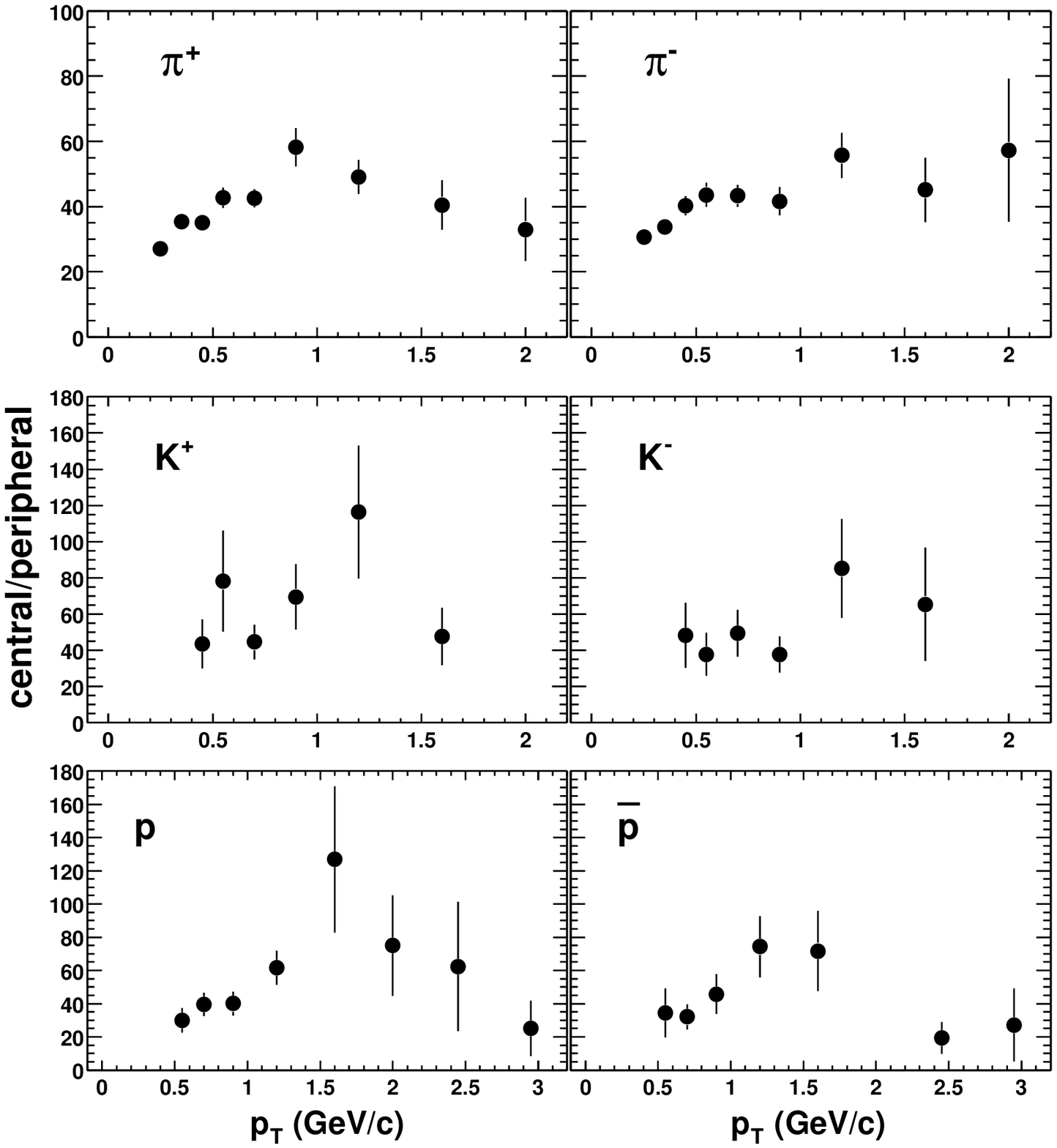}
\caption{\label{centToPerp}
The ratio of the most central to the most peripheral yields as a 
function of $p_{T}$ for pions (top), kaons (middle) and (anti)protons (bottom).}
\end{figure*}

As can be seen already from Figure~\ref{piKpCent} all the spectra
seem to be exponential; however, upon closer inspection,  small
deviations from an exponential form are apparent for the more
peripheral collisions.  The spectrum in the most peripheral 
collisions is noticeably power-law-like 
when compared to the more exponential-like spectrum in central collisions. This is especially apparent 
for the pions.  The effect can be
seen more clearly in the ratio of the spectra for a given particle
species in two different centrality classes.  Such ratios are shown in
Figure~\ref{centToPerp} for the 5\% central and the most peripheral
positive spectra (60-92\% centrality).  The ratios for protons and 
antiprotons as well as for $\pi^+$ have a maximum at intermediate 
$p_T$ and are lower both at low and high $p_T$.  The kaon shape change 
is not very significant, given the current statistics.

The change in slope at low-$p_T$ in central collisions compared to
peripheral is  consistent with a more substantial hydrodynamic,
pressure-driven transverse  flow existing in central collisions, since the
increased boost would tend to  deplete particles at the lowest $p_T$
(see Section~\ref{slope}).  This is observed at lower energies at the 
CERN SPS  \cite{wa80:pi0,wa98:pi0}.  It is in contrast to results
obtained at the ISR \cite{faessler} for p+p collisions at $\sqrt{s} = $ 63 GeV, where a shallow maximum or minimum exists at low $p_T$ (in the range $0.3-0.6$ GeV/c).

\subsubsection{FEED-DOWN CONTRIBUTION TO p AND $\overline{p}$ FROM INCLUSIVE 
$\Lambda$ and $\overline{\Lambda}$} 
Inclusive $\Lambda$ and $\overline{\Lambda}$ transverse momentum 
distributions have been measured
in the west arm of the PHENIX spectrometer using the tracking
detectors (DC, PC1) and a lead-scintillator electromagnetic
calorimeter (EMCal) \cite{PPG012}.  The invariant mass is
reconstructed from the weak decays $\Lambda \rightarrow p+\pi^{-}$ and
$\overline{\Lambda} \rightarrow \overline{p}+\pi^{+}$. 

The tracks from the tracking detectors are required to fall within
3$\sigma$ of EMCal  measured space-points.  The EMCal timing
resolution of the daughter  particles is $\approx$ 700 ps.  Using the
DC momentum and the EMCal time-of-flight,  the particle mass is
calculated, and protons, antiprotons, and pions are identified  using
2$\sigma$ momentum-dependent mass-squared cuts.  A clean particle
separation  is obtained using an upper momentum cut of 0.6 GeV/c and
1.4 GeV/c for pions and  protons, respectively.  The momentum is
determined assuming the primary decay vertex is positioned at the
event vertex and results in a momentum shift of 1-2\% based  on a
Monte Carlo study.

Using all combinations of pions and protons, the invariant  mass is
determined.  The mass distribution shows a $\Lambda$ peak on top of a
random  combinatorial background, which is determined by combining 
protons  and pions from different collisions with
the same centrality.  A signal-to-background ratio of 1/2 is obtained
after applying a decay kinematic cut on the daughter particles.  Fitting 
a  Gaussian function to the mass distribution in the range
$\rm{1.05} < m_{\rm{p\pi}} < \rm{1.20}~GeV/c^{2}$, 12000 $\Lambda$ and
9000 $\overline{\Lambda}$ are observed, with mass resolution 
$\delta m/m \approx $
2\%.  The reconstructed $\Lambda$ and $\overline{\Lambda}$
spectra are corrected for the acceptance,  pion decay-in-flight,
momentum resolution, and reconstruction efficiency \cite{PPG012}.  The
systematic uncertainty on the $p_T$ spectra is 13\% from the
corrections and  3\% from the combinatorial background subtraction.
The feed-down contributions  from heavier hyperons $\Sigma^{0}$ and
$\Omega$ are not measured but are estimated to be $<$ 5\%.

\begin{figure*} [hbt]
\includegraphics[scale=0.6]{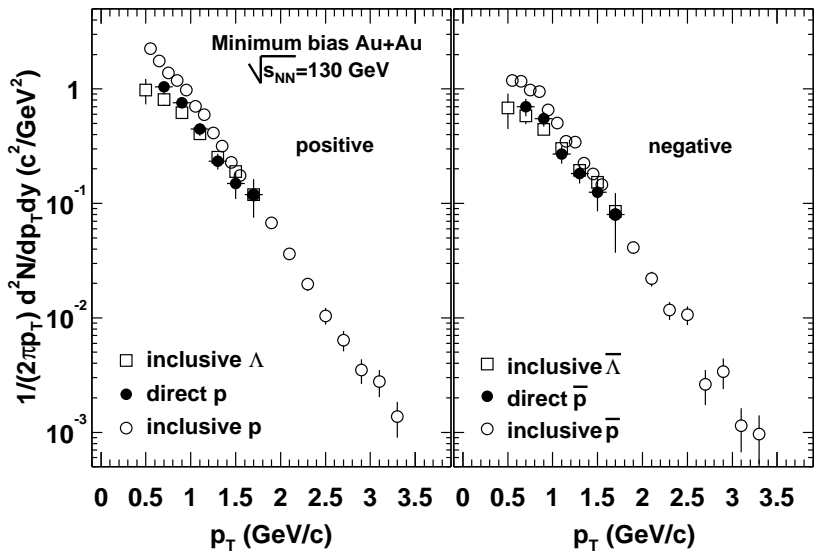}
\caption{\label{lambda_prot}
For minimum-bias collisions, the inclusive $\Lambda$, inclusive p, 
and direct p transverse momentum distributions are plotted together in the left panel.  
The equivalent comparison for inclusive $\overline{\Lambda}$, $\overline{p}$, and direct $\overline{p}$ 
transverse momentum distributions is in the right panel.}
\end{figure*}

In Figure~\ref{lambda_prot}, the transverse momentum spectra of inclusive 
protons (left) and antiprotons (right) are shown with the inclusive 
$\Lambda$ and $\overline{\Lambda}$ transverse momentum distributions.  The 
solid points are the (anti)proton spectra after the feed-down correction 
from $\Lambda$ and $\overline{\Lambda}$ weak decays.  From here forward, 
the data that are presented and discussed are not corrected for this 
feed-down effect; inclusive p and $\overline{p}$ yields are given.  More 
details on the $\Lambda$ and $\overline{\Lambda}$ measurement are 
included in \cite{PPG012}.  

\subsection{\label{data}YIELD AND $\langle p_T \rangle$}
The yield, dN/dy, and the average transverse momentum, $\langle p_{T}
\rangle$,
are determined
for each particle as described in the preceding section and have been 
previously published in \cite{PPG006}.  
For each centrality, the rapidity density dN/dy and average transverse momentum
$\langle p_{T} \rangle$  are tabulated in Tables~\ref{dndyTable} and
~\ref{meanptTable}, respectively.  

The $N_{part}$ and $N_{coll}$ in each centrality selection are 
determined using a Glauber-model calculation in \cite{PPG001}.
The resulting values of $N_{part}$ and $N_{coll}$ are also tabulated in Table~\ref{dndyTable}.
(See Appendix ~\ref{app:glauber} for more detail).  The errors on $N_{part}$ 
and $N_{coll}$ include the uncertainties in the model parameters as well as 
in the fraction of the total geometrical cross section ($92\% \pm 4\%$) seen by the 
interaction trigger.  The error due to model uncertainties is 
2\% \cite{PPG001}.  An additional 3.5\% error results from time 
dependencies in the centrality selection over the large data sample.

\begin{table*} [ht]
\caption{\label{dndyTable}
The dN/dy at midrapidity for hadrons produced at midrapidity
in each centrality class.  The errors are statistical only.  The
systematic errors are 13\%, 15\% and 14\% for pions, kaons, and
(anti)protons, respectively.  The $N_{part}$ and $N_{coll}$
in each centrality selection are from a Glauber-model 
calculation in \cite{PPG001}, also shown with systematic errors 
based on a 92$\pm$4\% coverage.}
\begin{ruledtabular}
\begin{tabular} {llllll}
                & 0-5\%      & 5-15\%     & 15-30\% & 30-60\% &
60-92\%\\
$N_{part}$  & $347.7\pm10$ & $271.3 \pm8.4$ & $180.2\pm6.6$ &
$78.5\pm4.6$ &$14.3\pm3.3$ \\
$N_{coll}$  & 1008.8 & 712.2 & 405.5 & 131.5
& 14.2 \\ \hline
$\pi^{+}$      & $276\pm3$ & $216 \pm2 $ & $141
\pm1.5 $ & $57.0 \pm0.6 $ & $9.6 \pm0.2 $ \\ $\pi^{-}$      &
$270\pm3.5$ & $200 \pm2.2 $ & $129 \pm1.4 $ & $53.3 \pm0.6 $ & $8.6
\pm0.2 $ \\ $K^+$          & $46.7\pm1.5$ & $35\pm1.3 $ & $22.2\pm0.8
$ & $8.3 \pm0.3 $ & $0.97 \pm0.11 $ \\ $K^-$          & $40.5\pm2.3$ &
$30.4\pm1.4 $ & $15.5 \pm0.7 $ & $6.2 \pm0.3 $ & $0.98 \pm0.1 $ \\ $p$
& $28.7\pm0.9$ & $21.6 \pm 0.6$ & $13.2 \pm0.4$ & $5.0 \pm0.2$ & $0.73
\pm0.06$ \\ $\overline{p}$ & $20.1\pm1.0$ & $13.8 \pm0.6$ & $9.2
\pm0.4$ & $3.6 \pm0.1 $ & $0.47 \pm0.05 $ \\
\end{tabular}
\end{ruledtabular}
\end{table*}
\begin{table*} [ht]
\caption{\label{meanptTable}
The $\langle p_T \rangle$ in MeV/c for hadrons produced at midrapidity
in each centrality  class.  The errors are statistical only.  The
systematic
uncertainties are 7\%, 10\%, and 8\% for pions, kaons, and
(anti)protons,
respectively. }
\begin{ruledtabular}
\begin{tabular} {llllll}
                & 0-5\%      & 5-15\%     & 15-30\% & 30-60\% &
60-92\% \\ \hline
$\pi^{+}$      
	& $390\pm 10$ 
	& $380\pm 10 $ 
	&  $380\pm 20 $ 
	& $360\pm 10 $ 
	& $310\pm 30 $ \\
$\pi^{-}$      
	& $380\pm20$ 
	& $390\pm10 $ 
	& $380\pm10 $
	& $370\pm20 $  
	& $320\pm20 $ \\
$K^+$          
	& $560\pm40$ 
	& $580 \pm40$ 
	& $570 \pm40$ 
	& $550 \pm40$ 
	& $470 \pm90$ \\
$K^-$          
	& $570\pm50$ 
	& $590 \pm40 $ 
	& $610 \pm40 $ 
	& $550 \pm50 $ 
	& $460 \pm90 $ \\
$p$              
	& $880\pm40$ 
	& $870 \pm30 $ 
	& $850 \pm30 $ 
	& $800 \pm30 $  
	& $710\pm80 $ \\
$\overline{p}$ 
	& $900\pm50$ 
	& $890 \pm40 $ 
	& $840 \pm40  $ 
	& $820 \pm40  $ 
	& $800 \pm100 $  \\
\end{tabular}
\end{ruledtabular}
\end{table*}

Pions dominate the charged particle
multiplicity, but a large number of kaons and (anti)protons are
produced.  The inclusive yield of antiprotons is nearly comparable to that of protons.
In the most central Au+Au collisions, the particle density at midrapidity (dN/dy) 
is $\approx$ 20 for antiprotons and 28 for protons, not corrected for feed-down from strange baryons. 

\begin{figure*} [htb]
\includegraphics[scale=0.6]{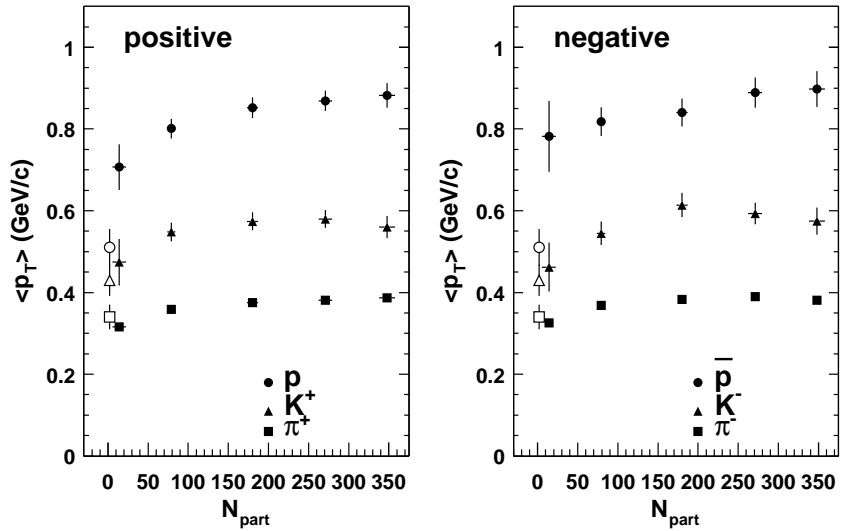}
\caption{\label{fig:meanpt}
The integrated mean $p_{T}$ for pions, kaons, and (anti)protons 
produced in the five different classes of event centrality \cite{PPG006}.  
The error bars are statistical only.  The systematic uncertainties are 
7\%, 10\%, and 8\% for pions, kaons, and (anti)protons,
respectively.  The open points are equivalent average transverse 
momenta from $pp$ and $p\overline{p}$ data, interpolated to 
$\sqrt{s}$=130 GeV.}
\end{figure*}

The average transverse momenta
increase with particle mass and with decreasing impact parameter.
The mean transverse momentum increases with the number of participant nucleons 
by 20$\pm$5\% for pions and protons, as shown in Figure~\ref{fig:meanpt}.  
The $\langle p_{T} \rangle$ of particles produced in $p+p$ and $p\overline{p}$ collisions, 
extrapolated to RHIC energies, are consistent with the most peripheral pion and kaon data; 
however, the $\langle p_{T} \rangle$ of protons produced in Au+Au collisions is significantly higher.
This dependence on the number of participant nucleons may be due to radial expansion.  

\subsection{TRANSVERSE MASS DISTRIBUTIONS}\label{slope}
Production of hadrons from a thermal source would make
transverse mass the natural variable for analysis.
Therefore we extract inverse slopes from the transverse
mass distributions  
by separately fitting a thermal distribution to each particle
species.  
The Boltzmann distribution is given in equation 12. 
\begin{equation}
\frac{d^2 N}{2\pi m_T dm_T dy} = A m_T e^{-m_T/T_{eff}}, \label{boltz}
\end{equation}
We use a simple exponential, however, with no powers of $m_T$ in
the prefactor, as shown in equation 10. This simplification
is acceptable as the difference in the inverse slope is
found to be less than 2\%.  The simple $m_T$ exponential
was also used in an equivalent analysis in Reference \cite{na44_78}.
The inverse slope, $T_{eff}$,
can be compared to other experiments, provided 
the same momentum range of the spectrum is used for fitting.

\begin{figure*} [htb]
\includegraphics[scale=0.6]{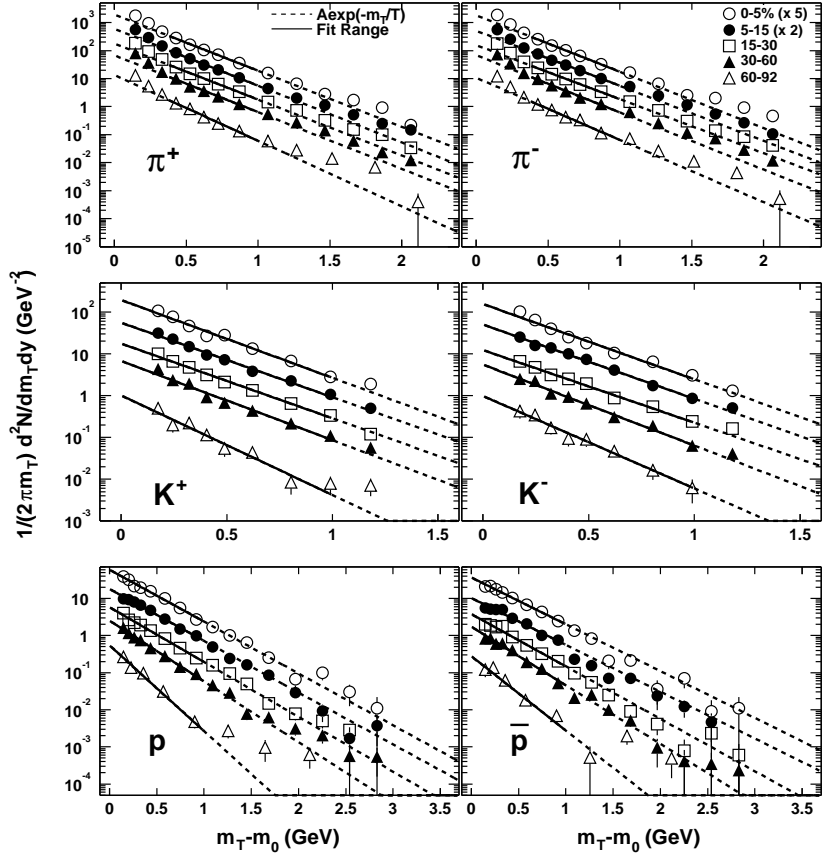}
\caption{\label{mtdist}
Transverse mass distributions of pions (top), kaons (middle),
and (anti)protons (bottom) in events with different centralities.
Positive particle distributions are on the left and negative particle
distributions are on the right.  The solid lines are $m_T$
exponential fits in the range $(m_T-m_0) < 1$ GeV for all particle
species while the dashed lines are the extrapolated fits.}
\end{figure*}

If the system develops collective motion, particles experience
a velocity boost from this motion, resulting in an additional
transverse kinetic energy component. This motivates use of
the transverse kinetic energy, i.e. transverse mass minus the 
particle rest mass, for plotting data.
Figure~\ref{mtdist} shows the transverse kinetic energy distributions
(i.e. transverse mass minus the particle rest mass) 
for all positive particles (left) and negative particles (right).  
Pions are in the top panel, kaons in the middle panel, and
(anti)protons in the bottom panel, with different symbols indicating
different centrality bins. The solid lines are $m_T$
exponential fits in the range $(m_T-m_0) < 1$ GeV for all particle
species while the dashed lines are the extrapolated fits.  The pion
spectra  follow an exponential for $0.38<(m_{T}-m_{0})<1.0$ GeV while
the kaons and  protons appear exponential over the entire measured
$m_{T}$ range.  The same is true for the negative particles in the
right  panel; however, the antiprotons have more curvature for
$(m_{T}-m_{0})<0.5$ GeV.   We extract  $T_{eff}$ by fitting
exponentials of the  form Equation ~\ref{mtexp} to the transverse mass
spectra in  the range $(m_{T}-m_{0})<$1 GeV.

This range is chosen common for all particle species and minimizes
contributions from hard processes.  Caution must be taken when
comparing $T_{eff}$ values as the local slope of the  transverse mass
spectra varies somewhat over $m_{T}$ for pions and  antiprotons even
within this fit range.  The resulting values of  $T_{eff}$ for all
particles and centralities are tabulated in Table~\ref{teffTable} in
units of MeV.  The inverse slopes increase and then saturate for  more
central collisions for all particles except antiprotons.  The fact
that the inverse slope is different for mesons and baryons and for
central and peripheral events  is consistent with the  mean $p_{T}$
trends discussed above.

\begin{table*} [ht]
\caption{\label{teffTable}
The resulting inverse slopes in MeV after fitting  an $m_T$
exponential to the spectra in the range $m_{T}-m_{0}<$1 GeV  in each
event centrality classes.  The pion resonance region is excluded in
the fits.  The equivalent $p_T$ fit range for each particle is shown 
accordingly.  The errors are statistical only.}
\begin{ruledtabular}
\begin{tabular} {llllll}
                & 0-5\%      & 5-15\%     & 15-30\% & 30-60\% &
60-92\%\\ \hline
$\pi^{+}$ in $0.5<p_T<1.05$ GeV/c 
& 216.8 $\pm$ 5.7	
& 214.3 $\pm$ 4.6
& 217.4 $\pm$ 4.7
	           & 214.4 $\pm$ 5.2
        	   & 176.9 $\pm$ 9.5 \\
$\pi^{-}$ in $0.5<p_T<1.05$ GeV/c          
& 215.8 $\pm$ 6.5
& 221.2 $\pm$ 5.6
         & 225.3 $\pm$ 5.8
           & 212.8 $\pm$ 5.7
           & 215.8 $\pm$ 16.8 \\
$K^{+}$ in $0.45<p_T<1.35$ GeV/c           
& 233.2 $\pm$ 10.8
& 243.6 $\pm$ 9.8
           & 242.4 $\pm$ 9.2
           & 228.7 $\pm$ 10.2
           & 182.3 $\pm$ 19.0 \\
$K^{-}$ in $0.45<p_T<1.35$ GeV/c            
& 241.1 $\pm$ 15.8
& 244.5 $\pm$ 10.2
           & 250.0 $\pm$ 12.3
           & 224.2 $\pm$ 11.1
           & 196.4 $\pm$ 22.3 \\
$p$ in $0.55<p_T<1.85$ GeV/c	  
& 310.8 $\pm$ 14.8
& 311.0 $\pm$ 12.3
	  & 293.8 $\pm$ 11.4
	  & 265.3 $\pm$ 10.9
	  & 200.9 $\pm$ 14.8 \\
$\overline{p}$ in $0.55<p_T<1.85$ GeV/c	 
& 344.2 $\pm$ 25.3
& 344.0 $\pm$ 20.9
	  & 307.6 $\pm$ 17.1
	  & 275.1 $\pm$ 14.0
	  & 217.0 $\pm$ 28.3 \\
\end{tabular}
\end{ruledtabular}
\end{table*}

We compare to 
published inverse slopes of transverse 
mass distributions at midrapidity from $m_T$ exponential fits 
in the region $(m_T - m_0) < 1.2$ GeV,
listed in Table~\ref{cern_teff}.
The comparison includes
NA44~\cite{na44_78,na44_kpi,na44_prot,na44_ppb} and WA97 
~\cite{wa97_pbpb,wa97} at the SPS at $\sqrt{s_{NN}} = 17$ GeV; 
and, at $\sqrt{s_{NN}} = 23$ GeV at the ISR, Alper 
et al.~\cite{alper} and Guettler et al.~\cite{guettler}.  
These data are chosen as they match the ($m_T - m_0$) range used
in fitting our data.
For pions, the low-$p_T$ region of ($m_T-m_0) <0.3$ GeV, 
populated by decay of baryonic resonances, is 
systematically excluded from the fits. The effective temperatures 
are given in Table~\ref{cern_teff} with the references 
noted accordingly.

\begin{table*}
\caption{\label{cern_teff}
Inverse slope parameters (in MeV) of hadrons 
for p+p, p+nucleus, and central S+S, S+Pb, and Pb+Pb colliding systems at CERN
energies. Pb+Pb is at $\sqrt{s_{NN}} = 17$ GeV, and the other systems at 23
GeV. The errors are statistical and systematic, respectively.}  
\begin{ruledtabular}
\begin{tabular}{cccccccc}  
Hadron &Pb+Pb &S+Pb &S+S  &p+Pb  &p+S  &p+Be &p+p\\ \hline 
$\pi^+$ &156$\pm$6$\pm$23\footnotemark[1] 
&165$\pm$9$\pm$10\footnotemark[2]
&148$\pm$4$\pm$22\footnotemark[1] 
&145$\pm$3$\pm$10\footnotemark[2] 
&139$\pm$3$\pm$10\footnotemark[2]  
&148$\pm$3$\pm$10\footnotemark[2] 
&139$\pm$13$\pm$21\footnotemark[3]\\ 
$K^+$ 
&234$\pm$6$\pm$12\footnotemark[1]
&181$\pm$8$\pm$10\footnotemark[2]  
&180$\pm$8$\pm$9\footnotemark[1]
&172$\pm$9$\pm$10\footnotemark[2] 
&163$\pm$14$\pm$10\footnotemark[2] 
&154$\pm$8$\pm$10\footnotemark[2] 
&139$\pm$15$\pm$7\footnotemark[3] \\ 
$p$     
&289$\pm$7$\pm$14\footnotemark[4]
&256$\pm$4$\pm$10\footnotemark[5]  
&208$\pm$8$\pm$10\footnotemark[1]
&203$\pm$6$\pm$10\footnotemark[5] 
&175$\pm$30$\pm$10\footnotemark[5] 
&156$\pm$4$\pm$10\footnotemark[5] 
&148$\pm$20$\pm$7\footnotemark[3] \\ 
$\Lambda$ 
&289$\pm$8$\pm$29\footnotemark[6]
&---  &--- 
&203$\pm$9$\pm$20\footnotemark[7]
&---  &---  &--- \\ 
$\overline \Lambda$ 
&287$\pm$13$\pm$29\footnotemark[6]
&---   &---  
&180$\pm$15$\pm$18\footnotemark[7]
&--- &---  &---  \\ 
\end{tabular}
\end{ruledtabular}
\footnotetext[1]{Reference \cite{na44_78} (NA44 Collaboration).}
\footnotetext[2]{Reference \cite{na44_kpi} (NA44 Collaboration).}
\footnotetext[3]{Reference \cite{alper,guettler} (ISR).}
\footnotetext[4]{Reference \cite{na44_ppb} (NA44 Collaboration).}
\footnotetext[5]{Reference \cite{na44_prot} (NA44 Collaboration).}
\footnotetext[6]{Reference \cite{wa97_pbpb} (WA97 Collaboration).}
\footnotetext[7]{Reference \cite{wa97} (WA97 Collaboration).}
\end{table*}
Radial flow imparts a radial velocity boost on top of the
thermal distribution.
Heavy particles are boosted to higher $p_{T}$, depleting the
cross section at lower $p_{T}$ and yielding a higher
inverse slope.  Therefore, the observed inverse slope dependence on
both centrality and particle mass implies more radial expansion
in more central collisions.  
At CERN SPS, the $T_{eff}$ depends on both mass and system size 
(the number of participating nucleons in the collision), indicating 
collective expansion. The $T_{eff}$ values at RHIC shown in Table IV
are somewhat larger.

\begin{figure*} [htb]
\includegraphics[scale=0.8]{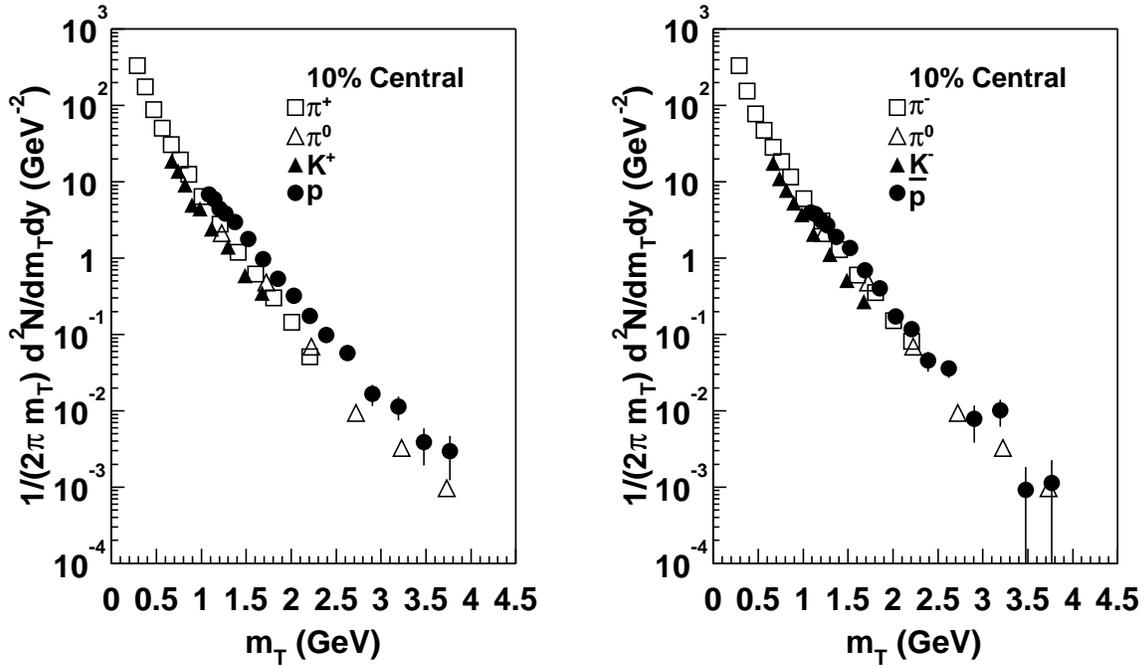}
\caption{\label{AllHadrons_mt}
The transverse mass distributions of inclusive identified 
hadrons produced in 10\% central events, including the $\pi^{0}$ as 
measured in the Electromagnetic Calorimeter in PHENIX and published 
in \protect\cite{PPG003}.}
\end{figure*}

In p-p collisions at similar $\sqrt{s}$ at the ISR, 
hadron spectra were analyzed in transverse
mass, $m_T$, rather than transverse kinetic energy $m_T - m_0$
\cite{pp_mt1, pp_mt2}. To facilitate a direct comparison, figure
~\ref{AllHadrons_mt} shows the PHENIX hadron spectra,
including $\pi^{0}$ from the 10\% most central Au+Au collisions.
The spectra approach one another, but do not fall upon a
universal curve, and thereby fail the usual definition of scaling.

It has been suggested that at transverse mass significantly larger 
than the
rest mass of the particle, thermal emission and radial flow may 
not be the only physics affecting the particle spectra. If heavy 
ion collisions can be described as collisions of two sheets of colored
glass in which the gluon occupation number is sufficiently large
to saturate, scaling of different hadron spectra with transverse 
mass is also predicted \cite{cgc}. For Au+Au collisions at different
impact parameters, the saturation scale differs, and some
differences in the spectra may be expected. Nevertheless, the
authors observe that the
level of $m_T$ scaling in our data is in qualitative agreement with
expectations from gluon saturation \cite{cgc}. Single particle spectra 
alone, however, 
are not sufficient to disentangle saturation from flow effects.

\begin{figure*} [htb]
\includegraphics[scale=0.6]{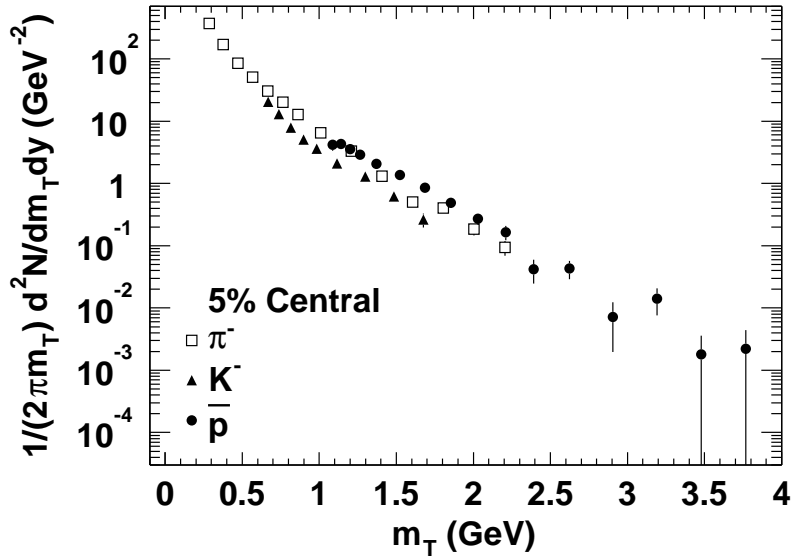}
\caption{\label{mtscaleC0}
The transverse mass distributions of the inclusive identified 
hadrons produced in the 5\% central events.}
\end{figure*}

It is often stated that $m_T$ scaling holds in pp collisions at
similar $\sqrt{s}$ to RHIC [see data, for example, in references
\cite{pp_mt1} and \cite{pp_mt2}]. Scaling 
in $m_T$, i.e. spectra following a universal curve in $m_T$, might 
be expected if the hadrons are emitted from a source in thermal 
equilibrium. 
It is instructive to note that 
reference \cite{pp_mt1} states ``Although the curves for different
particles do come together, there is no real evidence for any 
universal behavior in this variable." Thus, scaling at the ISR
was never claimed by the original authors. In central Au+Au
collisions, the slopes and
yields of $\pi$, K and p approach each other as well, but
figures~\ref{AllHadrons_mt} and~\ref{mtscaleC0} 
also do not support a truly universal behavior in $m_T$. 
Therefore the apparent puzzle of how the data could
exhibit both $m_T$ scaling and the mass-dependent $p_T$ boost
characteristic of radial flow is no puzzle at all, as 
any ``$m_T$ scaling" is only very approximate.

\begin{figure*} [htb]
\includegraphics[scale=0.6]{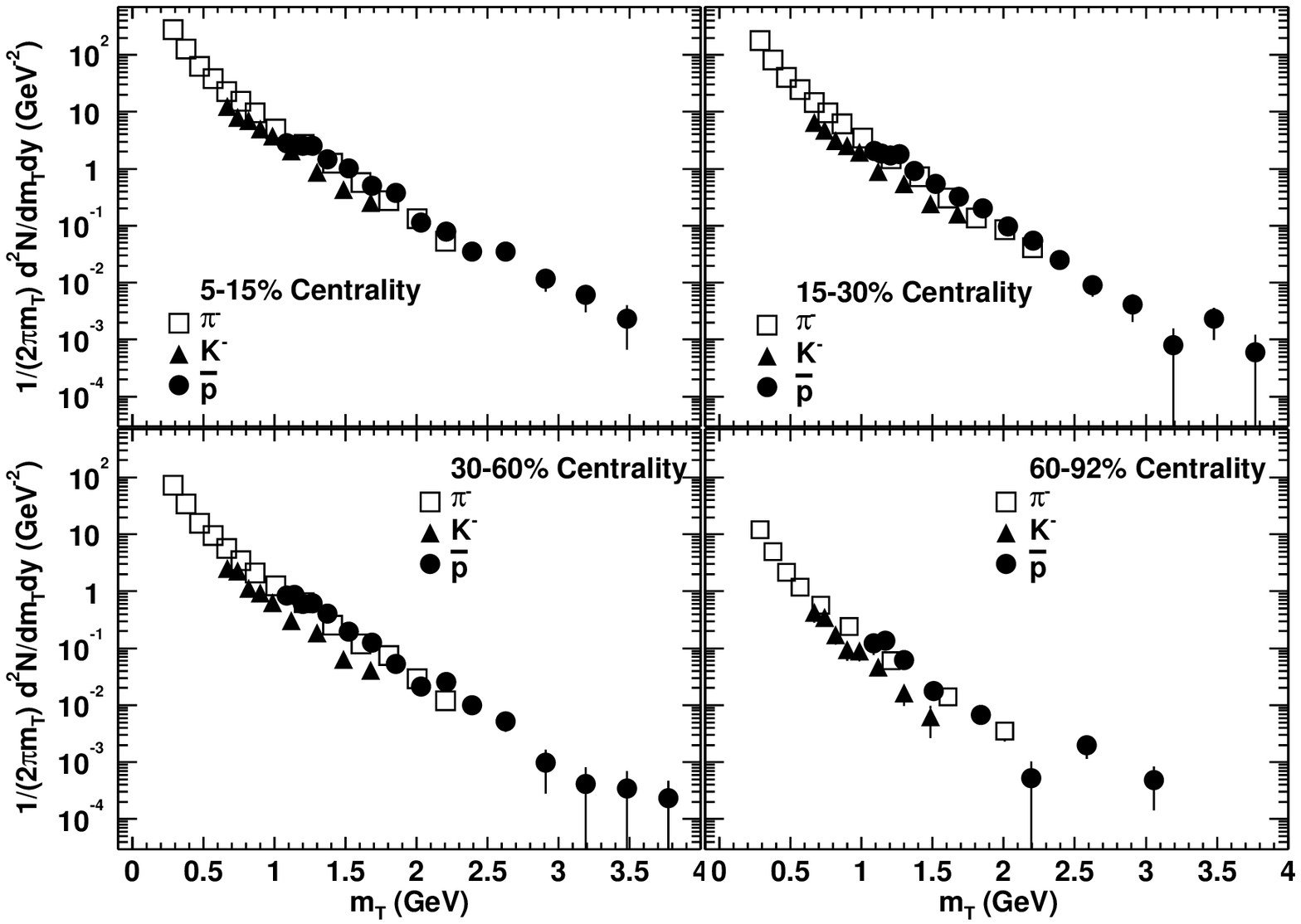}
\caption{\label{mtscaleC1-C45}
The transverse mass distributions of the inclusive identified 
hadrons produced in events with different centralities.}
\end{figure*}

\subsection{SUMMED CHARGED PARTICLE MULTIPLICITY}
\begin{figure*}[htb]
\begin{center}
\includegraphics[scale=0.6]{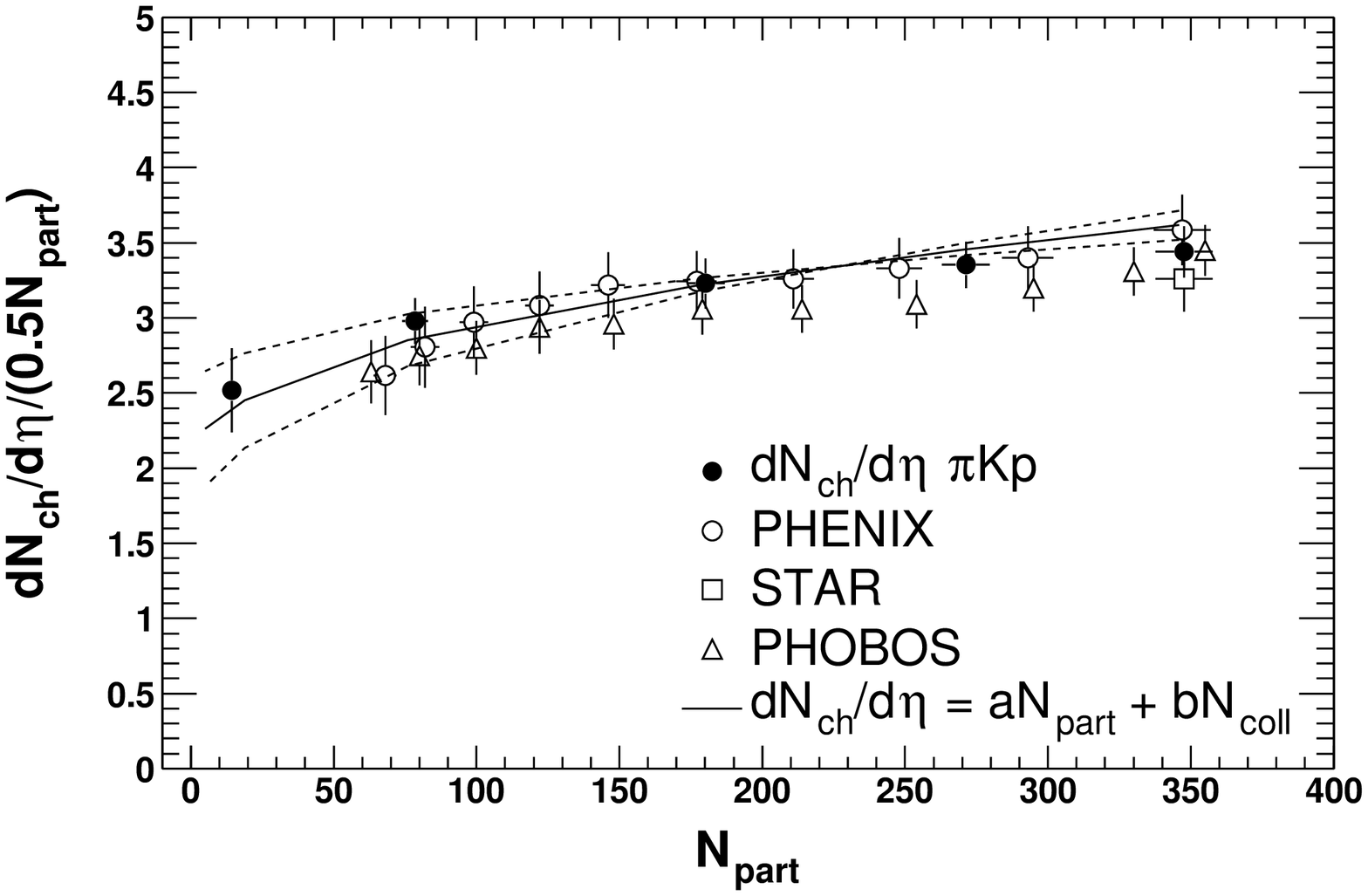}
\end{center}
\caption{\label{compare}
Both the total charged multiplicity (open) in References
\protect\cite{pubDndeta,star_dndeta,phobos_dndeta2} and the total 
identified charged multiplicity (closed) scaled by the number of 
participant pairs are plotted together as a function of the number of
participants.  The lines correspond to the fit of a linear parameterization of 
$N_{part}$ and $N_{coll}$ to the PHENIX charged multiplicity measurement (open circles) with $a = 0.88\pm0.28$ and $b = 0.34\pm0.12$ as 
described in \protect\cite{pubDndeta}.}
\end{figure*}

As a consistency check we compare the measured rapidity densities
as given in Section ~\ref{data} to previously published pseudorapidity
densities of charged particles.  The measured dN/dy for each hadron species is converted to
dN/d$\eta$, and the total dN/d$\eta$ is calculated by summation. 
Figure~\ref{compare} shows dN/d$\eta$ per participant nucleon pair, 
compared to the measurement made by PHENIX
using the pad chambers alone \cite{PPG001} as well as to
PHOBOS and STAR yields in central collisions \cite{phobos_dndeta2,star_dndeta}.
We note that the lines correspond to the fit of a linear parameterization of 
$N_{part}$ and $N_{coll}$ to the PHENIX measurement (open circles) with $a = 0.88\pm0.28$ and $b = 0.34\pm0.12$ as described in \cite{pubDndeta}.  For the 5\% central collisions, we measure $598\pm30$, and is comparable to 
the STAR result of $567\pm38$ \cite{star_dndeta}, the PHOBOS result of $555\pm37$ \cite{phobos_dndeta}, and the PHENIX pad chamber result of $622\pm41$ \cite{PPG001}.
The agreement is excellent, allowing the results of this analysis to be used to
decompose the particle type dependence of the charge particle
multiplicity increase with centrality.

\section{\label{models} COMPARISON WITH MODELS}
\subsection{HYDRODYNAMIC-INSPIRED FIT \label{hydrofits}}
The charged particle pseudorapdity distributions are incompatible 
with a static thermal source, but the 
flat distribution observed in \cite{phobos_dndeta} 
reflects
the strong longitudinal motion in the initial 
state.  Consequently, the longitudinal momentum distribution is not 
an unambiguous sign of collective motion.  Transverse momentum is, 
however, generated in the collision, so 
collective expansion may be more easily inferred from 
transverse momentum distributions.

Following the arguments of the previous section, we analyze the
particle $m_T$ spectra. A parameterization of
the $m_T$ distribution of particles emitted from a 
hydrodynamic expanding hadron source is used. 
In order to determine the freeze-out  temperature and collective flow
without confusion from hard scattering processes, a limited
$p_T$ range is used in the  fits.  We include only particles with
$(m_{T} - m_{0}) < $ 1 GeV  in the fit.  Pions with $(m_{T} - m_{0}) <
0.38$ GeV are excluded  to avoid resonance decays.  All particles are
assumed to decouple from the expanding hadron source \cite{sollfrank} 
at the same freeze-out temperature, $T_{fo}$. This procedure allows 
us to extract $T_{fo}$ and the magnitude of the collective boost in the 
transverse direction.

The inverse slope includes the local temperature of a section of the
hadronic matter along with its collective velocity.  The simple
exponential fit of Equation~\ref{mtexp} treats each particle spectrum
as a static thermal source, and a collective expansion velocity cannot
be extracted reliably from a single particle spectrum.  However, the
relative sensitivity to the temperature and collective radial flow
velocity differs for different particles.  By using the information
from all the  particles, the expansion velocity can be inferred.  We
fit all particle species simultaneously with a functional form for a
boosted thermal source based on  relativistic
hydrodynamics\cite{sollfrank}.

Use of this form assumes that
\begin{itemize}
\item all particles decouple kinematically on a freeze-out
hypersurface at the  same freeze-out temperature $T_{fo}$,
\item the particles collectively expand with a velocity profile
increasing linearly with the radial position in the source (i.e., 
Hubble expansion where fluid elements do not pass through one another), and
\item the particle density distribution is independent of the radial
position.
\end{itemize}
Longitudinally boost invariant expansion of the particle source is also
 assumed.

The transverse velocity profile is parameterized as:
\begin{equation}
\beta_{T}(\xi) = \beta_{T}^{max}\xi^{n} , \label{prof}
\end{equation}
where $\xi = \frac{r}{R}$, and  R is the maximum radius of the
expanding source at freeze-out  ($0<\xi<1$) \cite{esumi_55}. The
maximum surface velocity is given by $\beta_{T}^{max}$, and for a linear velocity
profile, n = 1.  The average of the transverse velocity is
equal to:
\begin{equation}
\langle \beta_{T} \rangle = \frac { \int{\beta_{T}^{max}\xi^{n}\xi d\xi}}{
\int{\xi d\xi} } = \frac{2}{2+n} \beta_{T}^{max}. \label{geoave}
\end{equation}
Each fluid element is locally thermalized and receives a transverse
boost $\rho$ that depends on the radial position as:
\begin{equation}
\rho = \tanh^{-1} \left( \beta_{T}\left( \xi \right) \right)
. \label{rho}
\end{equation}

The $m_T$ dependence of the invariant yield $\frac{dN}{m_T
dm_T}$ is determined by integrating over the rapidity, azimuthal
angle, and radial distribution of fluid elements in the source.  This
procedure, discussed in Appendix ~\ref{deriv}, yields
\begin{eqnarray}
\lefteqn{\frac{d^2 N}{m_T dm_T dy} =} \\
\nonumber & A \int_0^1{m_{T} f(\xi) K_1 \left( \frac{ m_{T} \cosh(\rho) }{ T_{fo} }
	\right)  I_0 \left( \frac{p_T \sinh(\rho)}{T_{fo}} \right) \xi
	d\xi} .& \label{hydro}
\end{eqnarray}
The parameters determined by fitting Equation 16
to the
data are the freeze-out temperature $T_{fo}$, the normalization A, and
the maximum surface velocity $\beta_{T}^{max}$ using a flat particle density 
distribution (i.e., $f(\xi) = 1$).

To study the parameter correlations, we make a grid of combinations of
temperature and velocity, and perform a chi-squared minimization to
extract the normalization, A, for each particle type. The fit is done
simultaneously for all particles in the range $(m_{T} - m_{0}) < 1$
GeV.  In addition to this upper limit in the fit, the pion fit range
includes a lower  limit of $(m_{T}-m_{0}) > 0.38\ \rm{GeV}$ to avoid
the resonance contribution  to the low $p_{T}$ region (see Section
~\ref{reso}).

\begin{figure*} [htb]
\includegraphics[scale=0.6]{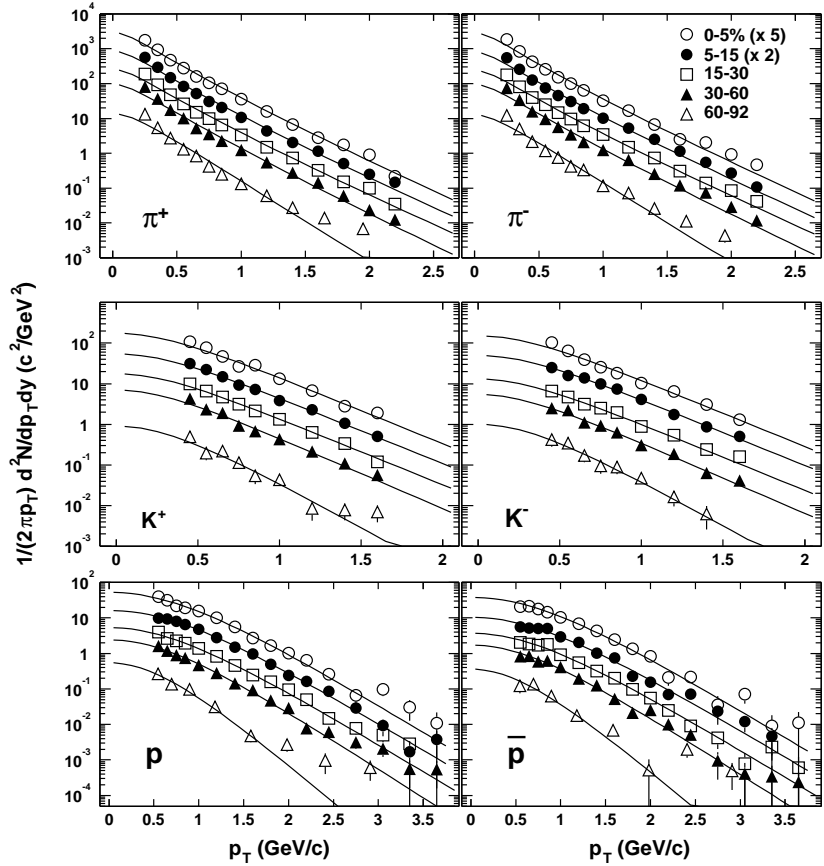}
\caption{\label{fits}
The parameterization and the $p_T$ hadron spectra for all
five centrality selections.}
\end{figure*}

The radial flow velocity and freeze-out temperature for all centralities are
determined in the same way.  The results are plotted together with the
spectra in Figure~\ref{fits}.  The  hydrodynamic
form clearly describes the spectra better than the simple 
exponential in Figure~\ref{mtdist}. The values for
$T_{fo}$ and $\beta_{T}^{max}$ are tabulated in Table~\ref{fitResults}.  

\begin{table} [ht]
\caption{\label{fitResults}
The minimum $\chi^{2}$ and the parameters $T_{fo}$ and
$\beta_{T}^{max}$ for each of the five centrality selections.   The best fit
parameters are determined by averaging all parameter pairs  within the
1$\sigma$ contour.  The errors correspond to the standard deviation of
the parameter pairs within the 1$\sigma$ $\chi^{2}$ contour.  It is important to note that the fit range in Figure~\ref{fits} is the same as was 
used to fit $m_T$ exponentials to the spectra in Figure~\ref{mtdist}.}
\begin{ruledtabular}
\begin{tabular} {ccccc}
Centrality (\%) & $\chi^{2}$/dof & $T_{fo}$ (MeV) & $\beta_{T}^{max}$ &
$<\beta_T>$\\ \hline  0-5 & 34.0/40 & $121\pm4$ & $0.70\pm0.01$ &
0.47$\pm$0.01\\  5-15 & 34.7/40 & $125\pm2$ & $0.69\pm0.01$ &
0.46$\pm$0.01\\ 15-30 & 36.2/40 & $134\pm2$ & $0.65\pm0.01$ &
0.43$\pm$0.01\\ 30-60 & 68.9/40 & $140\pm4$ & $0.58\pm0.01$ &
0.39$\pm$0.01\\  60-92 & 36.3/40 & $161\pm^{19}_{12}$ &
$0.24\pm^{0.16}_{0.2}$ & 0.16$\pm^{0.16}_{0.2}$ \\ 
\end{tabular}
\end{ruledtabular}
\end{table}

\begin{figure} [htb]
\resizebox{\columnwidth}{!}{\includegraphics{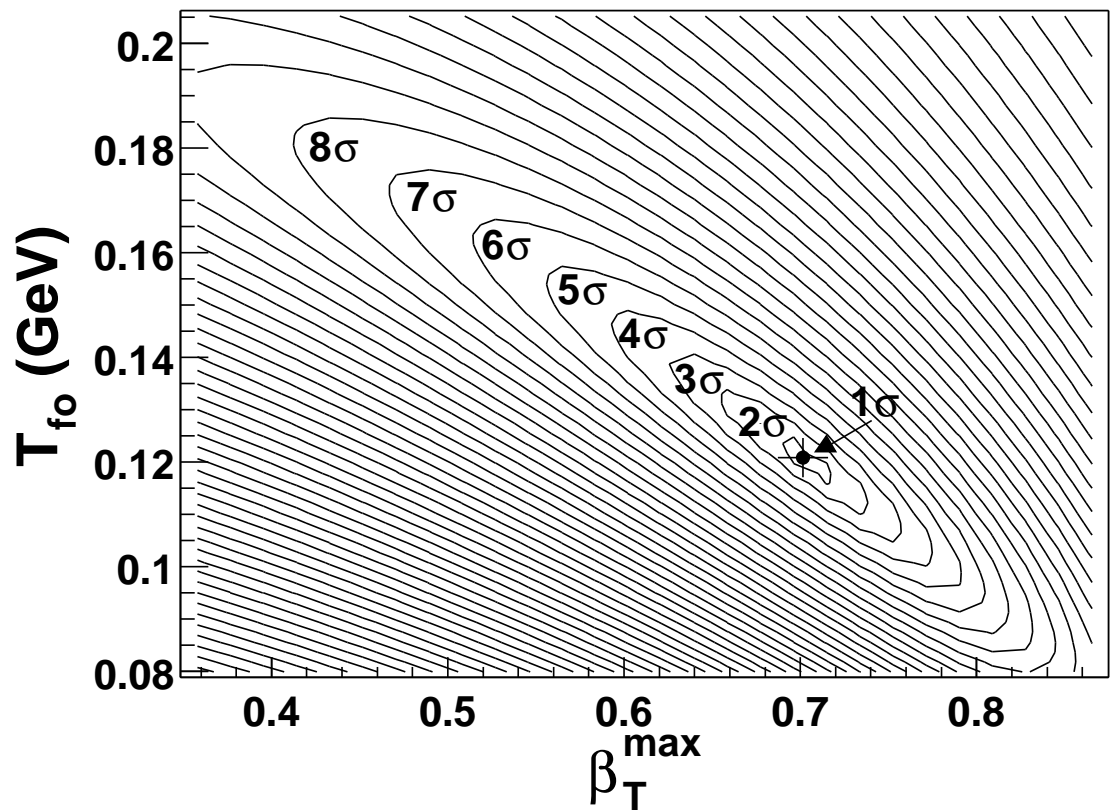}}
\resizebox{\columnwidth}{!}{\includegraphics{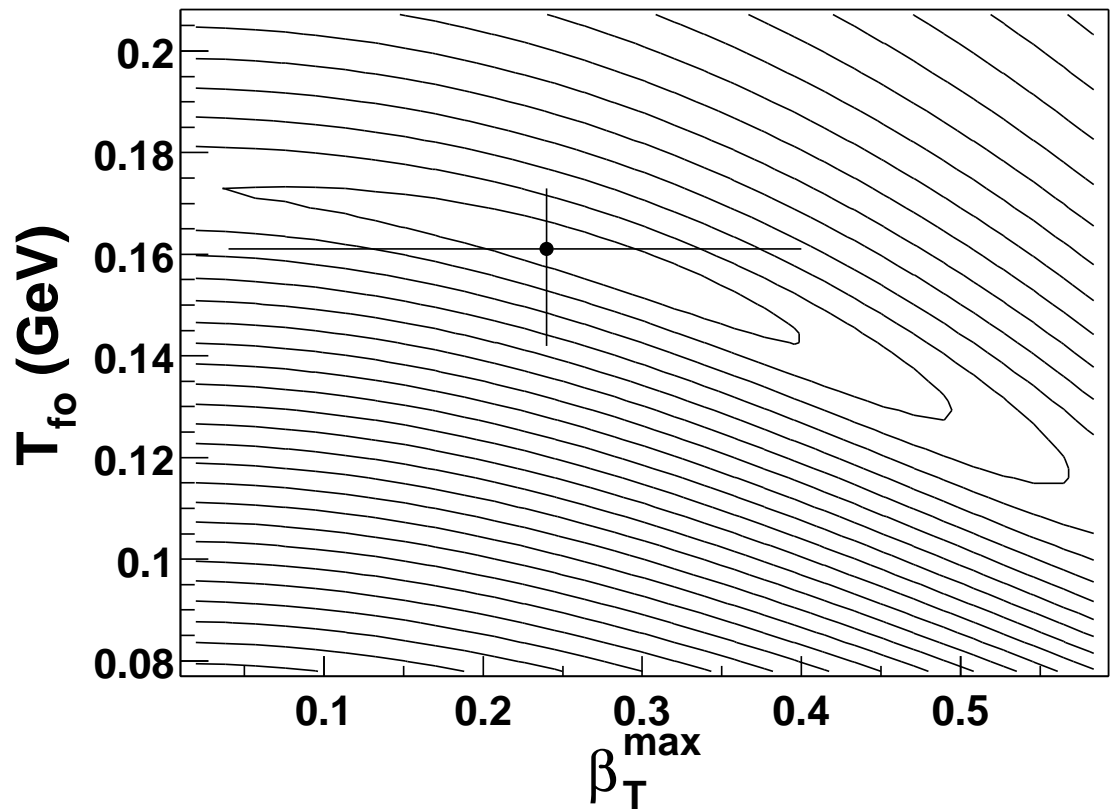}}
\caption{\label{Fig:contours}
The $\chi^{2}$ contours in the parameter space $T_{fo}$ and
$\beta_{T}^{max}$ that result after simultaneously fitting hadrons in the
$0-5\%$ centrality (top) and $60-92\%$ (bottom).  The n-sigma contours
are labeled accordingly in the top panel.  The fit results are
tabulated in Table~\ref{fitResults}.}
\end{figure}

Figure~\ref{Fig:contours} shows $\chi^{2}$ contours for the 
temperature and velocity parameters for the 5\%
most central collisions.  The n-sigma contours are labeled up to
8$\sigma$.  The $\chi^{2}$ contours indicate strong anti-correlation
of the two parameters.  If the freeze-out temperature  decreases, the
flow velocity increases.  The minimum $\chi^{2}$ is $34$  and the
total number of degrees of freedom (dof) is 40.  The parameters that
correspond to  this minimum are $T_{fo} = 121\pm4$MeV and $\beta_{T}^{max} =
0.70\pm0.01$.    The quoted errors are the 1$\sigma$ contour widths of
$\Delta \beta_{T}^{max}$ and $\Delta T_{fo}$.  Within
3$\sigma$,  the $T_{fo}$ range is $106-141$ MeV and the $\beta_{T}^{max}$
range  is $0.75-0.64$.

As a linear velocity profile ($n=1$ in Equation~\ref{prof}) is
assumed, the mean flow velocity in the  transverse plane is $\langle
\beta_{T} \rangle = 2\beta_{T}^{max}/3$.  If a different particle  density
distribution (for instance, a Gaussian function for $f(\xi)$) were used, then the average
should be  determined after weighting accordingly \cite{esumi_55}.

A similar analysis for Pb+Pb collisions at 158 A GeV, was reported 
by the NA49 Collaboration in \cite{na49}.  Using the same hydrodynamic 
parameterization, simultaneous fits of several hadron species for the 
highest energy results in $T_{fo} = 127\pm1$ MeV and $\beta_T^{max} = 
0.48\pm0.01$ 
with $\chi^2/NDF = 120/43$ for positive particles and $T_{fo}=114\pm2$ MeV 
and $\beta_T^{max} = 0.50\pm0.01$ with $\chi^2/NDF = 91/41$ for 
negative particles 
(statistical errors only).  Pions and deuterons are excluded from the 
fits to avoid dealing with resonance contributions to the pion yield and 
formation of deuterons by coalescence.  The $\phi$ meson is included 
in the fit together with the negative particles.  Previously, NA49
used a different parameterization to fit the charged hadron and deuteron spectra,
as well as the $m_T$ dependence of measured HBT source radii, resulting in 
overlapping $\chi^2$ contours with $T_{fo} = 120\pm12$ MeV and 
$\beta_T^{max} = 0.55\pm0.12$ \cite{na49_old}.
    
\subsubsection{\label{parab}VELOCITY AND PARTICLE DENSITY PROFILE}
In order to use $\beta_{T}^{max}$ and $T_{fo}$ from the fits described 
above, one needs to know their sensitivity to the assumed velocity 
and particle density profiles in the emitting source.
The choice of a linear velocity profile within the source is
motivated by the profile observed in a full hydrodynamic
calculation \cite{kolb}, which shows a nearly perfect  linear
increase of $\beta$(r) with r.  Nevertheless, we also used a parabolic
profile to check the sensitivity of the results to details of the
velocity profile. For a parabolic velocity profile ($n = 2$ in
Equation~\ref{prof}), $\beta_T^{max}$ increases by $\approx$ 13\% and
$T_{fo}$ increases by $\approx$ 5\%.

A Gaussian density profile used with a linear velocity profile
increases $\beta_{T}^{max}$ by $\approx$ 2\%, with a neglible difference in
the temperature $T_{fo}$.  As a test of the assumption that all the
particles freeze out at a common temperature, the simultaneous fits
were repeated without the kaons.  The difference in $T_{fo}$ is within
the measured uncertainties.

\subsubsection{\label{reso}INFLUENCE OF RESONANCE PRODUCTION}
The functional forms given by Equations~\ref{mtexp} and~\ref{hydro}
do not include particles arising from resonance or weak decays.  As
resonance decays are known to result in pions at low transverse
momenta \cite{restheory,res814,resna44}, we place a $p_T$ threshold of
500 MeV/c on pions included in the hydrodynamic fit. A similar
approach was followed by NA44, E814, and other experiments at lower
energies, which performed in-depth studies of resonance decays feeding 
hadron spectra.  However, these were for systems with higher 
baryon density, so we performed a
cross check on possible systematic uncertainties arising from the pion
threshold used in the fits. To estimate the effect of resonance decays
were they not excluded from the fit, we calculate resonance
contributions following Wiedemann \cite{wiedemann96}.

In order to reproduce the relative yields of different particle types, a 
chemical freeze-out temperature -- different from the kinetic
freeze-out temperature -- and a baryonic chemical potential are
introduced.  Direct production and resonance contribution are
calculated for pions and (anti)protons assuming a kinetic freeze-out
temperature of 123 MeV, a transverse flow velocity of $0.612$
(equivalent to $\langle \beta_T \rangle = 0.44$), a baryon chemical
potential of 37 MeV, and a chemical freeze-out temperature (when
particle production stops) of 172 MeV. These parameters are chosen 
as they provide a reasonable description of the (anti)proton and pion spectra and 
yields (10\% most central) and are in good agreement with chemical freeze-out analyses
\cite{chemfo}.  Most spectra from resonance decays show a steeper
fall-off than the direct production, which should lead to a smaller
apparent inverse slope, depending on what fraction of the low $p_T$
part of the spectrum is included in the fits.

To measure the effect of resonance production on the spectral shape, 
the local slope is determined.  For a given $m_T$ bin number i, the local slope is
defined as
\begin{equation}
T_{local} \left( i \right) = - \frac{m_T (i+1) - m_T
(i-1)}{\mathrm{log}[N(i+1)] - \mathrm{log}[N(i-1)]},
\end{equation}
which is identical to the inverse slope independent of $m_T$ for an
exponential.

\begin{figure} [htb]
\resizebox{\columnwidth}{!}{\includegraphics{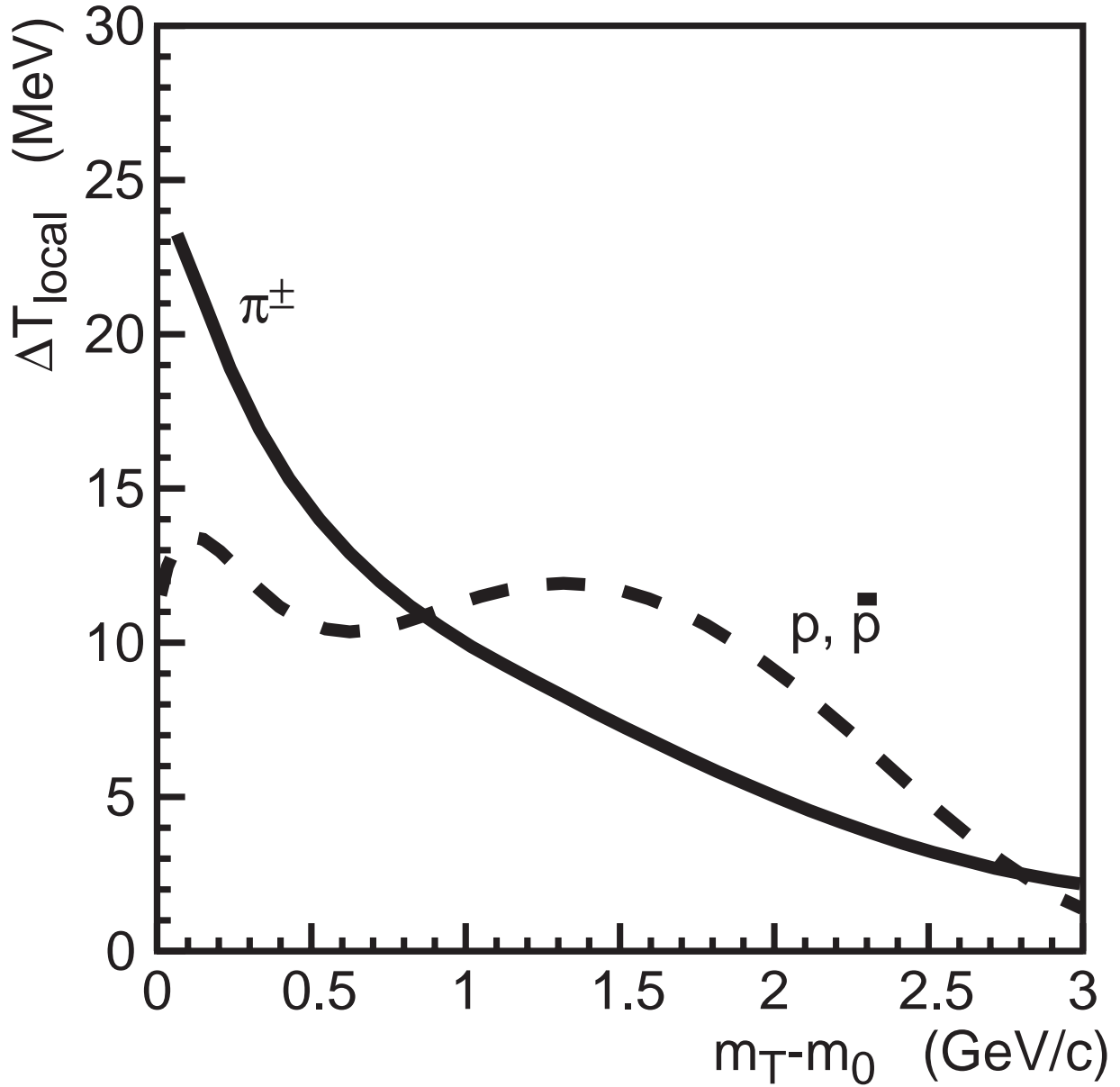}}
\caption{\label{resonances}
The difference in the local slope for direct and inclusive
pions (solid) and (anti)protons (dashed).}
\end{figure}

The difference in the local slope,
\begin{equation}
\Delta T_{local} = T_{local}^{direct} - T_{local}^{incl},
\end{equation}
is determined for direct and inclusive pions and (anti)protons.  The
differences are plotted as a function of $m_T -m_0$ in Figure
~\ref{resonances}.  The difference in local slope for protons is below
13 MeV for the full transverse mass range; the non-monotonic behavior
for protons is caused by the relatively strong transverse flow. For
pions, $\Delta T_{local}$ decreases monotonically with $m_{T}$ and is
below 10 MeV above $m_{T} = 1\mathrm{GeV}/c$. A fit of an exponential
to the pion spectra for $(m_{T}-m_{0}) > 0.38\ \rm{GeV}$ (which
corresponds to $p_T>0.5$ GeV/c) yields a difference in inverse slope
of 16~MeV with and without resonances.

\subsection{COMPARISON WITH HYDRODYNAMIC MODELS}
Hydrodynamic parameterizations as used in the previous
Section rely upon many simplifying assumptions.  Another approach to the
study of collective flow is to compare the data to hydrodynamic models.
Such models assume rapid equilibration  in the collision and describe
the subsequent motion of the matter using the laws of
hydrodynamics. Large pressure buildup is found, and we investigate
this ansatz by checking the consistency of the data with
calculations using a  reasonable set of initial conditions. We compare
to two separate models, the hydrodynamics model of Kolb and Heinz
\cite{peter1,peter2,peter3} and the ``Hydro to Hadrons" (H2H) model of
Teaney and Shuryak \cite{derekPhD,derek1}.  The H2H model consists of
a hydrodynamics calculation, followed by a hadronic cascade after
chemical freeze-out. The cascade step utilizes the Relativistic
Quantum Molecular Dynamics (RQMD) model, developed for lower energy
heavy ion collisions \cite{RQMD}.

In both models, initial conditions are tuned to reproduce the shape of
the transverse momentum spectra measured in the most central
collisions, along with the charged particle yield. Each model also
includes the formation and decay of resonances.  

\begin{figure*}[hbt]
\begin{center}
\includegraphics[scale=0.6]{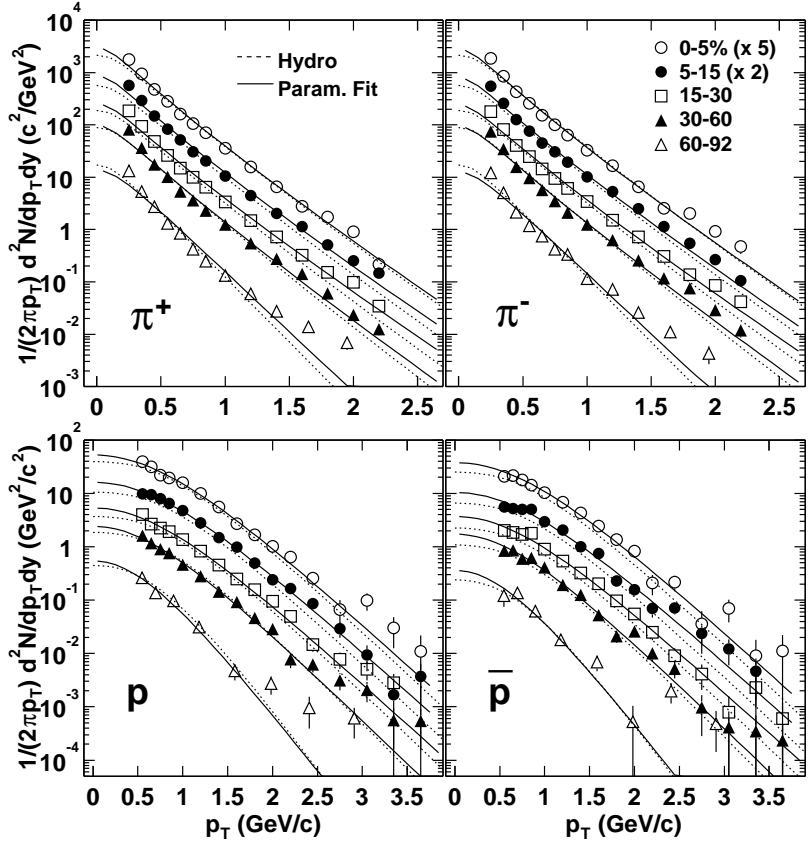}
\end{center}
\caption{\label{hydroCompare}
The hydrodynamics calculation with initial parameters  tuned
to match the most central spectra in the $p_T$ range $0.3-2.0$ GeV/c.}
\end{figure*}

In the Kolb and Heinz model, the initial parameters are the entropy density, 
baryon number density, the equilibrium time, and the freeze-out temperature 
which controls the duration of the expansion.  The chemical freeze-out 
temperature is the temperature at which  
particle production ceases.  The initial entropy or energy density  
and maximum temperature are fixed to match the measured multiplicity for 
the most central collisions using a parameterization that is tuned to 
produce the measured d$N_{ch}$/d$\eta$ dependence on both $N_{part}$ 
and $N_{coll}$.  A kinetic freeze-out temperature
of $T_{fo} = $ 128 MeV is used.  Spectra from the Kolb-Heinz
hydrodynamic model are shown in Figure~\ref{hydroCompare} for pions
(upper) and for protons (lower) as dotted lines.  The solid lines are
the results from the fits described in the previous sections. The
figure thus allows two comparisons. The similarity of the dashed and
solid lines shows that the hydrodynamic-inspired parameterization used
to fit the data  results in a $p_T$ distribution similar to this
hydrodynamic  calculation.  Comparing the dashed lines to the data
points shows that the hydrodynamic model agrees quite well for most
of the centrality ranges.  It is important to note that the model
parameters  are uncertain at the level of 10\%, and, more importantly,
the  application of hydrodynamics to peripheral collisions may be less
reasonable than for central collisions, as hydrodynamic calculations 
assume strong rescattering and a sufficiently large system size (discussed 
in \cite{peter3}).

In reference \cite{derekPhD,derek2}, the PHENIX $\overline{p}$
spectrum shape is well described by the H2H model with the LH8
equation of state. The cascade step in the H2H model removes the
requirement that all particles freeze out at a common temperature.
Thus the freeze-out temperature and its profile are predicted, 
rather than input parameters. 
Furthermore, following the hadronic interactions
explicitly with RQMD removes the need to rescale the particle ratios
at the end of the calculation, as they are fixed by the hadronic cross
sections rather than at some particular freeze-out temperature.  The
LH8 equation of state includes a phase transition with a latent heat
of 0.8 GeV. In \cite{derekPhD,derek2}, the $\Omega$ and the $\phi$ are
shown to decouple from the expanding system at $T =$ 160 MeV, and they
receive a flow velocity boost of 0.45c. Pions and kaons decouple at $T
=$ 135 MeV with flow velocity $= 0.55$c, while protons have $T=$ 120
MeV and flow velocity $\ge$ 0.6. These temperatures and flow velocities
are consistent with the values extracted from the data for the most
central events. However, the average initial energy density exceeds
the experimental estimate using formation time $\tau_0 = $1 fm/c.

\begin{figure} 
\resizebox{\columnwidth}{!}{\includegraphics{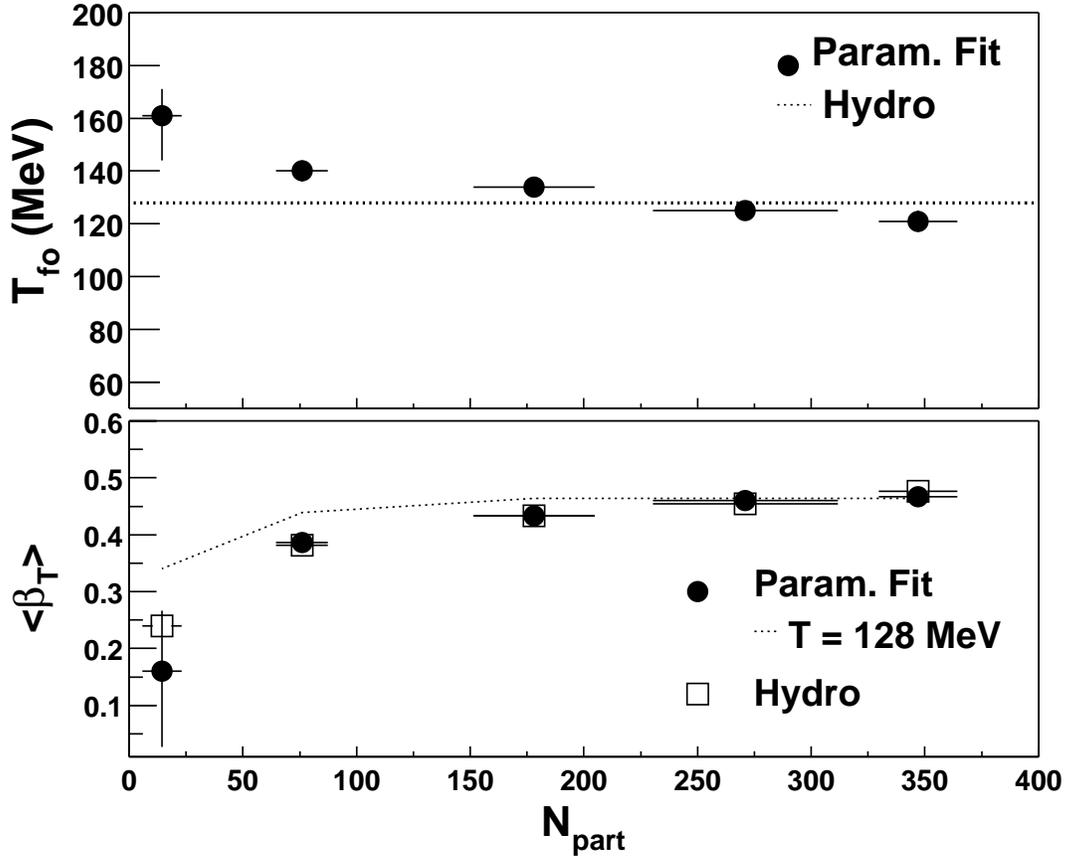}}
\caption{\label{parCent}
The expansion parameters $T_{fo}$ and $\beta_{T}^{max}$ as a
function of the number of participants.  As a comparison, the results
from a hydrodynamic model calculation are also shown (Ref. \cite{peter1,peter2,peter3}).  The
dashed line  corresponds to a fixed temperature of 128 MeV for all
centralities  in the parameterized fit to the data.}
\end{figure}

In Figure~\ref{parCent}, radial flow from the fits of the previous
section are  shown as a function of the number of participants for
$T_{fo}$ (top) and $\langle\beta_T\rangle$ (bottom). There is a slight
decrease of $T_{fo}$, while $\langle\beta_T\rangle$ increases with $N_{part}$,
saturating at $0.45$.  The value of $\langle\beta_T\rangle$ from Kolb and Heinz is
also shown, and agrees with the data reasonably well.  In the plot of 
$\langle \beta_T \rangle$, the dashed line indicates the results of 
fitting the parameterization to the data while keeping $T_{fo}$ fixed 
at 128 MeV to agree with the value used by Kolb and Heinz. Radial flow 
values for central collisions remain unchanged,
while those in peripheral collisions increase. Even with the extreme
assumption that all collisions freeze out at the same temperature,
regardless of centrality, the trend in centrality dependence of the
radial flow does not change.

\subsection{HYDRODYNAMIC CONTRIBUTIONS AT HIGHER $p_T$}
We use the parameters extracted from the fit to the charged hadron
spectra in the low $p_T$ region to extrapolate the effect of the soft
physics to higher $p_T$. This yields a prediction for the 
spectra of hadrons should a collective expanding thermal source be the only
mechanism for particle production in heavy ion collisions. Comparing
this prediction to the measured spectrum of charged particles or neutral pions should
indicate the $p_T$ range over which soft thermal processes dominate
the cross section. Where the data deviate from the hydrodynamic
extrapolation, other contributions, as e.g. from hard processes  or
non-equilibrium production become visible.  The approach described
here differs from hydrodynamic fits to the entire hadron spectrum, as
we fix the parameters from the low $p_T$ region alone, where soft
physics should be dominant.

\begin{figure*} 
\includegraphics[scale=0.5]{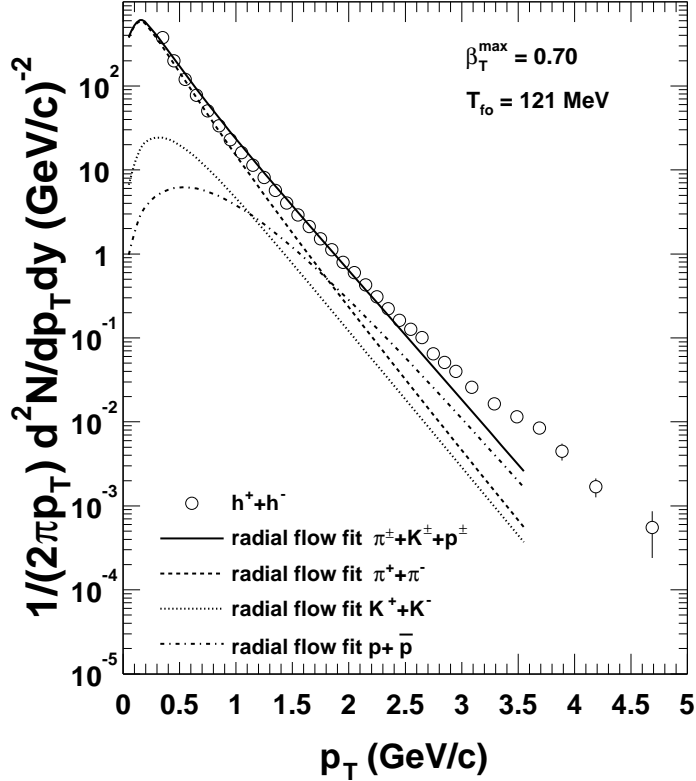}
\caption{\label{compareToHm}
The high $p_T$ hadron spectra in Reference \cite{PPG003}
compared to the fit results assuming radial flow from the $\pi$Kp
spectra in the 5\% central events.}
\end{figure*}

The hadron spectrum is calculated using the fit parameters from the
low $p_T$ region fits shown in the preceding section, and extrapolated
to  higher $p_T$.  Figure~\ref{compareToHm} shows the calculated spectrum
for each particle type, and the sum of the
extrapolated spectra is compared to the measured charged hadrons ($h^+
+ h^-$) in the 5\% most central collisions.  
As non-identified charged hadrons are measured in $\eta$ rather
than in $y$, the
extrapolated spectra are converted to units of $\eta$. This conversion
is most important in the low $p_T$ region.  
No additional scale factor is applied -- the
extrapolation and data are compared absolutely. Below $\approx$ 2.5
GeV/c $p_T$, the agreement is very good, while at higher $p_T$ the
data begin to exceed the hydrodynamic extrapolation.

Other hydrodynamic calculations have been successful in describing the
distributions over the full $p_T$ range \cite{bron} with different
parameter values.  There  are clear
indications that particle production from a hydrodynamic  source, if
invoked to explain the spectra at low $p_T$, will have a
non-negligible influence even at relatively large $p_T$.  
Furthermore, the range of $p_T$ populated by hydrodynamically boosted
hadrons is species dependent. This is clearly visible in Figure~\ref{compareToHm}, 
which shows that the extrapolated proton spectra have a flatter $p_T$
distribution than the extrapolated pions and kaons. The yield of the 
``soft" protons reaches, and even exceeds, that of the extrapolated 
``soft" pions at 2 GeV/c $p_T$.
Therefore the transition from soft to hard processes must also be species
dependent, and the boost of the protons causes the region where  hard 
processes dominate the inclusive charged particle spectrum to be at 
significantly higher transverse momenta in central Au+Au than in p+p 
collisions. Our analysis suggests this occurs not lower than $p_T$ = 3 GeV/c.

\subsection{HADRON YIELDS AS A FUNCTION OF CENTRALITY}
The previous discussion focused on the hadron spectra; now we
turn to the centrality dependence of the pion, kaon, proton,
and antiproton yields, which
can shed further light on the
importance of different mechanisms in particle production. 
It is instructive to see whether yields of the different hadrons 
scale with the number of participant nucleons,
$N_{part}$, the number of binary nucleon-nucleon collisions,
$N_{coll}$, or some combination of the two.

The total yields of the hadrons may be expected to be dominated
by soft processes, and the wounded nucleon model of soft interactions
suggests that the yields should scale as the number of participants, 
$N_{part}$. If
each
participant loses a certain fraction of its incoming energy,
like e.g. in string models, where each pair of participants
(or wounded nucleons) contributes a color flux tube,
the total energy of the fireball formed at central rapidity
would be proportional to the number of participants $N_{part}$.
If, furthermore, the fireball is locally  thermalized and particle
production is determined at a single temperature, the multiplicity
would scale with $N_{part}$.  On the other hand, at very high $p_T$, 
particle production may be dominated by hard processes and scale with
$N_{coll}$\cite{dima,wangdndy}.

In order to investigate the existence of scaling, the 
multiplicities are parameterized as:
\begin{equation}
\frac{dN}{dy} = C \cdot (N_{part})^{\alpha_{part}}
\label{eq:dndypart}
\end{equation}
and
\begin{equation}
\frac{dN}{dy} = C^{\prime} \cdot (N_{coll})^{\alpha_{coll}} .
\label{eq:dndycoll}
\end{equation}
Fit results for these parameterizations are shown in 
Table~\ref{table:fitdndy}. As can be seen, the exponents $\alpha_{part}$
are $> 1$ for all species, while $\alpha_{coll}$ is consistently $< 1$.
The production of all particles increases more strongly than with
$N_{part}$, but not as strongly as with $N_{coll}$. Small differences
between the different particle species are apparent: The (anti-)proton
yield  increases more strongly than the pion yield, and the kaon yield
shows  the strongest centrality dependence.  Remarkably, the 
yield fraction scaling beyond linear with $N_{part}$ is larger 
for kaons, protons, and antiprotons than for pions. Perhaps it is
not surprising that the yields do not scale simply with $N_{part}$;
the collective flow seen in the $p_T$ spectra already shows that the 
nucleon-nucleon collisions cannot be independent.

\begin{table}[bt]
\caption{\label{table:fitdndy}
Fit parameters for each particle species using equations
\protect\ref{eq:dndypart} and \protect\ref{eq:dndycoll}.  }
\begin{center}
\begin{ruledtabular}
\begin{tabular}{ccc} 
particle  &  $\alpha_{part}$    &       $\alpha_{coll}$ \\ \hline
$\pi^{+}$&  $1.06 \pm 0.01 $ & $0.79 \pm 0.01$ \\ $\pi^{-}$&  $1.08
\pm 0.01 $ & $0.80 \pm 0.01$ \\ $K^{+}$  &  $1.18 \pm 0.02$ & $0.88
\pm 0.02$  \\ $K^{-}$  &  $1.20 \pm 0.03$ & $0.89 \pm 0.02$  \\
$p^{+}$  &  $1.16 \pm 0.02$ & $0.86 \pm 0.02$  \\ $p^{-}$  &  $1.14
\pm 0.03$ & $0.84 \pm 0.02$  \\ 
\end{tabular}
\end{ruledtabular}
\end{center}
\end{table}

\begin{figure}[tb]
     \centering
\resizebox{\columnwidth}{!}{\includegraphics{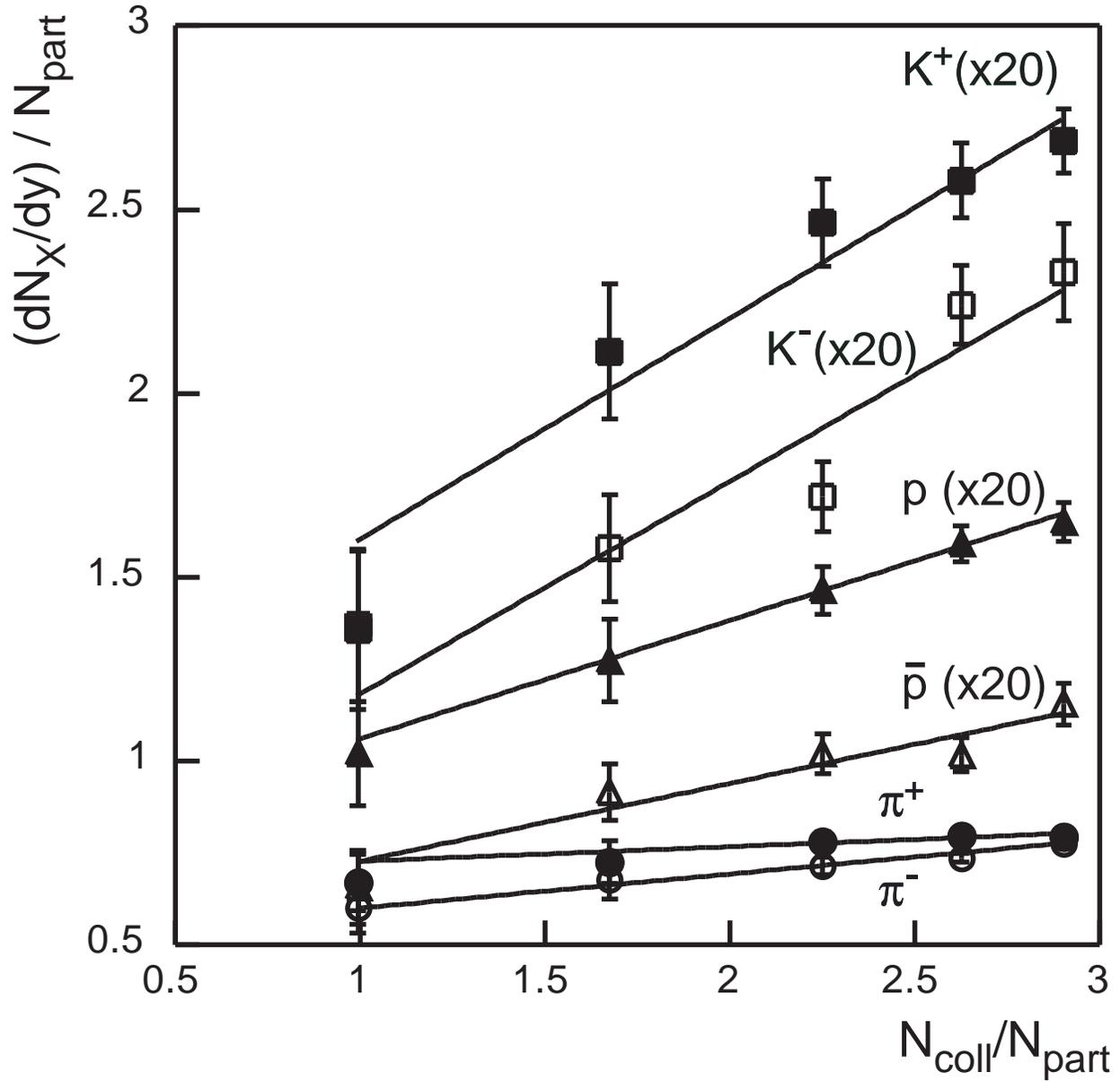}}
     \caption{\label{fig:scale}
$dN/dy$ per participant of different particle species as
a function of the number of collisions per participant.  Kaon and
(anti-)proton multiplicities are scaled by a factor of 20.}
\end{figure}

We next check whether the simple model of hadron yields can be brought 
into agreement with the data by adding a component of the yields
scaling as the number of binary collisions, $N_{coll}$. Such an
admixture inspires simple two-component models \cite{dima,wangdndy}.
The nonlinearity of dN/dy on the number of participants is illustrated 
by the ratio (dN/dy)/$N_{part}$,  shown in Figure~\ref{fig:scale} 
as a function of centrality. The yields are seen to depend linearly 
on $N_{coll}/N_{part}$.  
As seen already from the exponents in Table~\ref{table:fitdndy}, the 
increase with centrality is strongest for kaons, intermediate for 
(anti-)protons, and weakest for pions. 
This indicates that protons and antiprotons have a larger component
scaling with $N_{coll}$ than pions.

We fit the yields per participant with Equation ~\ref{eq:xmult2}.
As in \cite{dima,wangdndy} we parameterize the multiplicity using 
two free parameters:  $n_{pp}$, the multiplicity in p+p collisions, 
and $x$, the relative strength of the 
component scaling with $N_{coll}$. 
\begin{eqnarray}
     R \equiv {dN/dy \over N_{part} } & = & (1-x) \cdot n_{pp}
     \frac{1}{2} + x \cdot  n_{pp} \frac{N_{coll} }{ N_{part} }
     \label{eq:xmult2a} \nonumber \\
      & = & n_{pp} \left[ \frac{1}{2} + x \left( \frac{ N_{coll} }{
N_{part} } - \frac{1}{2} \right) \right].
     \label{eq:xmult2}
\end{eqnarray}
The results of the fit are shown as solid lines in Figure~\ref{fig:scale}.
The fit parameter values are given in Table~\ref{fittable}.
All hadron species are well fit. 
The importance of the component scaling as $N_{coll}$ is largest for
kaons and smallest for pions.

\begin{table} [ht]
\caption{\label{fittable}
Values of the parameters $n_{pp}$ and $x$ from fitting
Equation~\ref{eq:xmult2} to the observed dN/dy per $N_{part}$. }
\begin{ruledtabular}
\begin{tabular} {ccc}
                & $n_{pp}$   & $x$ \\ \hline
$\pi^{+}$      
	& $1.41\pm 0.11 $ 
	& $0.028\pm 0.020 $ \\
$\pi^{-}$      
	& $1.10 \pm 0.11 $  
	& $0.085 \pm 0.030 $ \\
$K^+$          
	& $0.130 \pm 0.021$ 
	& $0.232  \pm 0.076$ \\
$K^-$          
	& $0.089 \pm 0.020 $ 
	& $0.326 \pm 0.132 $ \\
$p$              
	& $0.089 \pm 0.013 $  
	& $0.181 \pm 0.062 $ \\
$\overline{p}$ 
	& $0.062 \pm 0.010  $ 
	& $0.172 \pm 0.068 $  \\
\end{tabular}
\end{ruledtabular}
\end{table}

We check the consistency of the fits in Figure~\ref{fig:scale}
with known hadron yields in p+p collisions by
extrapolating the fits down to two participants (and one binary
nucleon-nucleon collision).
Isospin differences between p+p and Au+Au are ignored. 
The check is done by separately extrapolating the fitted 
fraction of yield which scales with $N_{coll}$ and the 
fraction scaling with $N_{part}$ down to one nucleon-nucleon 
collision and two participant nucleons, and summing the result. 
One obtains particle ratios of $K/\pi = (8.7 \pm 2.6) \%$
and $\bar{p}/\pi = (4.9 \pm 0.8) \%$. 
These values fall between those measured at lower $\sqrt{s}$ 
at the ISR \cite{ISR1} and those at higher $\sqrt{s}$ at the
Tevatron \cite{e735}, as expected since the RHIC energy
lies in between. Thus the Au+Au data are shown to scale down 
to p+p reasonably.

\begin{figure}[tb]
\centering
\includegraphics{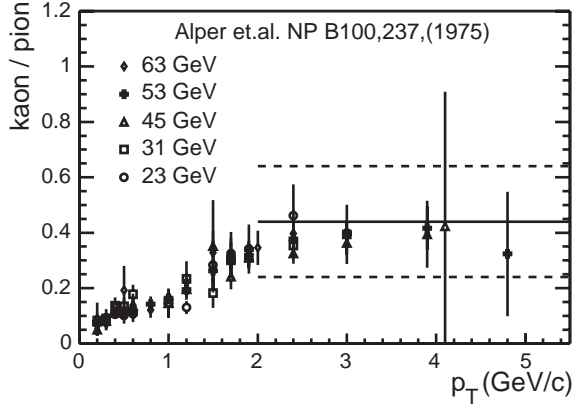}
\caption{\label{f:rkpipt}
Kaon to pion ratio as a function of $p_{T}$. The different
points are
measured in $p+p$ collisions
(data from ref.~\protect\cite{ISR2}). The solid line is the asymptotic value
for
high $p_{T}$ in p+p derived from the hard scattering component of the
fits using Equation ~\ref{eq:xmult2} to the measured centrality
dependence of dN/dy  in Au-Au collisions at
$\sqrt{s_{NN}} = 130$ GeV. The dashed lines indicate the
corresponding uncertainty.}
\end{figure}

\begin{figure}[tb]
\centering
\includegraphics{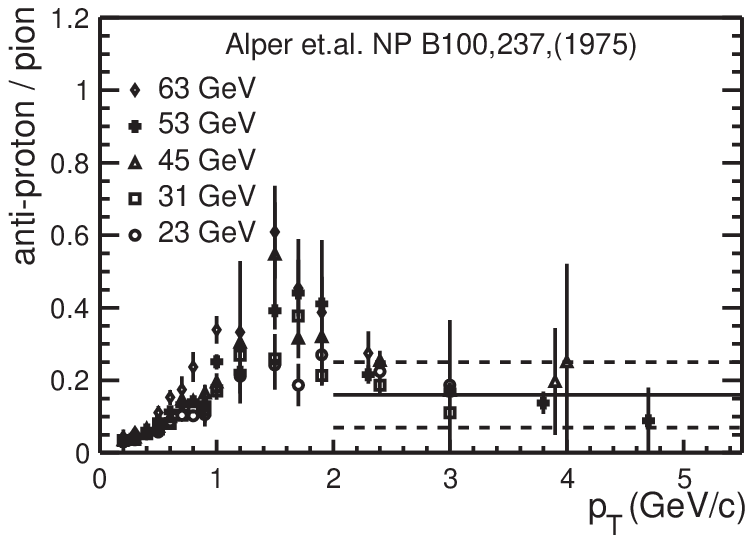}
\caption{\label{f:rpbarpipt}
Antiproton to $\pi^-$ ratio as a function of $p_{T}$.
The different points are measured in $p+p$ collisions
(data from ref.~\protect\cite{ISR2}). The solid line is the asymptotic value
for high $p_{T}$ in p+p derived from the hard scattering component of the
fits using Equation ~\ref{eq:xmult2} to the measured centrality
dependence of dN/dy  in Au-Au collisions at $\sqrt{s_{NN}} = 130$ GeV. 
The dashed lines indicate the corresponding uncertainty.
}
\end{figure}

One may expect that the particle ratios at very high $p_T$ 
should be dominated by hard scattering, and therefore scale 
with the number of binary collisions. 
Consequently, we look at ratios of
the $N_{coll}$ scaling components alone,
extrapolated down to one binary collision.
The values are compared to 
measurements of hadron ratios at the ISR \cite{ISR2} in
Figures~\ref{f:rkpipt} and~\ref{f:rpbarpipt}. 
The ratio of the  extrapolated
Au+Au yield fractions scaling as $N_{coll}$ are shown as solid lines for 
$p_{T} \ge 2 \, \mathrm{GeV}/c$. 
The agreement with the p+p data at high $p_{T}$ 
is quite good.

\begin{figure}[tb]
\centering \includegraphics{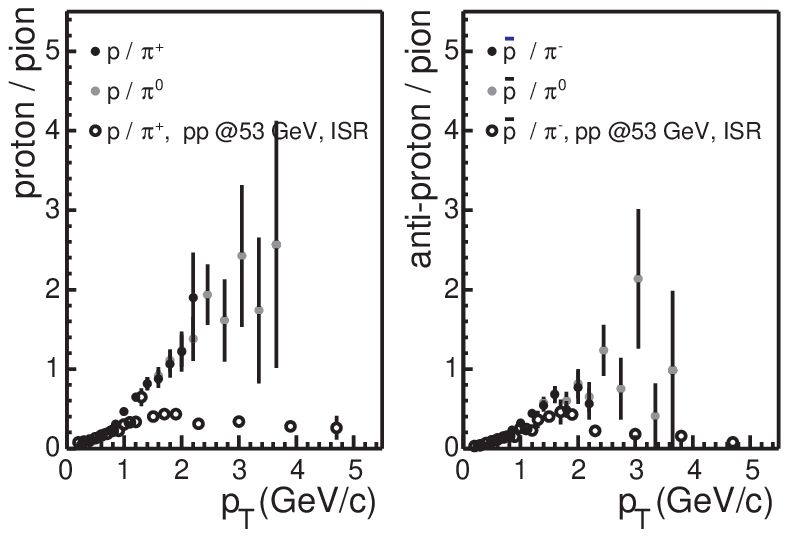}
\caption{\label{f:rppiptauau} 
a) Proton to pion ratio as a function of $p_T$.  b)
Antiproton to pion ratio as a function of $p_T$.  The open circles
represent measurements in $p+p$ collisions (data from
ref.~\protect\cite{ISR2}). The filled circles show the 10 $\%$ most
central Au+Au collisions. The neutral pion spectra are from the data
published in \protect\cite{PPG003}.  }
\end{figure}

Finally, we directly compare p/$\pi$ and $\overline{p}/\pi$ ratios in
central Au+Au collisions with p+p, as a function of $p_T$. 
These ratios from the $10 \% $ most central data,
using the charged particle measurement from this paper and  neutral
pions from \cite{PPG003}, are shown in Figure~\ref{f:rppiptauau}. The
ratios show a steady increase up to 2.5 GeV/c in $p_T$.
Even though the simple extrapolation of the $N_{coll}$ scaling
yield fraction agreed with p+p, the ratios of the full yield
significantly exceed those in the ISR measurements  \cite{ISR2}.
According to Gyulassy and collaborators, 
\cite{gyulassy}, this result may give insight  into baryon number
transport and the interplay between soft and  hard processes.

Of course, splitting the observed yields into portions which
scale with $N_{part}$ and $N_{coll}$ is by no means a unique
explanation of the data. The spectra and yields can also be
well reproduced by thermal models, which break such simple
scalings due to the multiple interactions suffered by the
constituents.

Simple thermal models that ignore transverse and longitudinal 
flow~\cite{Cleymans} are able to describe the centrality dependence of 
the mid-rapidity $\pi^{\pm}$, $K^{\pm}$, $p$, and $\overline{p}$ yields by 
tuning the chemical freeze-out temperature $T_{ch}$, the baryon chemical 
potential $\mu_{B}$ and by introducing a strangeness saturation factor 
$\gamma_{s}$. It was found that $\mu_{B}$ is independent of centrality, 
while both $\gamma_{s}$ and $T_{ch}$ increase from peripheral to central 
collisions. Within the same model, the centrality dependence of the particle 
yields at lower energy ($\sqrt{s_{NN}} = 17 $ GeV ~\cite{NA49a,NA49b}) are 
described by constant $T_{ch}$ and $\mu_{B}$. The strong centrality 
dependence in kaon production at both energies is accounted for by the 
increase in the strangeness saturation factor $\gamma_{s}$. Although the 
integrated particle yields are very well described, such simple thermal 
models do not attempt a comparison to the single particle spectra, which 
clearly indicate centrality dependent flow effects not included in the model.

Thermal models which include hydrodynamical parameters on a freeze-out 
hypersurface to account for longitudinal and transverse flow can 
reproduce the absolutely normalized particle spectra by introducing 
only two thermal parameters $T_{ch}$ and $\mu_{B}$ ~\cite{WB1,WB2}. 
In this approach, the thermal parameters are independent of centrality, 
while the geometric parameters are adjusted to reproduce the 
spectra. Good agreement with the data is obtained up to 
$p_{T} \approx 2 - 3$ GeV/c, however an explicit comparison with the 
centrality dependence of the integrated mid-rapidity yields has not yet 
been made.

This section shows that the yields of all hadrons increase more rapidly 
than linearly
with the number of participants, but the increase is weaker than
scaling with the number of binary collisions. The excess beyond
linear scaling with $N_{part}$ is strongest for kaons, intermediate
for (anti-)protons, and weakest for pions. The centrality dependence
of the total yields can be well fit with a sum of these two kinds
of scaling. At high $p_T$, the baryon and anti-baryon yields
greatly exceed expectations from p+p collisions. Thermal models,
which do not invoke strict scaling rules, can successfully
reproduce the data as well, providing that they include the radial 
flow required by the $p_T$ spectra.

\section{SUMMARY AND CONCLUSION}
We have presented the spectra and yields of identified hadrons
produced in $\sqrt{s_{NN}} =$ 130 GeV Au + Au collisions. The yields
of pions increase approximately linearly with the number of
participant nucleons, while the yield increase is faster than linear
for kaons, protons, and antiprotons.

Hydrodynamic analyses of the particle spectra are performed:
the spectra are fit with a hydrodynamic-inspired
parameterization to extract freeze-out temperature and radial flow
velocity of the particle source. The data are also compared to two
full hydrodynamics calculations. The simultaneous fits of pion, kaon,
proton, and antiproton spectra show that radial flow in central
collisions at RHIC exceeds that at lower energies and
increases with centrality of the collision. 
The hydrodynamic models are consistent with the measured
spectral shapes, extracted freeze-out temperature $T_{fo}$, and the
flow velocity $\beta_T$ in central collisions.

Extrapolating the fits to estimate thermal particle
production at higher $p_T$ 
allows us to study the soft-hard physics boundary by comparing 
to measured spectra at high
$p_T$. The yield of the ``soft" protons reaches, and even
exceeds, that of the extrapolated ``soft" pions at 2 GeV/c $p_T$.
The sum of the extrapolated ``soft" spectra agree with the 
measured inclusive data to $p_T$ $\approx 2.5 - 3$ GeV/c. 
The transition from soft to hard processes must be species
dependent, and the admixture of boosted nucleons implies that 
hard processes do not dominate the inclusive charged
particle spectra until approximately 3 GeV/c. 

\appendix
\section{\label{app:glauber}Determining $N_{part}$ and $N_{coll}$}
As only the fraction of the total cross section is measured 
in both the ZDC and BBC detectors, a model-dependent calculation 
is used to map collision centrality to the number of 
participant nucleons, $N_{part}$, and the number of nucleons 
undergoing binary collisions, $N_{coll}$.  A discussion of 
this calculation at RHIC can be found elsewhere \cite{dima}.

%
\begin{table}
\caption{\label{Tab:minBiasData}
Invariant yields for $\pi^{\pm}$, $K^{\pm}$, and (anti)p measured in
minimum bias events at midrapidity and normalized to one rapidity unit.  
The errors are statistical only.}
\begin{ruledtabular}
\begin{tabular} {cccc}
$p_{T}$ (GeV/c) & $\pi^{\pm}$ & $K^{\pm}$ & p($\overline{p}$) \\ \hline
0.25 &   112$\pm$2&                           &                       \\
     &   109$\pm$2&			     &				\\
0.35 &   56 $\pm$1 &			     &				\\
         &   49.9 $\pm$0.9 &			     &				\\
0.45 &   28.0 $\pm$0.5 &6.1$\pm$0.4 &				\\
         &   24.1 $\pm$0.5 &4.6$\pm$0.4 &				\\
0.55 &   15.7 $\pm$0.3 &4.0$\pm$0.3 &2.3$\pm$	0.1\\
         &   14.6 $\pm$0.3 &3.2$\pm$0.2 &1.2$\pm$0.1\\
0.65 &   9.1  $\pm$0.2&2.8$\pm$0.2 &1.8$\pm$	0.1\\
         &   8.7  $\pm$0.2 & 2.1 $\pm$0.2 &1.17$\pm$	0.09\\
0.75 &   5.8  $\pm$0.1 &1.7$\pm$0.1 &1.38$\pm$	0.08\\
         &   5.6  $\pm$0.2 & 1.6$\pm$0.1 &0.98$\pm$0.07\\
0.85 &   3.8  $\pm$0.1&1.30$\pm$0.08 &1.18$\pm$0.07\\
         &   3.6$\pm$0.1 &1.17$\pm$0.09 &0.95$\pm$0.07\\
0.95 &   2.40 $\pm$0.08	 &0.87$\pm$0.06 &0.98$\pm$	0.06\\
         &   2.28  $\pm$0.08	 &0.69$\pm$0.06 &0.65$\pm$	0.05\\
1.05 &   1.61  $\pm$0.06	 &0.62$\pm$0.04 &0.70$\pm$	0.04\\
         &   1.61  $\pm$0.06	 &0.53$\pm$0.05 &0.50$\pm$	0.04\\
1.15 &   1.03  $\pm$0.04	 &0.43$\pm$0.03 &0.60$\pm$	0.04\\
         &   1.17  $\pm$0.05	 &0.38$\pm$0.04 &0.35$\pm$	0.03\\
1.25 &   0.71$\pm$0.03	 &0.33$\pm$0.03 &0.41$\pm$	0.03\\
         &   0.76  $\pm$0.04	 &0.27$\pm$0.03 &0.34$\pm$0.03\\
1.35 &   0.46  $\pm$0.02	 &0.20$\pm$0.02 &0.32$\pm$0.02\\
         &   0.54$\pm$0.03	 &0.16$\pm$0.02 &0.22$\pm$0.02\\
1.45 &   0.35$\pm$0.02	 &0.17$\pm$0.02 &0.23$\pm$	0.02\\
         &   0.31 $\pm$0.02	 &0.13$\pm$0.02 &0.18$\pm$	0.02\\
1.55 &   0.24  $\pm$0.02	 &0.10$\pm$0.01 &0.17$\pm$	0.02\\
         &   0.22 $\pm$0.02	 &0.10$\pm$0.01 &0.15$\pm$	0.02\\
1.65 &   0.16 $\pm$0.01	 &0.08$\pm$0.01 &	      		\\
         &   0.15  $\pm$0.01	 &0.07$\pm$0.01 &	      		\\
1.70     &                               &                      &0.119$\pm$	0.008\\
         &                               &                      &0.080$\pm$	0.007\\
1.75 &   0.11  $\pm$0.01	 &			&	      		\\
         &   0.11  $\pm$0.01	 &			&	      		\\
1.85 &   0.079  $\pm$0.008	 &			&	      		\\
         &   0.092  $\pm$0.009	 &			&	      		\\
1.90      &                              &                      &0.068$\pm$	0.006\\
         &                               &                      &0.041$\pm$	0.005\\
1.95 &   0.063  $\pm$0.007	 &			&	      		\\
         &   0.066  $\pm$0.008	 &			&	      		\\
2.05 &   0.036  $\pm$0.005	 &			&	      		\\
         &   0.034  $\pm$0.005	 &			&	      		\\
2.10     &                               &                      &0.036$\pm$	0.004\\
         &                               &                      &0.022$\pm$	0.003\\
2.15 &   0.026  $\pm$0.004	 &			&	      		\\
         &   0.025  $\pm$0.004	 &			&	      		\\
2.25 &   0.015  $\pm$0.003	 &			&	      		\\
         &   0.020  $\pm$0.004  &                         &	      		\\
2.30     &                               &                      &0.020$\pm$	0.003\\
         &                               &                      &0.012$\pm$	0.002\\
2.50     &                               &                      &0.010$\pm$	0.002\\
         &                               &                      &0.011$\pm$	0.002\\
2.70     &                               &                      &0.006$\pm$	0.001\\
         &                               &                      &0.0026$\pm$	0.0009\\
2.90     &                               &                      &0.0035$\pm$	0.0008\\
         &                               &                      &0.003$\pm$	0.001\\
3.10     &                               &                      &0.0028$\pm$	0.0007\\
         &                               &                      &0.0011$\pm$	0.0005\\
3.30     &                               &                      &0.0014$\pm$	0.0005\\
         &                               &                      &0.0010$\pm$	0.0004\\
\end{tabular}
\end{ruledtabular}
\end{table}

Using a Glauber model combined with a simulation of the BBC and ZDC 
responses, $N_{part}$ and $N_{coll}$ are determined in each 
centrality.  The model provides 
the thickness of nuclear matter in the direct path of each 
oncoming nucleon, and uses the inelastic nucleon-nucleon cross section 
$\sigma^{inel}_{N+N}$ to determine whether or not a nucleon-nucleon 
collision occurs.  We assume the following:
\begin{itemize}
\item The nucleons travel in straight-line paths, parallel to the 
velocity of its respective nucleus.
\item An inelastic collision occurs if the relative distance between 
two nucleons is less than $\sqrt{\sigma^{inel}_{N+N}/\pi}$.
\item Fluctuations are introduced by using the simulated detector response for both the ZDC and 
BBC.
\end{itemize}

In this calculation, the Woods-Saxon nuclear density distribution ($\rho(r)$) 
is used for each nucleus with three parameters,  the nuclear radius R 
$ = 6.38^{+0.27}_{-0.13}$ fm, diffusivity d $ = 0.53\pm0.01$ 
fm \cite{Glauber}, and the inelastic nucleon-nucleon cross section 
 $\sigma^{inel}_{N+N}$ $ =$ 40 $\pm$ 3 mb,
\begin{equation}
\rho(r) = \frac{\rho_0}{1+e^\frac{r-r_n}{d}}.
\end{equation}  

\section{\label{xsec}Invariant Cross Sections}
Tabulated here are the measured invariant yields of pions, kaons, 
and (anti)protons produced in Au+Au collisions at 130 GeV.  The first set 
of tables (Tables ~\ref{Tab:minBiasData}-~\ref{Tab:protData}) are the 
invariant cross sections plotted in Figures~\ref{fig1} and~\ref{piKpCent}.
The second set of tables (Tables ~\ref{Tab:minBiasSameBin}-~\ref{protSameBin}) 
are the invariant cross sections in equal $p_{T}$ bins as used in 
Figure~\ref{centToPerp}.

\begin{table*}
\caption{\label{Tab:pionData}
Pion invariant yields in each event centrality and $p_{T}$ bin measured at midrapidity, normalized to one rapidity unit.  For each measured $p_{T}$ bin, the positive pion cross section is the top row and the negative pion cross section is the bottom row.  Errors are statistical only.}
\begin{ruledtabular}
\begin{tabular} {cccccc} 
$p_{T}$ (GeV/c) & 0-5\% & 5-15\% & 15-30\% & 30-60\% & 60-92\% \\ \hline
0.25      & 355 $\pm$	9
          &282  $\pm$   6
          &186  $\pm$   4
          &81   $\pm$  2
          &13.2   $\pm$  0.5 \\
          & 371 $\pm$	10
          & 275 $\pm$	7
          &180  $\pm$   4
          &74   $\pm$  2
          & 12.1  $\pm$   0.5\\
0.35  & 188 $\pm$	5
          & 146 $\pm$	3
          & 93  $\pm$   2
          & 36.6  $\pm$   0.8
          & 5.3  $\pm$  0.2\\
          & 169 $\pm$	5
          &128  $\pm$   3
          & 82  $\pm$   2
          &34.3   $\pm$  0.9
          & 5.0   $\pm$  0.2\\
0.45  &  95 $\pm$	3
          & 74  $\pm$   2
          & 48  $\pm$   1
          & 17.5  $\pm$   0.5
          & 2.7   $\pm$  0.1 \\
          & 86  $\pm$   3
          &63   $\pm$  2
          & 40  $\pm$   1
          & 15.7  $\pm$   0.5
          & 2.1   $\pm$  0.1\\
0.55      & 56  $\pm$   2
          &41   $\pm$  1
          & 26.0  $\pm$   0.7
          &10.1   $\pm$  0.3
          & 1.32   $\pm$  0.09\\
          & 51  $\pm$   2
          &38   $\pm$  1
          & 24.5  $\pm$   0.8
          &9.6    $\pm$ 0.3
          & 1.18   $\pm$  0.09\\
0.65  &  32$\pm$	1
          &25.6  $\pm$  0.8
	  & 15.0  $\pm$   0.5
	  &5.3    $\pm$ 0.2
          & \\ 
          & 30  $\pm$   1
          & 22.6  $\pm$   0.8
	  & 14.5  $\pm$   0.5
	  & 5.7   $\pm$  0.2
          & \\
0.70 & & & & & 0.62$\pm$     0.04\\
     & & & & & 0.57$\pm$    0.04\\
0.75  & 21.1 $\pm$    0.9
          & 15.4 $\pm$    0.6
	  & 9.9  $\pm$   0.4
	  & 3.6 $\pm$   0.1
          & \\
          &  20 $\pm$	 1
          & 15.3 $\pm$   0.6
	  & 9.5   $\pm$  0.4
	  & 3.5   $\pm$  0.2
          & \\
0.85  &  14.0 $\pm$    0.7
& 10.3 $\pm$    0.4
& 6.4  $\pm$   0.3
&2.3   $\pm$  0.1
& \\
& 12.8 $\pm$    0.8
&9.6   $\pm$  0.5
& 6.1  $\pm$   0.3
&2.1   $\pm$  0.1
& \\
0.90 & & & & & 0.19 $\pm$    0.02\\
     & & & & & 0.24 $\pm$    0.02\\
1.00 &    7.1 $\pm$    0.3
&5.3$\pm$	   0.2
&3.4 $\pm$	   0.1
&1.25 $\pm$	   0.05
& \\
&6.5 $\pm$	   0.3
&5.0 $\pm$	   0.2
&3.4 $\pm$	   0.1
&1.25 $\pm$	   0.05
& \\
1.20 &    3.2 $\pm$    0.2
&2.2 $\pm$	   0.1
&1.47 $\pm$	   0.07
&0.55 $\pm$	   0.03
&0.064 $\pm$	   0.006 \\
&3.3 $\pm$	   0.2
&2.6 $\pm$	   0.1
&1.51 $\pm$	   0.08
&0.63 $\pm$	   0.04
&0.061 $\pm$	   0.007 \\
1.40  &  1.3 $\pm$    0.1
&1.01 $\pm$	   0.07
&0.72 $\pm$	   0.04
&  0.27 $\pm$     0.02
& \\
&1.3 $\pm$	   0.1
&1.25 $\pm$	   0.09
&0.72 $\pm$	   0.05
&0.26 $\pm$	   0.02
& \\
1.60 &   0.55 $\pm$    0.07
&0.57 $\pm$	   0.05
&0.33 $\pm$	   0.03
&0.14 $\pm$	   0.01
&0.015 $\pm$	   0.003 \\
& 0.51 $\pm$     0.07
&0.57  $\pm$   0.05
&0.30  $\pm$   0.03
&0.12  $\pm$   0.01
& 0.014 $\pm$    0.003\\
1.80 &   0.35 $\pm$    0.05
&0.25 $\pm$	   0.03
&0.15 $\pm$	   0.02
&0.060 $\pm$	   0.008
& \\
&0.40 $\pm$	   0.06
&0.27 $\pm$	   0.03
&0.14 $\pm$	   0.02
&0.075 $\pm$	   0.009
& \\
2.00 &   0.18 $\pm$     0.03
&0.13 $\pm$	   0.02
&0.10 $\pm$	   0.01
&0.023 $\pm$	   0.004
&0.005 $\pm$	   0.001 \\
&0.18 $\pm$	   0.04
&0.13 $\pm$	   0.02
&0.09 $\pm$	   0.01
&0.029 $\pm$	   0.005
& 0.004$\pm$	    0.001\\
2.20 &   0.043 $\pm$     0.02
&0.07 $\pm$	   0.01
&0.034 $\pm$	   0.007
&0.012 $\pm$	   0.003
& \\
&0.09 $\pm$	   0.03
&0.05 $\pm$	   0.01
&0.042 $\pm$	   0.009
&0.012 $\pm$	   0.003
& \\ 
\end{tabular}
\end{ruledtabular}
\end{table*}

\begin{table*}
\caption{\label{Tab:kaonData}
Kaon invariant yields in each event centrality and $p_{T}$ bin, measured at midrapidity and normalized to one rapidity unit.  The top row in each $p_{T}$ bin is $K^{+}$, and the bottom row is $K^{-}$.  Errors are statistical only.}
\begin{ruledtabular}
\begin{tabular} {cccccc}
$p_{T}$ (GeV/c) & 0-5\% & 5-15\% & 15-30\% & 30-60\% & 60-92\% \\ \hline
0.44 &&&&&   0.5 $\pm$	   0.1\\
 &&&&&   0.4 $\pm$	   0.1 \\       
0.45 &    21$\pm$	3
&16 $\pm$    2
&10 $\pm$    1
&4.3  $\pm$   0.5
& \\
&21$\pm$     3
&13 $\pm$    2
&7 $\pm$	   1
&2.5 $\pm$	   0.4
& \\
0.54 &&&&&    0.20 $\pm$     0.07 \\ 
 &&&&&   0.3  $\pm$   0.1 \\ 
0.55 &    15$\pm$	2
&11 $\pm$    1
&6.6  $\pm$   0.7
&2.4  $\pm$   0.3
& \\
&13$\pm$     2
&8  $\pm$   1
&4.8 $\pm$	   0.6
&2.3 $\pm$	   0.3
& \\ 
0.65&    9 $\pm$    1
&7.5  $\pm$   0.7
&4.7 $\pm$	   0.4
&1.9 $\pm$	   0.2
& \\
&8 $\pm$    1
&7.0  $\pm$   0.8
&3.1 $\pm$	   0.4
&1.1 $\pm$	   0.2
& \\ 
0.69 &&&&&    0.18 $\pm$     0.03 \\
 &&&&&   0.14  $\pm$   0.03 \\
0.75&    5.3 $\pm$    0.7
&4.6  $\pm$   0.5
&3.1  $\pm$   0.3
&0.9  $\pm$   0.1
& \\
&5.1 $\pm$    0.8
&5.0  $\pm$   0.6
&2.5 $\pm$	   0.3
&0.9 $\pm$	   0.1
& \\ 
0.85&    5.7 $\pm$    0.7
&3.6  $\pm$   0.4
&2.1 $\pm$	   0.2
&0.67 $\pm$	   0.08
& \\
&3.6 $\pm$    0.6
&3.7  $\pm$   0.4
&2.0 $\pm$	   0.2
&0.64 $\pm$	   0.09
& \\
0.89 &&&&&   0.07 $\pm$	    0.02 \\ 
 &&&&&   0.09 $\pm$	   0.02 \\ 
0.95&    3.0 $\pm$    0.4
&2.3  $\pm$   0.2
&1.5  $\pm$   0.2
&0.51  $\pm$   0.07
& \\
&2.4 $\pm$    0.4
&2.3  $\pm$   0.3
&1.0 $\pm$	   0.2
&0.35 $\pm$	   0.06
& \\ 
1.05&    2.3 $\pm$    0.3
&1.5  $\pm$   0.2
&1.2 $\pm$	   0.1
&0.37 $\pm$	   0.05
& \\
&1.8 $\pm$    0.3
&1.8  $\pm$   0.2
&0.8 $\pm$	   0.1
&0.27 $\pm$	   0.05
& \\ 
1.15&    1.6 $\pm$    0.3
&1.3  $\pm$   0.2
&0.62 $\pm$	   0.09
&0.29 $\pm$	   0.04
& \\
&1.4 $\pm$    0.3
&0.9  $\pm$   0.2
&0.7 $\pm$	   0.1
&0.23 $\pm$	   0.04
& \\ 
1.17  &&&&&  0.012$\pm$	   0.004 \\
  &&&&&  0.015$\pm$	   0.005 \\ 
1.25 &    1.1 $\pm$    0.2
&1.0  $\pm$   0.1
&0.66 $\pm$	   0.09
&0.15 $\pm$	   0.03
& \\
&1.2 $\pm$    0.2
&0.8  $\pm$   0.1
&0.41 $\pm$	   0.08
&0.15 $\pm$	   0.03
& \\ 
1.35&    0.6 $\pm$    0.1
&0.6  $\pm$   0.1
&0.35 $\pm$	   0.05
&0.12 $\pm$	   0.02
& \\
&0.9 $\pm$    0.2
&0.5  $\pm$   0.1
&0.22 $\pm$	   0.05
&0.05 $\pm$	   0.02
& \\ 
1.45&    0.5 $\pm$    0.1
&0.45  $\pm$   0.08
&0.32 $\pm$	   0.05
&0.10 $\pm$	   0.02
& \\
&0.4 $\pm$    0.1
&0.36  $\pm$   0.08
&0.26 $\pm$	   0.05
&0.07 $\pm$	   0.02
& \\ 
1.55 &    0.4 $\pm$    0.1
&0.26  $\pm$   0.06
&0.14 $\pm$	   0.03
&0.07 $\pm$	   0.02
& \\
&0.4 $\pm$    0.1
&0.28  $\pm$   0.07
&0.21 $\pm$	   0.04
&0.04 $\pm$	   0.01
& \\ 
1.57 &&&&&   0.008$\pm$	   0.002 \\
 &&&&&   0.004$\pm$	   0.002 \\
1.65 &    0.4 $\pm$    0.1
&0.24  $\pm$   0.05
&0.10  $\pm$	  0.03
&0.05  $\pm$	  0.01
& \\
&0.12 $\pm$    0.06
&0.23  $\pm$   0.06
&0.12 $\pm$	   0.03
&0.04 $\pm$	   0.01
& \\ 
\end{tabular}
\end{ruledtabular}
\end{table*}

\begin{table*}
\caption{\label{Tab:protData}
(Anti)proton invariant yields in each event centrality and $p_{T}$ bin, measured at midrapidity and normalized to one rapidity unit.  The top row in each $p_{T}$ is the proton cross section, and the bottom row the antiproton.  The errors are statistical only.}
\begin{ruledtabular}
\begin{tabular} {cccccc}
$p_{T}$ (GeV/c) & 0-5\% & 5-15\% & 15-30\% & 30-60\% & 60-92\% \\ \hline
0.545 &&&&&   0.26 $\pm$     0.06 \\
 &&&&&   0.12 $\pm$    0.05 \\ 
0.55 &   8 $\pm$    1
&4.9 $\pm$	   0.5
&4.0 $\pm$	   0.4
&1.6 $\pm$    0.2
& \\
&4.2 $\pm$    0.8
&2.8 $\pm$	   0.5
&2.0 $\pm$	   0.3
&0.8 $\pm$    0.1
& \\ 
0.65 &   6.3 $\pm$    0.7
&4.7 $\pm$	   0.4
&2.7 $\pm$	   0.2
&1.2 $\pm$    0.1
& \\
&4.3 $\pm$    0.7
&2.6 $\pm$	   0.4
&1.9 $\pm$	   0.3
&0.9 $\pm$    0.1
& \\
0.695 &&&&&   0.14 $\pm$   0.02 \\
 &&&&&   0.14 $\pm$    0.03 \\ 
0.75 &   4.4 $\pm$    0.5
&4.0 $\pm$	   0.4
&2.2 $\pm$	   0.2
&0.90 $\pm$    0.08
& \\
&3.6 $\pm$    0.5
&2.5 $\pm$	   0.3
&1.7 $\pm$	   0.2
&0.60 $\pm$    0.08
& \\ 
0.85 &   3.9 $\pm$    0.4
&3.3 $\pm$	   0.3
&1.9 $\pm$	   0.2
&0.75 $\pm$    0.07
& \\
&2.9 $\pm$    0.5
&2.5 $\pm$	   0.3
&1.8 $\pm$	   0.2
&0.62$\pm$    0.08
& \\ 
0.895  &&&&&  0.10$\pm$	   0.02 \\
  &&&&&  0.06$\pm$	   0.01 \\ 
1.00 &   3.1 $\pm$    0.2
&2.4 $\pm$	   0.2
&1.37 $\pm$	   0.09
&0.46 $\pm$    0.03
& \\
&2.1 $\pm$    0.2
&1.5 $\pm$	   0.1
&0.91 $\pm$	   0.08
&0.41 $\pm$    0.04
& \\ 
1.18  &&&&&  0.031$\pm$	   0.005 \\
  &&&&&  0.018$\pm$	   0.004 \\ 
1.20 &   2.0 $\pm$    0.2
&1.4 $\pm$	   0.1
&0.82 $\pm$	   0.06
&0.28 $\pm$    0.02
& \\
&1.4$\pm$    0.2
&1.0 $\pm$	   0.1
&0.54 $\pm$	   0.05
&0.19 $\pm$    0.02
& \\ 
1.40 &   1.1 $\pm$    0.1
&0.74 $\pm$	   0.07
&0.46 $\pm$	   0.04
&0.14 $\pm$    0.01
& \\
&0.9 $\pm$    0.1
&0.50 $\pm$	   0.06
&0.32 $\pm$	   0.04
&0.13 $\pm$    0.02
& \\ 
1.58 &&&&&   0.005$\pm$	   0.002 \\
 &&&&&   0.007$\pm$	   0.002 \\ 
1.60 &   0.54 $\pm$    0.07
&0.49 $\pm$	   0.05
&0.25 $\pm$	   0.03
&0.09 $\pm$    0.01
& \\
&0.49 $\pm$    0.08
&0.37 $\pm$	   0.05
&0.20 $\pm$	   0.03
&0.053 $\pm$    0.009
& \\ 
1.80 &   0.34 $\pm$    0.05
&0.25 $\pm$	   0.03
&0.16 $\pm$	   0.02
&0.047 $\pm$    0.007
& \\
&0.27 $\pm$    0.06
&0.11 $\pm$	   0.02
&0.10 $\pm$	   0.02
&0.021 $\pm$    0.005
& \\
1.98 &&&&&   0.003 $\pm$	   0.001 \\
 &&&&&   0.0005 $\pm$	   0.0005 \\ 
2.00 &   0.20 $\pm$    0.04
&0.12 $\pm$	   0.02
&0.10 $\pm$	   0.01
&0.029 $\pm$    0.005
& \\
&0.16 $\pm$    0.04
&0.08 $\pm$	   0.02
&0.05 $\pm$	   0.01
&0.026 $\pm$    0.006
& \\ 
2.20 &   0.13 $\pm$    0.03
&0.08 $\pm$	   0.02
&0.05 $\pm$	   0.01
&0.008 $\pm$    0.002
& \\
&0.04 $\pm$    0.02
&0.03 $\pm$	   0.01
&0.025 $\pm$	   0.007
&0.010 $\pm$    0.003
& \\ 
2.41 &&&&&   0.0010 $\pm$	   0.0006 \\
 &&&&&   0.0020 $\pm$	   0.0008 \\ 
2.425 &   0.06 $\pm$    0.02
&0.05 $\pm$	   0.01
&0.017 $\pm$	   0.005
&0.007 $\pm$    0.002
& \\
&0.05 $\pm$    0.02
&0.04 $\pm$	   0.01
&0.011 $\pm$	   0.004
&0.006$\pm$    0.002
& \\ 
2.675 &   0.014 $\pm$    0.008
&0.019 $\pm$	   0.006
&0.010 $\pm$	   0.003
&0.004 $\pm$    0.001
& \\
&0.005 $\pm$    0.005
&0.018     0.007
&0.002     0.002
&0.0012 $\pm$    0.0008
& \\ 
2.908 &&&&&   0.0006 $\pm$	   0.0004\\
 &&&&&   0.0005 $\pm$	   0.0003 \\ 
2.925 &   0.017 $\pm$    0.007
&0.006 $\pm$	   0.003
&0.006 $\pm$	   0.002
&0.003 $\pm$    0.001
& \\
&0.020 $\pm$    0.009
&0.009 $\pm$	   0.004
&0.003 $\pm$	   0.002
&0.0009 $\pm$    0.0006
& \\ 
3.175 &   0.013 $\pm$    0.006
&0.002 $\pm$	   0.002
&0.003 $\pm$	   0.002
&0.0012 $\pm$    0.0007
& \\
&0.006 $\pm$    0.004
&0.003 $\pm$	   0.002
&0.0010 $\pm$	   0.001
&
& \\ 
3.425 &   0.004 $\pm$    0.003
&0.003 $\pm$	   0.002
&0.002 $\pm$	   0.001
&0.0007 $\pm$    0.0005
& \\
&0.002 $\pm$    0.002
&0.001 $\pm$	   0.001
&0.002 $\pm$	   0.001
&0.0003 $\pm$    0.0003
& \\ 
3.675 &   0.003 $\pm$    0.003
&
&
&0.0003 $\pm$    0.0003
& \\
&0.003 $\pm$    0.003
&
&0.0008 $\pm$     0.0008
&
& \\ 
\end{tabular}
\end{ruledtabular}
\end{table*}


\begin{table*}
\caption{\label{Tab:minBiasSameBin}
Minimum bias invariant yields for all particles in equal $p_{T}$ bins.  
For each $p_{T}$, the first line are the positive particle yields, and 
the second are the negative particle yields.  The units are $c^{2}/GeV^{2}$.}
\begin{ruledtabular}
\begin{tabular} {cccc}
$p_{T} (GeV/c)$ & $\pi^{\pm}$ & $K^{\pm}$ & (anti)p \\ \hline
0.25 & 112$\pm$2&  & \\
     & 109$\pm$2& & \\ 
0.35 & 56 $\pm$1& & \\
     & 49.9 $\pm$0.9& & \\ 
0.45 & 28.0 $\pm$0.5& 6.1$\pm$0.4 & \\
     & 24.1 $\pm$0.5& 4.6$\pm$0.4 & \\
0.55 & 15.7 $\pm$0.3& 4.0$\pm$0.3 & 2.3$\pm$0.1 \\
     & 14.6 $\pm$0.3& 3.2$\pm$0.2 & 0.38$\pm$0.02 \\
0.70 & 7.3  $\pm$0.1& 2.18$\pm$0.09 & 1.55$\pm$0.06 \\
     & 7.0  $\pm$0.1& 1.9$\pm$0.1 & 1.07$\pm$0.06 \\
0.90 & 3.06  $\pm$0.06& 1.07$\pm$0.05 & 1.08$\pm$0.04 \\
     & 2.89  $\pm$0.07& 0.91$\pm$0.05 & 0.79$\pm$0.04 \\
1.20 & 0.91  $\pm$0.02& 0.38$\pm$0.02 & 0.49$\pm$0.02 \\
     & 0.98  $\pm$0.02& 0.32$\pm$0.02 & 0.35$\pm$0.01 \\
1.60 & 0.208  $\pm$0.007& 0.104$\pm$0.006 & 0.157$\pm$0.007 \\
     & 0.193  $\pm$0.007& 0.093$\pm$0.006 & 0.119$\pm$0.007 \\
2.00 & 0.050  $\pm$0.003&                       & 0.051$\pm$0.003 \\
     & 0.053  $\pm$0.003&                       & 0.031$\pm$0.003 \\
2.45 & 0.0028 $\pm$0.0005  &                       & 0.013$\pm$0.001 \\
     & 0.0034 $\pm$0.0006  &                       & 0.009$\pm$0.001 \\
2.95 &                         &                       & 0.0036$\pm$0.0006 \\
     &                         &                       & 0.0022$\pm$0.0005 \\
3.55 &                         &                       & 0.0007$\pm$0.0002 \\
     &                         &                       & 0.0006$\pm$0.0002 \\ 
\end{tabular}
\end{ruledtabular}
\end{table*}

\begin{table*}
\caption{\label{pionsSameBin}
Pion invariant yields in each event centrality normalized to one rapidity unit at midrapidity.  
The first line corresponds to positive pions, and the second to negative pions.}
\begin{ruledtabular}
\begin{tabular} {cccccc} 
$p_{T}$ (GeV/c) & 0-5\% & 5-15\% & 15-30\% & 30-60\% & 60-92\% \\ \hline
0.25& 355$\pm$9
    & 282$\pm$6 
    & 186$\pm$4
    & 81$\pm$2 
    & 13.2$\pm$0.5 \\
    & 371$\pm$10 
    & 275$\pm$7 
    & 180$\pm$	  4
    & 74 $\pm$	 2
    & 12.1 $\pm$	 0.5\\
0.35& 188$\pm$5 
    & 146$\pm$3 
    & 93$\pm$	 2
    & 36.6$\pm$	 0.8
    & 5.3 $\pm$	0.2\\
    & 169$\pm$5 
    & 128$\pm$3
    & 82$\pm$	 2
    & 34.3$\pm$	 0.9
    & 5.0 $\pm$	0.2\\
 
0.45& 95$\pm$3
    & 74$\pm$2
    & 48$\pm$	 1
    & 17.5$\pm$	 0.5
    & 2.7 $\pm$	0.1\\
    & 86$\pm$3
    & 63$\pm$2
    & 40$\pm$	 1
    & 15.7$\pm$	 0.5
    & 2.1 $\pm$	0.1\\
0.55& 56$\pm$2
    & 41$\pm$1
    & 26.0$\pm$	 0.7
    & 10.1$\pm$	 0.3
    & 1.32 $\pm$	0.09\\
    & 51$\pm$2
    & 38$\pm$1
    & 24.5$\pm$	 0.8
    & 9.6 $\pm$	0.3
    & 1.18 $\pm$	0.09\\
0.70& 26.3$\pm$0.8
    & 20.2$\pm$0.5
    & 12.3$\pm$	 0.3
    & 4.4 $\pm$	0.1
    & 0.62 $\pm$	0.04\\
    & 24.7$\pm$0.9
    & 18.6$\pm$0.6
    & 11.8$\pm$	 0.4
    & 4.5 $\pm$	0.1
    & 0.57 $\pm$	0.04\\
0.90& 11.0$\pm$0.4
    & 7.9 $\pm$0.2
    & 5.1 $\pm$	0.2
    & 1.79 $\pm$	0.06
    & 0.19 $\pm$	0.02\\
    & 10.0$\pm$0.5
    & 7.9 $\pm$0.3
    & 5.0 $\pm$	0.2
    & 1.76 $\pm$	0.07
    & 0.24 $\pm$	0.02\\
1.20& 3.1$\pm$0.1
    & 2.37 $\pm$0.09
    & 1.50 $\pm$	0.05
    & 0.58 $\pm$	0.02
    & 0.064 $\pm$	0.006\\
    & 3.4$\pm$0.2
    & 2.7 $\pm$0.1
    & 1.67 $\pm$	0.07
    & 0.64 $\pm$	0.03
    & 0.061 $\pm$	0.007 \\
1.60& 0.62$\pm$0.05
    & 0.54 $\pm$0.03
    & 0.34 $\pm$	0.02
    & 0.142 $\pm$	0.009
    & 0.015  $\pm$	0.003\\
    & 0.63$\pm$0.06
    & 0.58 $\pm$0.04
    & 0.32 $\pm$	0.02
    & 0.129 $\pm$	0.009
    & 0.014 $\pm$	0.003\\
2.00& 0.17$\pm$0.02
    & 0.14 $\pm$0.01
    & 0.083 $\pm$	0.009
    & 0.027 $\pm$	0.003
    & 0.005 $\pm$	0.001\\ 
    & 0.20$\pm$0.03
    & 0.14 $\pm$0.02
    & 0.09 $\pm$	0.01
    &0.035  $\pm$   0.004
    & 0.004 $\pm$	0.001\\
2.45& 0.005 $\pm$0.003
    & 0.009 $\pm$0.003
    & 0.006 $\pm$	0.002
    & 0.0017 $\pm$	0.0006
    & \\
    & 0.011$\pm$0.005
    & 0.011 $\pm$0.004
    & 0.005 $\pm$	0.002
    & 0.0025 $\pm$	0.0009
    & \\ 
\end{tabular}
\end{ruledtabular}
\end{table*}

\begin{table*}
\caption{\label{kaonSameBin}
Kaon invariant yields in each event centrality normalized to one rapidity unit at midrapidity.  
The first line corresponds to positive kaons, and the second to negative kaons.}
\begin{ruledtabular}
\begin{tabular} {cccccc} 
$p_{T}$ (GeV/c) & 0-5\% & 5-15\% & 15-30\% & 30-60\% & 60-92\% \\ \hline
0.45 & 21$\pm$3
     &16$\pm$	 2
     &10$\pm$	 1
     &4.3 $\pm$	0.5
     &0.5 $\pm$	0.1 \\
     & 21$\pm$3
     & 13$\pm$	  2
     & 7 $\pm$	 1
     & 2.5 $\pm$	 0.4
     & 0.4 $\pm$	 0.1\\
0.55 & 15$\pm$2
     & 11$\pm$	  1
     & 6.6 $\pm$	 0.7
     & 2.4 $\pm$	 0.3
     & 0.20 $\pm$	 0.07\\
     & 13$\pm$2
     & 8 $\pm$	 1
     & 4.8 $\pm$	 0.6
     &2.3  $\pm$	0.3
     & 0.3 $\pm$	 0.1\\
0.70 & 8.0$\pm$0.7
     & 6.6 $\pm$	 0.5
     & 4.3 $\pm$	 0.3
     & 1.5 $\pm$	 0.1
     & 0.18 $\pm$	 0.03\\
     & 7.0$\pm$0.8
     & 6.5 $\pm$	 0.6
     & 3.1 $\pm$	 0.3
     & 1.1 $\pm$	 0.1
     & 0.14 $\pm$	 0.03\\
0.90 & 4.5$\pm$0.4
     & 3.1 $\pm$	 0.2
     & 1.9 $\pm$	 0.1
     & 0.62 $\pm$	 0.06
     & 0.06 $\pm$	 0.02\\
     & 3.3$\pm$0.4
     & 3.3 $\pm$	 0.3
     & 1.6 $\pm$	 0.2
     & 0.5 $\pm$	 0.06
     & 0.09 $\pm$	 0.02\\
1.20 & 1.4$\pm$0.1
     & 1.10 $\pm$	 0.08
     & 0.68 $\pm$	 0.05
     & 0.22 $\pm$	 0.02
     & 0.012 $\pm$	 0.004\\
     & 1.3$\pm$0.1
     & 0.97 $\pm$	 0.08
     & 0.50 $\pm$	 0.05
     & 0.17 $\pm$	 0.02
     & 0.015 $\pm$	 0.005\\
1.60 & 0.36$\pm$0.05
     & 0.29 $\pm$	 0.03
     & 0.17 $\pm$	 0.02
     & 0.062 $\pm$	 0.007
     & 0.008 $\pm$	 0.002\\
     & 0.29$\pm$0.05 
     & 0.27 $\pm$	 0.03
     & 0.20 $\pm$	 0.02
     & 0.058 $\pm$	 0.008
     & 0.004 $\pm$	 0.002\\ 
\end{tabular}
\end{ruledtabular}
\end{table*}

\begin{table*}
\caption{\label{protSameBin}
(Anti)proton invariant yields in each event centrality normalized to one rapidity unit at midrapidity.  
The first line corresponds to protons, and the second to antiprotons.}
\begin{ruledtabular}
\begin{tabular} {cccccc} 
$p_{T}$ (GeV/c) & 0-5\% & 5-15\% & 15-30\% & 30-60\% & 60-92\% \\ \hline
0.55 & 8$\pm$1
     & 4.9 $\pm$	 0.5
     & 4.0 $\pm$	 0.4
     & 1.6 $\pm$	 0.2
     & 0.26 $\pm$	 0.06\\
     & 4.2$\pm$0.8
     & 2.8$\pm$	 0.5
     & 2.0$\pm$	 0.3
     & 0.8$\pm$	 0.1
     & 0.12$\pm$	 0.05\\
0.70 & 5.4$\pm$0.4
     & 4.5$\pm$	 0.3
     & 2.5$\pm$	 0.2
     & 1.06$\pm$	 0.07
     & 0.14$\pm$	 0.02\\
     & 4.4$\pm$0.5
     & 2.9$\pm$	 0.3
     & 2.0$\pm$	 0.2
     & 0.80$\pm$	 0.08
     & 0.14$\pm$	 0.03\\
0.90 & 3.9$\pm$0.3
     & 3.1$\pm$	 0.2
     & 1.9$\pm$	 0.1
     & 0.71$\pm$	 0.05
     & 0.10$\pm$	 0.02\\
     & 2.8$\pm$0.3
     & 2.1$\pm$	 0.2
     & 1.4$\pm$	 0.1
     & 0.58$\pm$	 0.05
     & 0.06$\pm$	 0.01\\
1.20 & 1.9$\pm$0.1
     & 1.37$\pm$	 0.08
     & 0.78$\pm$	 0.04
     & 0.26$\pm$	 0.02
     & 0.031$\pm$	 0.005\\
     & 1.3$\pm$0.1
     & 0.96$\pm$	 0.07
     & 0.56$\pm$	 0.04
     & 0.21$\pm$	 0.02
     & 0.018$\pm$	 0.004\\
1.60 & 0.60$\pm$0.06 
     & 0.44$\pm$	 0.03
     & 0.27$\pm$	 0.02
     & 0.087$\pm$	 0.008
     & 0.005$\pm$	 0.002\\
     & 0.49$\pm$0.06
     & 0.34$\pm$	 0.03
     & 0.19$\pm$	 0.02
     & 0.062$\pm$	 0.007
     & 0.007$\pm$	 0.002\\
2.00 & 0.20$\pm$0.03
     & 0.15$\pm$	 0.02
     & 0.09$\pm$	 0.01
     & 0.025$\pm$	 0.003
     & 0.003$\pm$	 0.001\\
     & 0.15$\pm$0.03
     & 0.07$\pm$	 0.01
     & 0.055$\pm$	 0.009
     &0.019 $\pm$	0.003
     & 0.0005$\pm$	 0.0005\\
2.45 & 0.06$\pm$0.01
     & 0.040$\pm$	 0.007
     & 0.020 $\pm$	 0.004
     & 0.006$\pm$	 0.001
     & 0.0010$\pm$	 0.0006\\
     & 0.04$\pm$0.01
     & 0.028$\pm$	 0.006
     & 0.011$\pm$	 0.003
     & 0.005$\pm$	 0.001
     & 0.0020 $\pm$	 0.0008\\
2.95 & 0.015$\pm$0.005
     & 0.007$\pm$	 0.002
     & 0.005$\pm$	 0.002
     & 0.0023$\pm$	 0.0007
     & 0.0006$\pm$	 0.0004\\
     & 0.013$\pm$0.005
     & 0.008$\pm$	 0.003
     & 0.002$\pm$	 0.001
     & 0.0003$\pm$	 0.0003
     & 0.0005$\pm$	 0.0003\\
3.55 & 0.003 $\pm$0.002
     & 0.0012 $\pm$	 0.0007
     & 0.0014$\pm$	 0.0006
     & 0.0005$\pm$	 0.0002
     & \\
     & 0.002$\pm$0.001
     & 0.0011$\pm$	 0.0007
     & 0.0013$\pm$	 0.0006
     & 0.0002$\pm$	 0.0002
     & \\ 
\end{tabular}
\end{ruledtabular}
\end{table*}
 
\section{\label{deriv}Freeze-out Surface Assumptions}

The freeze-out surface is $\sigma\left(r,\phi,\eta \right)$,  where
the radius r is between zero and R, the radius  at freeze-out, the
azimuthal angle $\phi$ is between  zero and $2\pi$, and the
longitudinal space-time rapidity  variable $\eta$ varies between
$-\eta_{max}$ and $\eta_{max}$.  In the Bjorken scenario, the
freeze-out surface in space-time is hyperbolic,  with contours of
constant proper time  $\tau = \sqrt{t^2 - z^2}$.  Assuming
instantaneous  freeze-out in the radial direction and longitudinal
boost-invariance, the model-dependence factors out of Equation
~\ref{hydro} and is included in the normalization constant A.

At 130 GeV, the PHOBOS experiment measures the total  charged particle
pseudorapidity distribution to be  flat over 2 units of pseudorapidity
\cite{phobos_dndeta}.  The measured rapidity in PHOBOS is taken to be
the same as the  rapidity of the fireball, defined here as z.  The
rapidity  variables in the integrand vanish for $|z|>2$.  Therefore,
the integration over the fireball rapidity is  generally taken to be
from $-\infty$ to $+\infty$ using  the modified $K_{1}$ Bessel
function
\begin{equation}
K_{1}(m_{T}/T) = \int_{0}^{\infty} cosh(z) e^{-m_{T} cosh(z)/T} dz
  \end{equation}	 where the variable z is the fireball rapidity
  variable.
The $K_1$ bessel function can also result by  integration over the
measured rapidity y with the assumption that  the freeze-out is
instantaneous in the radial direction.  In this  case, no assumption
is made on the shape of the freeze-out  hypersurface.  This also
assumes that the total rapidity  distribution is measured in the
detector.  What results is the  single differential 1/$m_T$
dN/$dm_T$.\footnote{Private communication with U. Heinz.}


\begin{acknowledgments}

We thank the staff of the RHIC Project, Collider-Accelerator,
and Physics Departments at Brookhaven National Laboratory and
the staff of the other PHENIX participating institutions for
their vital contributions.  We acknowledge support from the
Department of Energy, Office of Science, Nuclear Physics
Division, the National Science Foundation, and Dean of the
College of Arts and Sciences, Vanderbilt University (U.S.A),
Ministry of Education, Culture, Sports, Science, and Technology
and the Japan Society for the Promotion of Science (Japan),
Russian Academy of Science, Ministry of Atomic Energy of
Russian Federation, Ministry of Industry, Science, and
Technologies of Russian Federation (Russia), Bundesministerium
fuer Bildung und Forschung, Deutscher Akademischer
Auslandsdienst, and Alexander von Humboldt Stiftung (Germany),
VR and the Wallenberg Foundation (Sweden), MIST and the Natural
Sciences and Engineering Research Council (Canada), Conselho
Nacional de Desenvolvimento Cient\'{\i}fico e Tecnol\'ogico and
Funda\c c\~ao de Amparo \`a Pesquisa do Estado de S\~ao Paulo
(Brazil), Natural Science Foundation of China (People's
Republic of China), Centre National de la Recherche
Scientifique, Commissariat {\`a} l'{\'E}nergie Atomique,
Institut National de Physique Nucl{\'a}ire et de Physique des
Particules, and Association pour la Recherche et le
D{\'e}veloppement des M{\'e}thodes et Processus Industriels
(France), Department of Atomic Energy and Department of Science
and Technology (India), Israel Science Foundation (Israel), 
Korea Research Foundation and Center
for High Energy Physics (Korea), the U.S. Civilian Research and
Development Foundation for the Independent States of the Former
Soviet Union, and the US-Israel Binational Science Foundation.
\end{acknowledgments}

\clearpage


\bibliography{ppg009}

\end{document}